\documentclass[twocolumn]{aastex63}
\usepackage{multirow,amsmath,amsthm,mathtools}
\usepackage{chngcntr,amssymb,cancel}
\usepackage[normalem]{ulem}
\usepackage{natbib}

\received{2020}
\revised{2020}
\accepted{}

\submitjournal{ApJ}
\shorttitle{Magnetic fields surrounding LkH$\alpha$ 101}
\shortauthors{Ngoc et al.}

\graphicspath{{./}{figures/}}

\begin{document}

\title{\LARGE{\textbf{Observations of magnetic fields surrounding LkH$\alpha$ 101 taken by the BISTRO survey with JCMT-POL-2}}}

\correspondingauthor{Pham Ngoc Diep}
\email{pndiep@vnsc.org.vn}

\author[0000-0002-5913-5554]{Nguyen Bich Ngoc}
\affiliation{Vietnam National Space Center, Vietnam Academy of Science and Technology, 18 Hoang Quoc Viet, Hanoi, Vietnam}

\author[0000-0002-2808-0888]{Pham Ngoc Diep}
\affiliation{Vietnam National Space Center, Vietnam Academy of Science and Technology, 18 Hoang Quoc Viet, Hanoi, Vietnam}

\author[0000-0002-6327-3423]{Harriet Parsons}
\affiliation{East Asian Observatory, 660 N. A'oh\={o}k\={u} Place, University Park, Hilo, HI 96720, USA}

\author[0000-0002-8557-3582]{Kate Pattle}
\affiliation{National University of Ireland Galway, University Road, Galway, Ireland H91 TK33}

\author[0000-0003-2017-0982]{Thiem Hoang}
\affiliation{Korea Astronomy and Space Science Institute, 776 Daedeokdae-ro, Yuseong-gu, Daejeon 34055, Republic of Korea}
\affiliation{University of Science and Technology, Korea, 217 Gajeong-ro, Yuseong-gu, Daejeon 34113, Republic of Korea}

\author[0000-0003-1140-2761]{Derek Ward-Thompson}
\affiliation{Jeremiah Horrocks Institute, University of Central Lancashire, Preston PR1 2HE, UK}

\author[0000-0002-6488-8227]{Le Ngoc Tram}
\affiliation{University of Science and Technology of Hanoi, Vietnam Academy of Science and Technology, 18 Hoang Quoc Viet, Hanoi, Vietnam}

\author[0000-0002-8975-7573]{Charles L. H. Hull}
\affiliation{National Astronomical Observatory of Japan, NAOJ Chile, Alonso de C\'ordova 3788, Office 61B, 7630422, Vitacura, Santiago, Chile}
\affiliation{Joint ALMA Observatory, Alonso de C\'ordova 3107, Vitacura, Santiago, Chile}
\affiliation{NAOJ Fellow}

\author[0000-0001-8749-1436]{Mehrnoosh Tahani}
\affiliation{Dominion Radio Astrophysical Observatory, Herzberg Astronomy and Astrophysics Research Centre, National Research Council Canada, P. O. Box 248, Penticton, BC V2A 6J9 Canada}

\author{Ray Furuya}
\affiliation{Tokushima University, Minami Jousanajima-machi 1-1, Tokushima 770-8502, Japan}
\affiliation{Institute of Liberal Arts and Sciences Tokushima University, Minami Jousanajima-machi 1-1, Tokushima 770-8502, Japan}

% ...

\begin{abstract}
We report the first high spatial resolution measurement of magnetic fields surrounding LkH$\alpha$ 101, a part of the Auriga-California molecular cloud. The observations were taken with the POL-2 polarimeter on the James Clerk Maxwell Telescope within the framework of the B-fields In Star-forming Region Observations (BISTRO) survey. Observed polarization of thermal dust emission at 850 $\mu$m is found to be mostly associated with the red-shifted gas component of the cloud. The magnetic field displays a relatively complex morphology. Two variants of the Davis-Chandrasekhar-Fermi method, unsharp masking and structure function, are used to calculate the strength of magnetic fields in the plane of the sky, yielding a similar result of $B_{\rm POS}\sim 115$ $\mathrm{\mu}$G. The mass-to-magnetic-flux ratio in critical value units, $\lambda\sim0.3$, is the smallest among the values obtained for other regions surveyed by POL-2. This implies that the LkH$\alpha$ 101 region is sub-critical and the magnetic field is strong enough to prevent gravitational collapse. The inferred $\delta B/B_0\sim 0.3$ implies that the large scale component of the magnetic field dominates the turbulent one. The variation of the polarization fraction with total emission intensity can be fitted by a power-law with an index of $\alpha=0.82\pm0.03$, which lies in the range previously reported for molecular clouds. We find that the polarization fraction decreases rapidly with proximity to the only early B star (LkH$\alpha$ 101) in the region. The magnetic field tangling and the joint effect of grain alignment and rotational disruption by radiative torques are potential of explaining such a decreasing trend.
\end{abstract}


\section{Introduction}\label{sec:intro}
Several factors are thought to play an important role in the formation and evolution of interstellar clouds and protostars, including magnetic fields (B-fields) and turbulence. However, their precise roles, in particular at different stages of the cloud evolution, are not well-understood. Models of cloud and star formation, such as the weak-field and strong-field models \citep{crutcher2012}, describe the different roles played by magnetic fields and turbulence. More observational constraints are required to test the proposed models and, therefore, to better understand the relative importance of magnetic fields and turbulence.

The alignment of dust grains with the magnetic field induces polarization of the light from background stars \citep{hall1949, hiltner1949} and of thermal dust emission \citep{hildebrand1988}. The polarization vectors of background starlight are parallel to the magnetic field, while those of thermal dust are perpendicular to the magnetic field. Thus, dust polarization has become a popular technique to measure the projected magnetic field direction and strength \citep{lazarian2007, andersson2015}.

B-fields In Star-forming Region Observations (BISTRO) is a large program of the James Clerk Maxwell Telescope (JCMT) that aims at mapping the magnetic fields in star-forming regions. BISTRO-1 looks at scales of 1,000 to 5,000 au within dense cores and filaments of the star-forming regions of the part of the sky known informally as the `Gould Belt' \citep{ward2017}. It has recently been speculated \citep{alves2020} that the Gould Belt is not a single, homogeneous ring, but rather a large-scale `gas wave' in the inter-stellar medium. Whatever the origin of the structure, it remains convenient to refer to the band of star-forming regions seen across the sky as the `Gould Belt', regardless of the cause of this band. The observations were carried out using the polarimeter, POL-2, placed in front of the JCMT Sub-millimeter Common User Bolometer Array-2 (SCUBA-2) \citep{holland2013}.

In this work, we study the magnetic fields in the densest region of the Auriga-California molecular cloud (AMC) around LkH$\alpha$ 101 using new data taken by POL-2. The observed region is identical to the one labeled LkH$\alpha$ 101-S (S stands for South) in Figure 1 of \citet{broekhoven2018}. Auriga-California is part of the Gould Belt with a total mass of \mbox{$\sim10^5$ \(\textup{M}_\odot\)}. It is located at $\sim 466\pm23$ pc away from Earth \citep{zucker2020}, with a spatial extent of 80 pc \citep{lada2009}. Using data from SCUBA-2 at 450 and \mbox{850 $\mu$m}, \citet{broekhoven2018} found that Auriga-California has 59 candidate protostars, out of which 35 in the LkH$\alpha$ 101 region. Auriga-California was also observed with \textit{Herschel}/PACS at 70 and \mbox{160 $\mu$m}, \textit{Herschel}/SPIRE at 250, 350, and 500 $\mu$m and by the Caltech Sub-millimeter Observatory (CSO) at 1.1 mm \citep{harvey2013}.

Together with the Orion Molecular Cloud (OMC), Auriga-California is one of the two nearby giant molecular clouds in the Gould Belt. Although being similar in size, mass, and distance, Auriga-California is very different from Orion. Additionally, observations by \cite{tahani2018} show that both Orion-A and Auriga-California have the same line-of-sight magnetic field morphology associated with them on large scales. Auriga-California is forming less massive stars and about twenty times fewer stars than the OMC. The OMC has 50 OB stars while Auriga-California has only one early B star, LkH$\alpha$ 101, a member of the embedded cluster in NGC 1579 \citep{herbig2004}. Using an H$_2$O maser associated with L1482 where LkH$\alpha$ 101 is located, \cite{omodaka2020} measured parallax corresponding to the distance to the filament of \mbox{532$\pm$28 pc}. Meanwhile, the distance to the B star LkH$\alpha$ 101 is estimated to be 567$\pm$68 pc with Gaia DR2 \citep{gaiadr2}. Therefore, measuring magnetic fields in Auriga-California is of particular interest to shed light on what governs the star formation efficiency in molecular clouds. The main purpose of the present BISTRO paper is to use polarization data taken with POL-2 to measure the strength and characterize the morphology of the magnetic field in the region.

The structure of the paper is as follows: in Section \ref{sec:obs}, we describe the observations, in Section \ref{sec:dis}, we discuss the extraction of polarization angle dispersion, velocity dispersion, and densities needed for estimating the magnetic field strengths. In Section \ref{sec:rslt}, we present the main results and interpretation of observational data. Our conclusions are presented in Section \ref{sec:conclusions}.

\section{Observations} \label{sec:obs}
\subsection{Data description and selection}

\begin{figure*}
\centering
\includegraphics[trim=0cm 0cm 0cm 0cm,width=6.5cm]{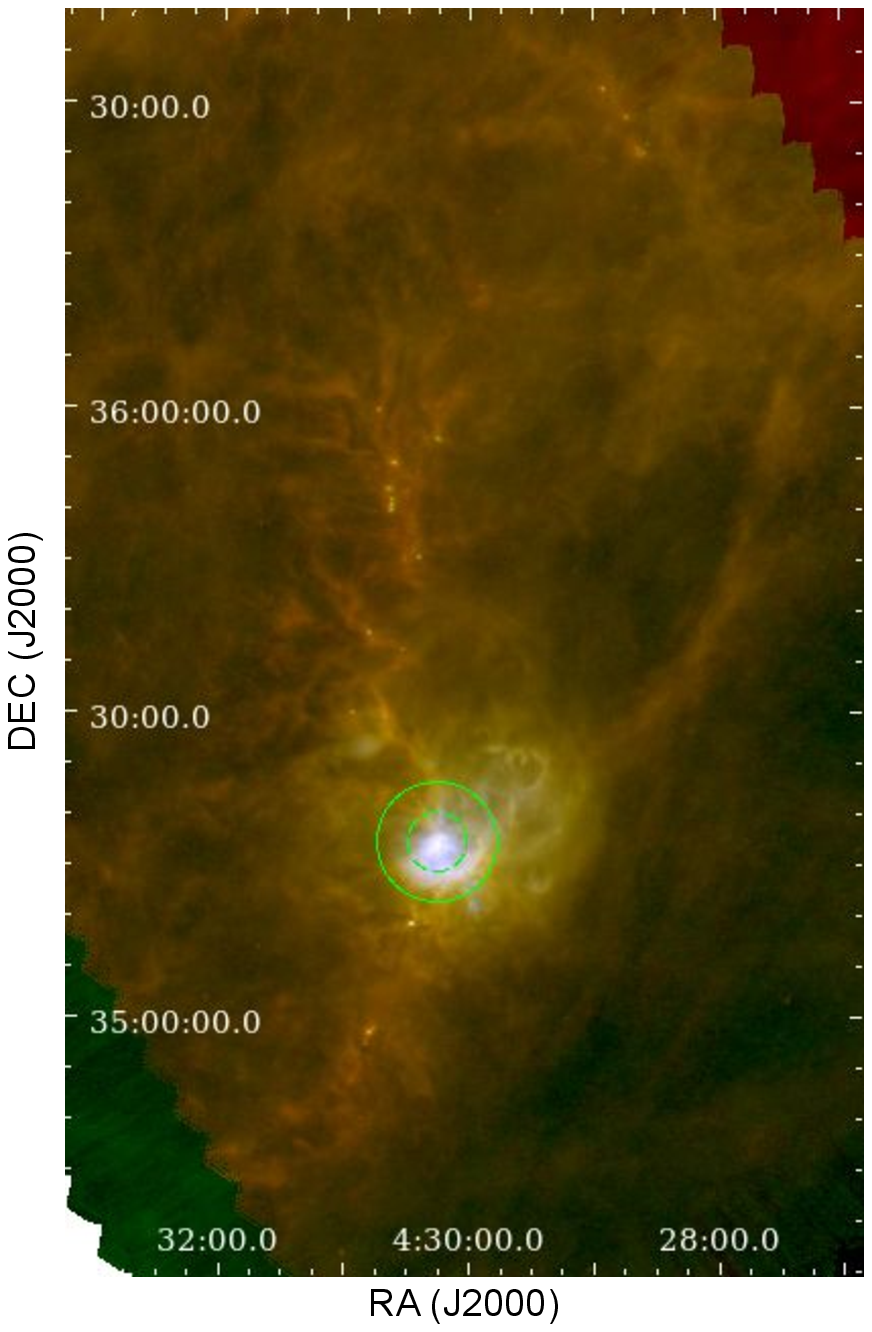}
\includegraphics[trim=1.cm 0.2cm 3.4cm 0.2cm,clip,width=8.5cm]{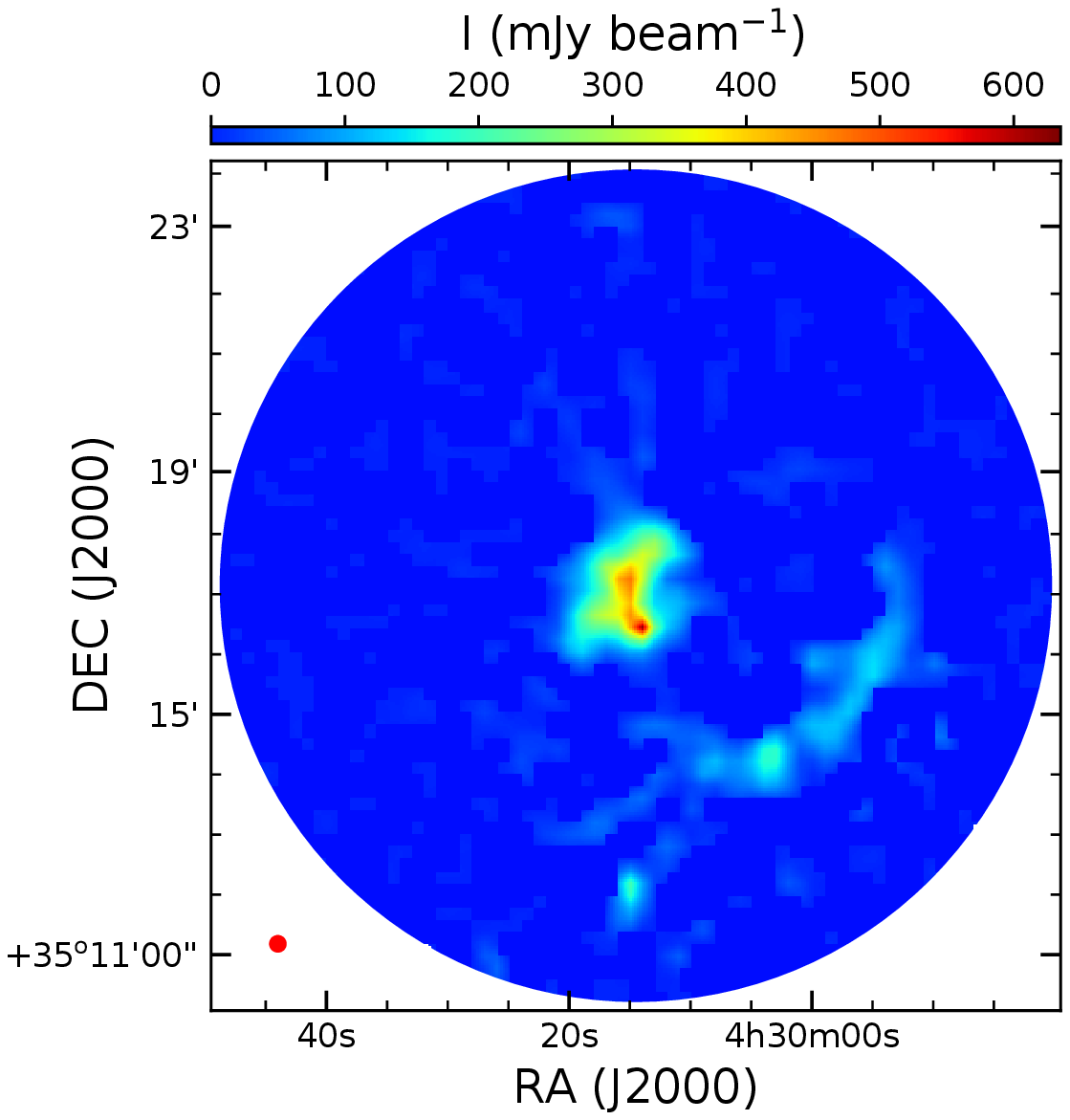}
\caption{Left: The LkH$\alpha$ 101 field observed by POL-2 overlaid on the \textit{Herschel} RGB image (R=250 $\mu$m, G=160 $\mu$m, \mbox{B=70 $\mu$m}) \citep{harvey2013}. The green outer circle represents the $6'$ radius field of POL-2 and the green inner circle indicates the inner $3'$ radius with the best sensitivity. Right: 850 $\mu$m intensity map. The red circle at the lower left corner of the panel shows the JCMT 850 $\mu$m beam size of $14''$.}
\label{fig1}
\end{figure*}

JCMT is a 15-m diameter telescope situated near the summit of Maunakea. It is the largest single dish telescope working at sub-millimeter wavelengths between 1.4 and 0.4 mm. The present work studies the polarized emission received from the LkH$\alpha$ 101 region at 850 $\mu$m. The beam size at this wavelength is 14.1$''$. The JCMT detector, SCUBA-2, consists of four arrays, of $40\times32$ bolometers each, covering a sky solid angle of $\sim45'\times45'$. The BISTRO program has been allocated two observation campaigns, called BISTRO-1 and BISTRO-2, each having 224 hours to map the regions towards Auriga, IC 5146, Ophiuchus, L1689B, Orion A$\&$B, Perseus B1, NGC 1333, Serpens, Taurus B211/213, L1495, Serpens Aquila, M16, DR15, DR21, NGC 2264, NGC 6334, Mon R2, and Rosette. Recently, a third campaign, BISTRO-3, has been approved which focuses on mapping various massive clouds, some nearby prestellar cores and the Galactic Center clouds. General descriptions of the BISTRO survey and the measurement of magnetic fields have been reported for Orion A \citep{ward2017,pattle2017}, Ophiuchus A/B/C \citep{kwon2018,soam2018,liu2019,pattle2019}, IC 5146 \citep{wang2019}, Perseus B1 \citep{coude2019}, M16 \citep{pattle2018}, and NGC 1333 \citep{doi2020jcmt}.

LkH$\alpha$ 101 was the last region to be observed within the framework of the BISTRO-1 program. The data were taken over nine days between 2017 and 2019 with 21 visits and a total integration time of about 14 hours. The last two observations were made on January $\mathrm{8^{th}}$, 2019. Data were read and reduced using Starlink \citep{currie2014}. In this analysis, we use gridded data with pixel size of $12''\times12''$, similar to the telescope beam. The Stokes parameter $Q$, $U$, and $I$ time-streams are reduced using \textit{pol2map}, a POL-2-specific implementation of the iterative map-making procedure \textit{makemap} \citep{chapin2013}. The instrumental polarization (IP) is mainly caused by the wind-blind; it is corrected for by using the IP model determined from POL-2 measurements with the wind-blind in place from observations of Uranus \citep{friberg2018}. In each pixel, the total intensity, $I$, the Stokes parameters $Q$, $U$, and their uncertainties, $\delta I$, $\delta Q$, and $\delta U$, are provided. The POL-2 data presented in this paper are available at http://dx.doi.org/10.11570/20.0011

Figure \ref{fig1} (left) shows the location of the POL-2 observed field in the LkH$\alpha$ 101 area. The 850 $\mu$m intensity map, $I$-map, is shown in the right panel of the figure. The center of the map is at $\mathrm{RA=4h30m15s}$, $\mathrm{DEC=35^\circ17'05''}$. Emission is observed to come from a region at the center of the map and a lane of dust in the south-western direction. It matches the densest area of the region shown in the top left panel of Figure 2 of \citet{broekhoven2018}.

\begin{figure*}[hbt!]
\centering
\begin{tabular}{cc}
    \includegraphics[trim=1.cm 0.2cm 2.5cm 0.2cm,clip,width=7.5cm]{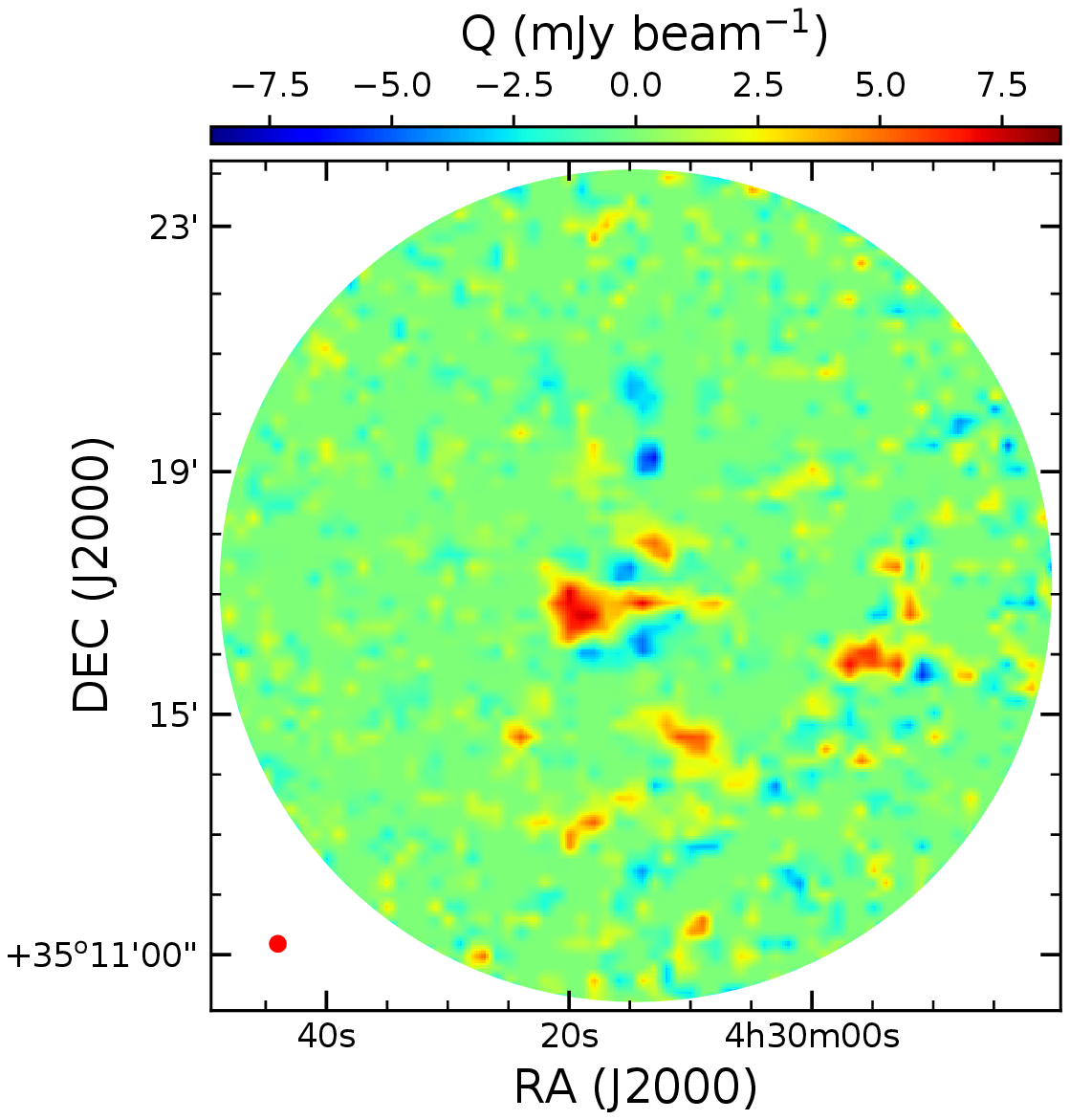}
    \includegraphics[trim=1.5cm 0.2cm 2.cm 0.2cm,clip,width=7.5cm]{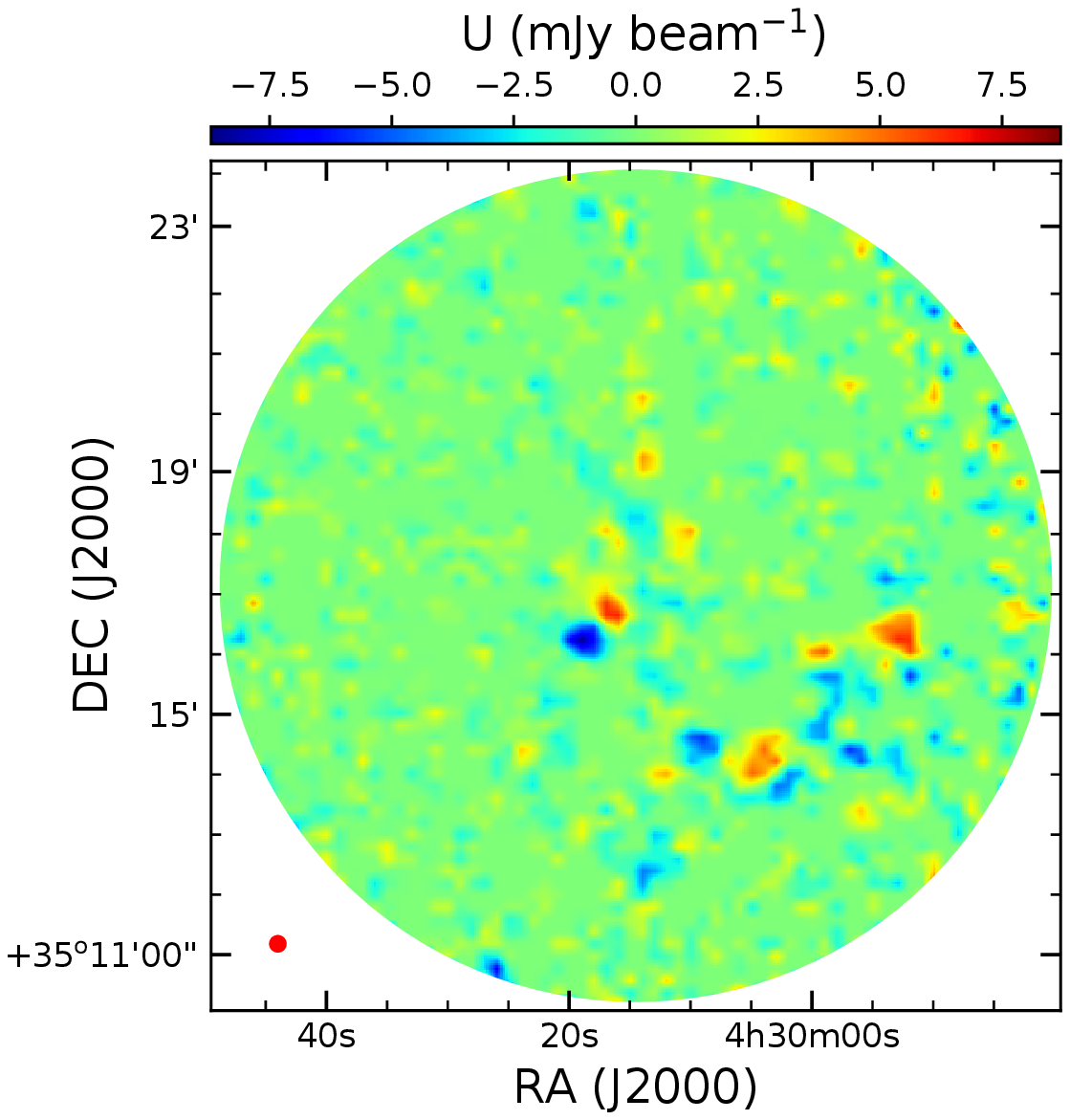}
\end{tabular}
\includegraphics[trim=1.cm 1.2cm 1.cm 1.75cm,clip,width=6.6cm]{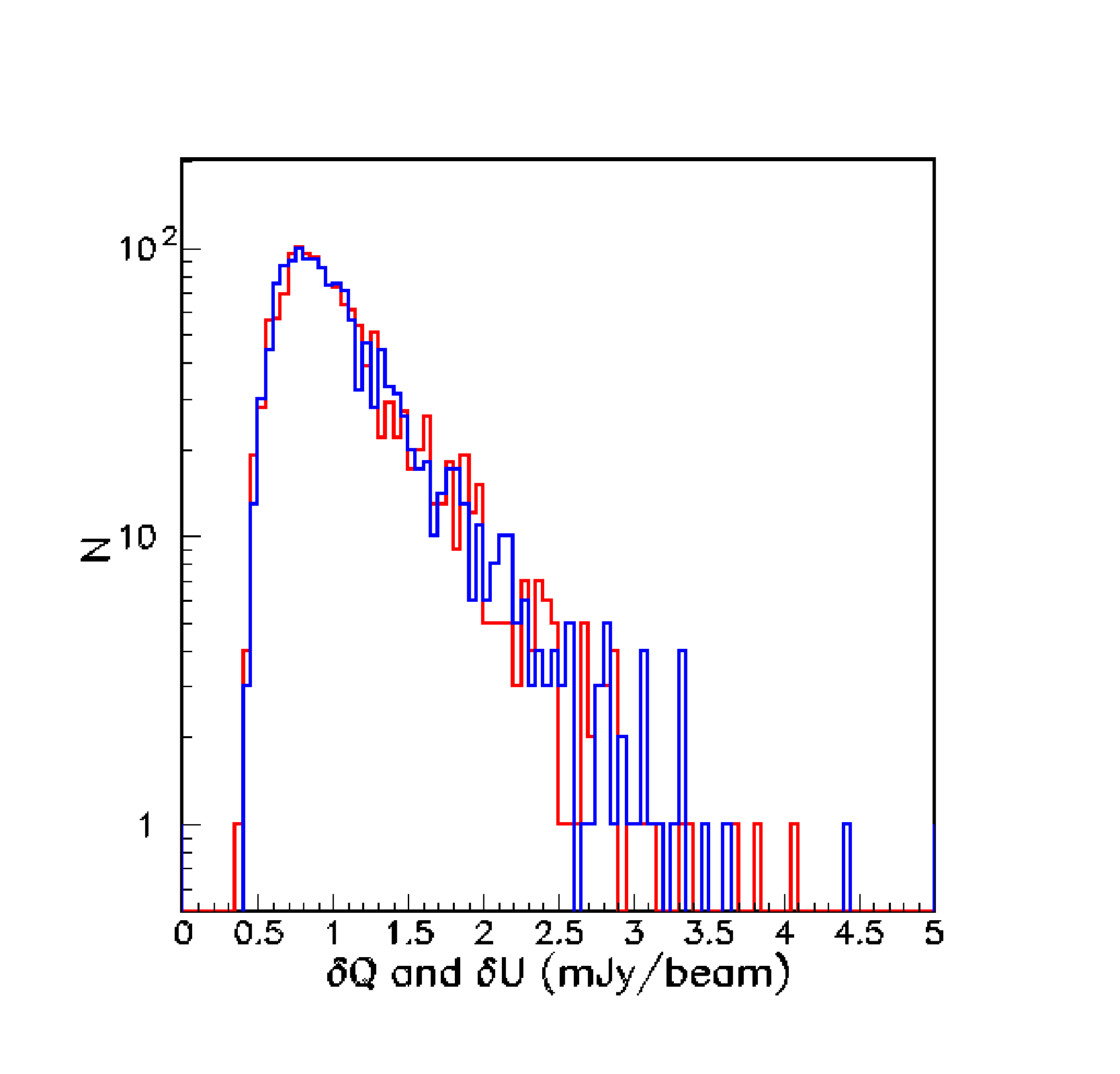}
\caption{Upper: $Q$ (left) and $U$ (right) maps; the red circles at the lower left corners indicate the JCMT beam size at 850 $\mu$m of $14''$. Lower: Distributions of $\delta Q$ (red) and $\delta U$ (blue).}\label{fig2}
\end{figure*}

Figure \ref{fig2} shows the maps of the measured Stokes $Q$ (upper left) and $U$ (upper right) and the distributions of their uncertainties, $\delta Q$ and $\delta U$ (lower), which are used to calculate other polarization parameters for further analysis. The distributions of $\delta Q$ and $\delta U$ have the same mean and RMS values of 1.1 mJy beam$^{-1}$ and 0.5 mJy beam$^{-1}$ respectively. The mean value of $\delta Q$ and $\delta U$ measured in the region is at the level expected for BISTRO survey (\mbox{$\sim$3 mJy beam$^{-1}$} for $4''$-pixel map).

The de-biased polarized intensity, $PI$, is calculated using the following formula
\begin{eqnarray}
PI=\sqrt{Q^2+U^2-\delta PI^2}
\end{eqnarray} \citep{montier2015a,montier2015b} where the uncertainty on $PI$ is
\begin{eqnarray}
\delta PI=\sqrt{\frac{Q^2\delta Q^2+U^2\delta U^2}{Q^2+U^2}}.
\end{eqnarray}
The polarization angle is defined as
\begin{eqnarray}
\theta=0.5\tan^{-1}\left(\frac{U}{Q}\right)
\end{eqnarray} and its uncertainty
\begin{eqnarray}
\delta \theta=0.5\times\frac{\sqrt{U^2\delta Q^2+Q^2\delta U^2}}{(Q^2+U^2)}.
\end{eqnarray}
We note that the orientation of the magnetic field line is perpendicular to the polarization angle with the assumption that the polarization of the emission comes from elongated grains that interact with an underlying magnetic field. The angle of the magnetic field line is east of north ranging from 0$^\circ$ to 180$^\circ$. Finally, the polarization fraction, $P$, and its uncertainty, $\delta P$, are calculated as 
\begin{eqnarray}
P (\%)=100\times\frac{PI}{I}
\end{eqnarray}
and
\begin{eqnarray}
\delta P(\%)=100\times \sqrt{\frac{\delta PI^2}{I^2}+\frac{\delta I^2(Q^2+U^2)}{I^4}}. 
\end{eqnarray}

\begin{figure*}[!htb]
\centering
  \begin{tabular}{cc}
    \includegraphics[trim=1.4cm 6.3cm 3.2cm 5.75cm,clip,width=6.5cm]{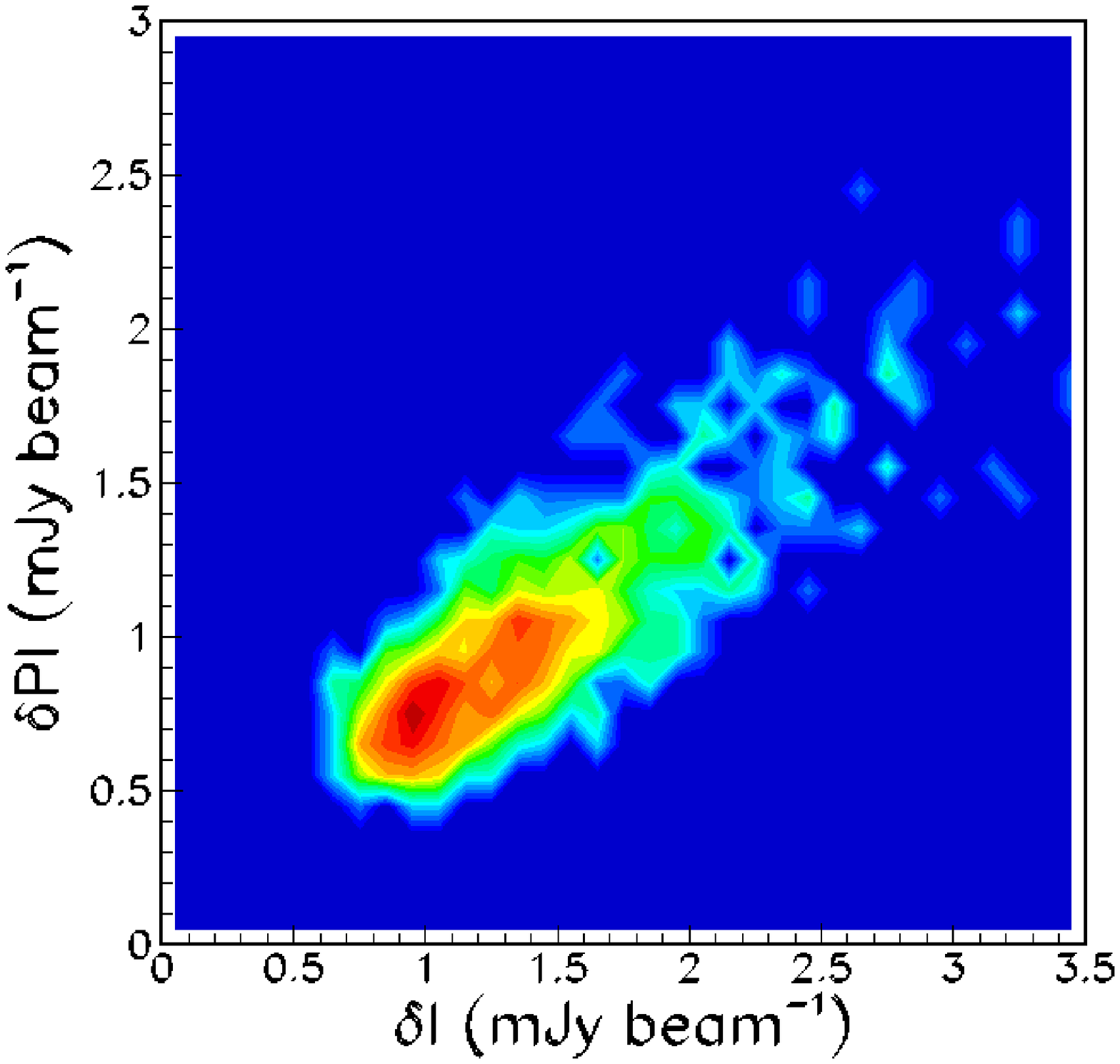}&
    \includegraphics[trim=1.3cm 6.3cm 3.3cm 5.75cm,clip,width=6.5cm]{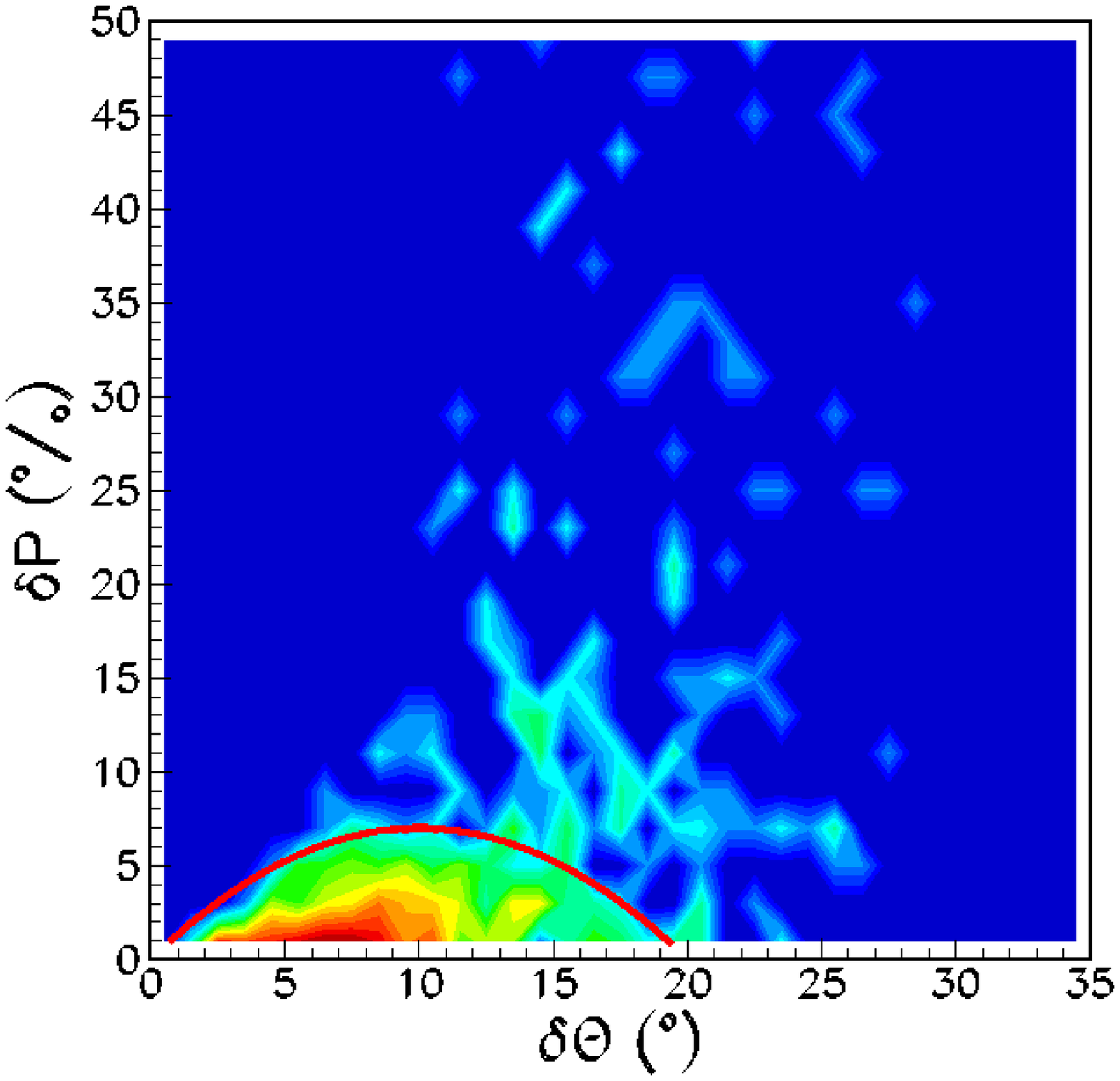}
  \end{tabular}
\caption{Correlations of measurement uncertainties (the color represents the number of data points in logarithmic scales): $\delta PI$ vs $\delta I$ (left) and $\delta P$ vs $\delta\theta$ (right). Strong correlation is found for $\delta PI$ vs $\delta I$ (left panel). Pixels having well-defined polarization fraction ($\delta P<\sim 7\%$) and polarization angles ($\delta\theta<\sim20^\circ$) are encompassed by a parabola (red curve in the right panel).} \label{fig5}
\end{figure*}

\begin{figure*}[!htb]
\centering
\begin{tabular}{ccc}
\includegraphics[trim=1.7cm 6.4cm 3.1cm 6.5cm,clip,width=6.1cm]{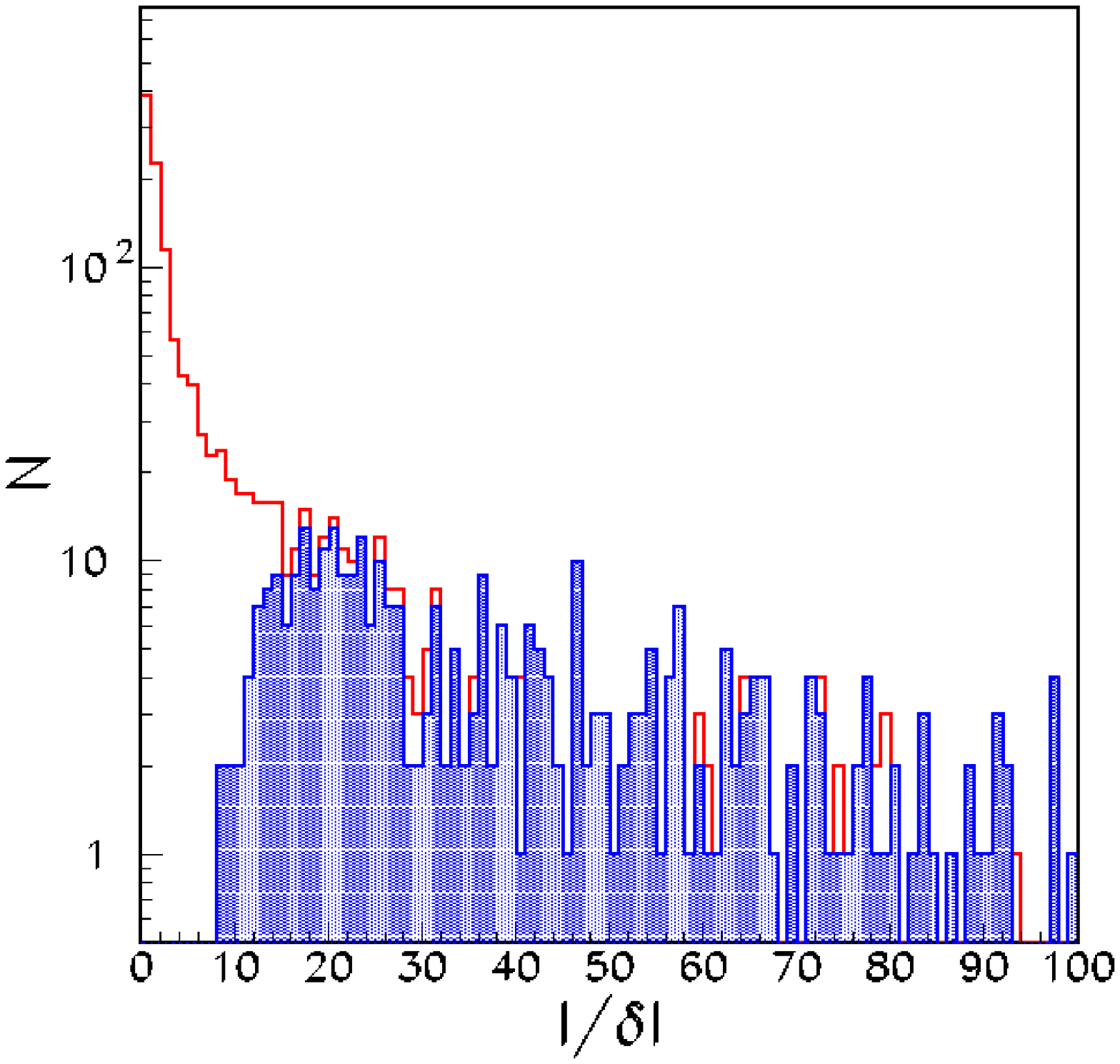}&
\includegraphics[trim=2.4cm 6.4cm 3.5cm 6.5cm,clip,width=5.7cm]{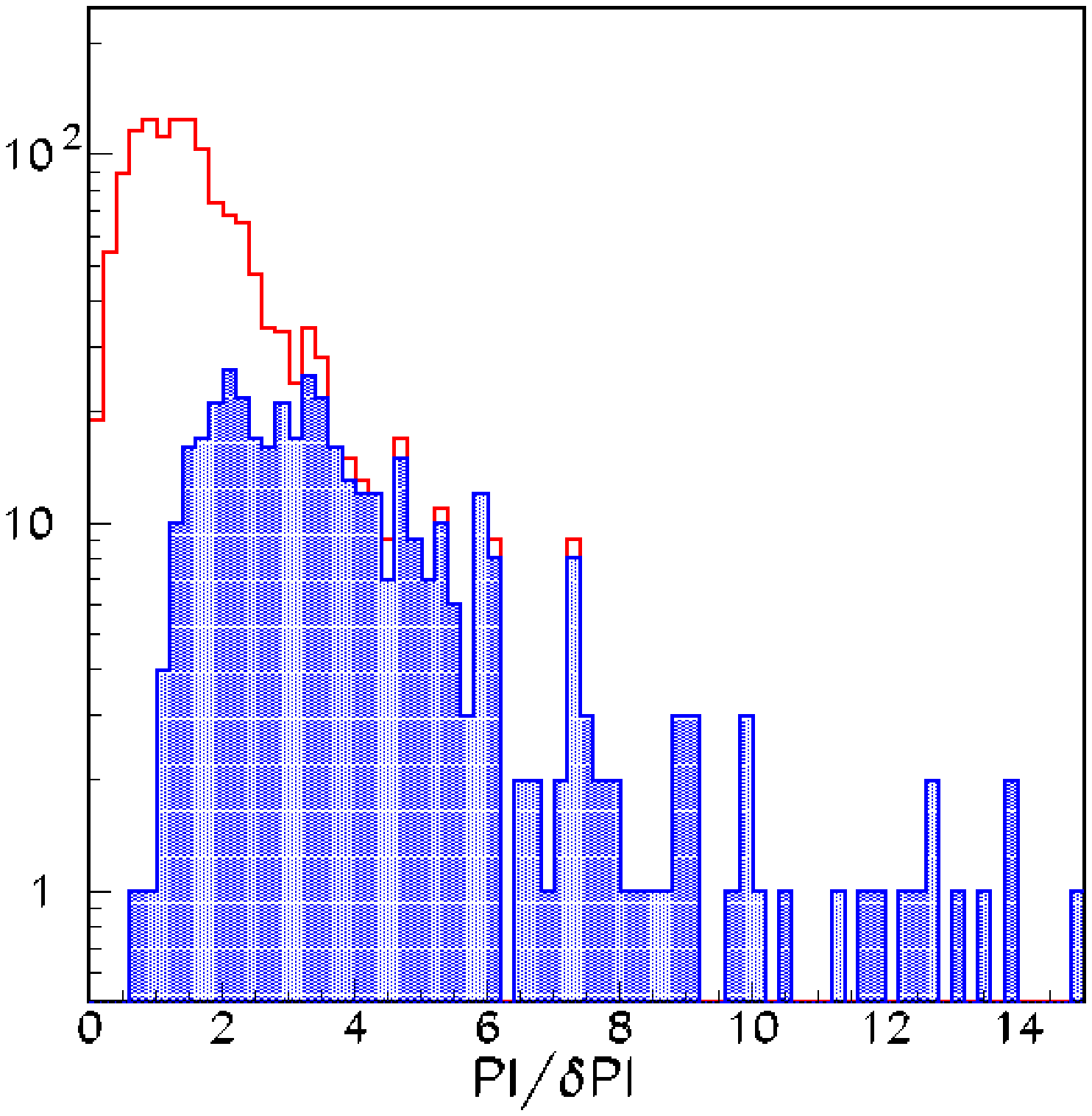}&
\includegraphics[trim=2.4cm 6.4cm 3.5cm 6.5cm,clip,width=5.7cm]{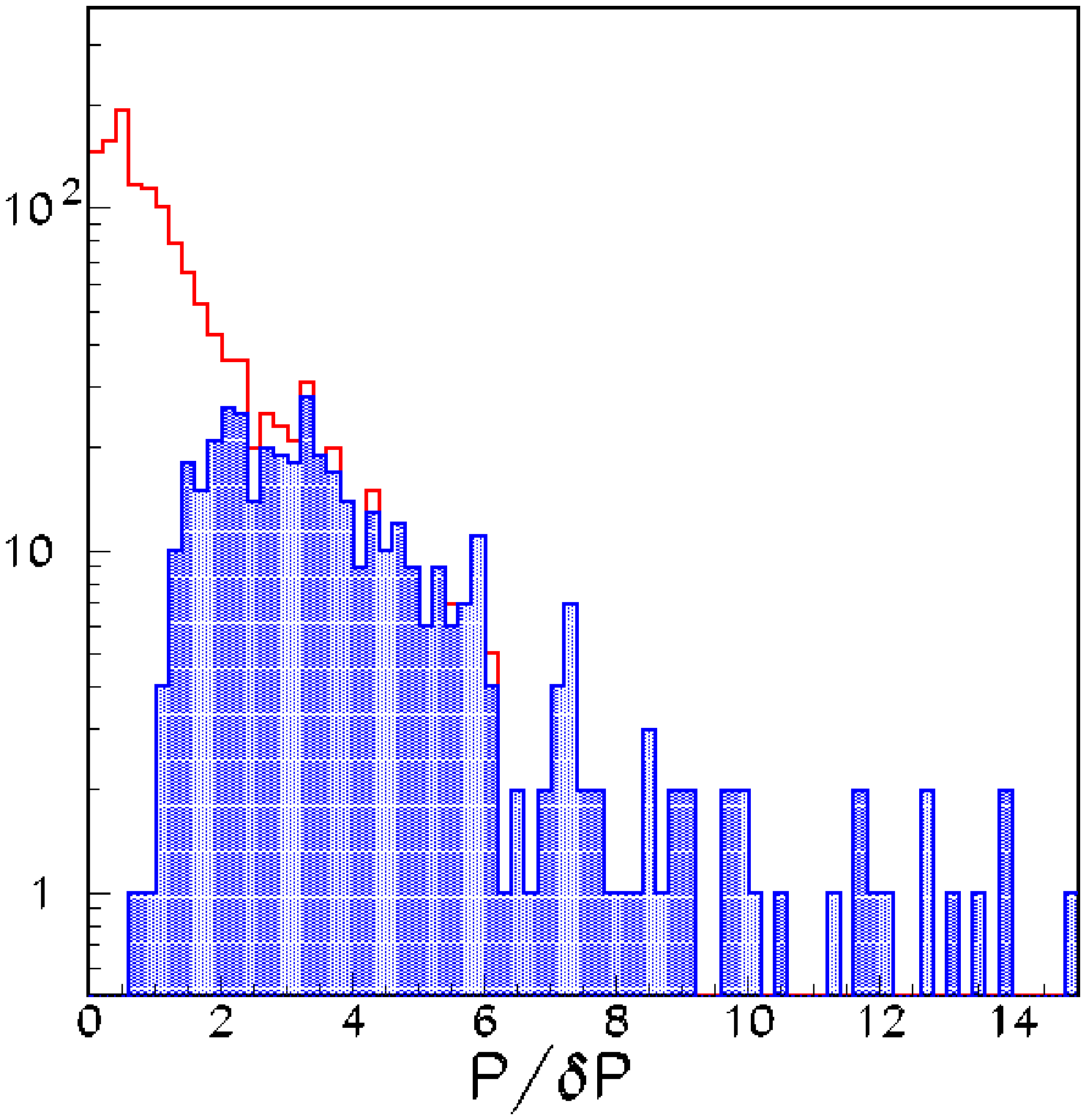}
  \end{tabular}
\caption{Distributions of $I/\delta I$ (left), $PI/\delta PI$ (center), and $P/\delta P$ (right): all pixels (red histograms) and pixels retained after employing $\delta P$-$\delta\theta$ cut having high S/N (blue histograms).} \label{fig6}
\end{figure*}

\begin{figure*}[!htb]
\centering
\includegraphics[trim=0.35cm 0.43cm 2.5cm 0.cm,clip,width=12.cm]{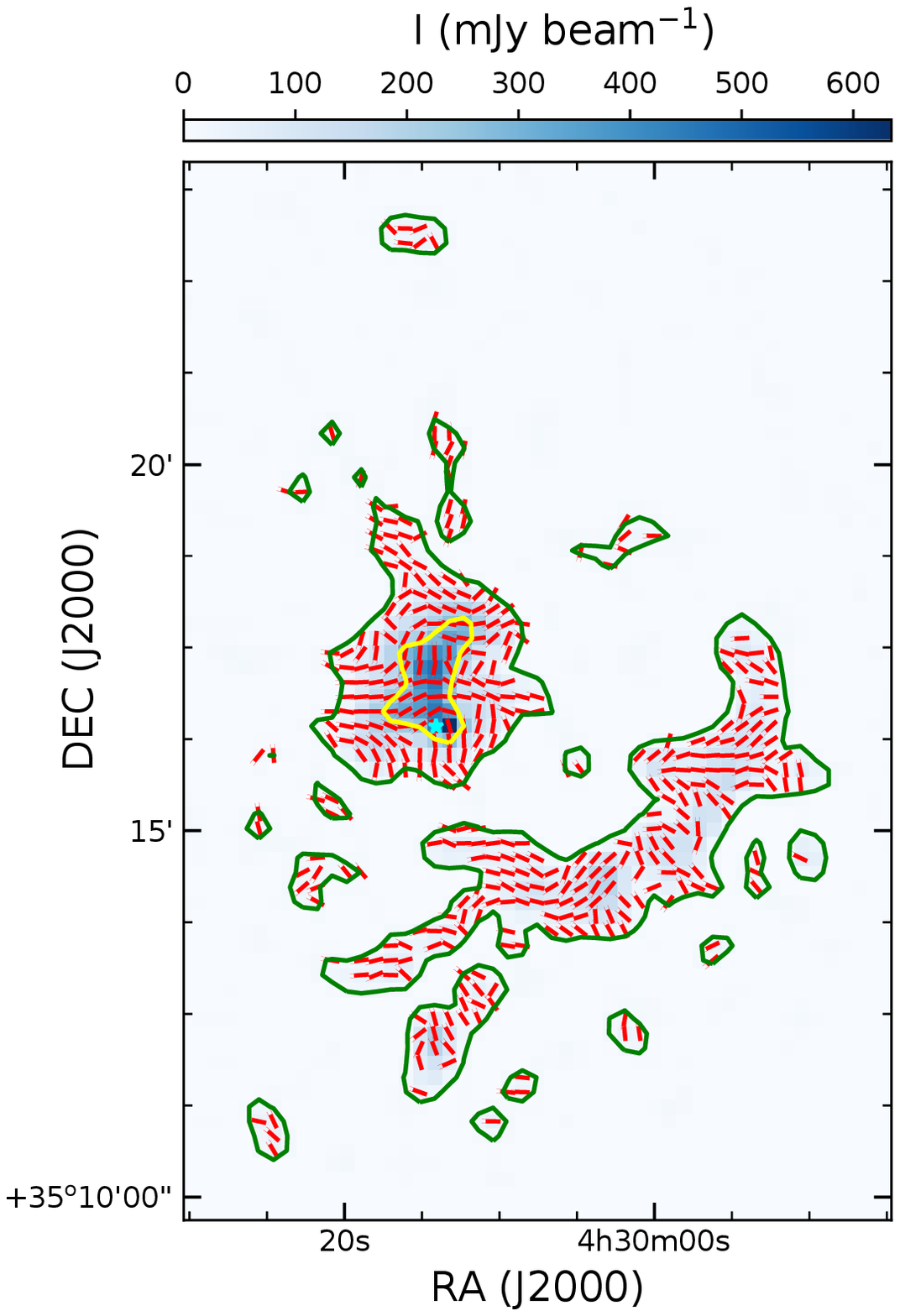}
\caption{Map of the inferred magnetic field orientation (line segments) overlaid on the 850 $\mu$m emission intensity map (color scale). The green contours correspond to 15 mJy beam$^{-1}$ and yellow contour 250 mJy beam$^{-1}$. The line segments shown are those remaining after the $\delta P$-$\delta\theta$ cut. Location of the B star LkH$\alpha$ 101 is at the cyan star marker at $\mathrm{RA=4h30m14.4s}$, $\mathrm{DEC=35^\circ16'24''}$.} \label{fig7}
\end{figure*}

In an attempt to select measurements to be retained for further analysis, we plot in Figure \ref{fig5} correlations relating the uncertainties of the measured parameters (no cut has been applied). From the left panel of the figure it is clear that $\delta I$ and $\delta PI$ are strongly correlated. When $I$ is well-measured so is $PI$; the converse also holds. It is not the case for other parameters in which the correlations are weak (see Figure \ref{figA3}) except for the case of $\delta P$ vs $\delta\theta$ (Figure \ref{fig5} right) where we find that most of the good measurements are confined within a parabola shown as a red solid curve. The equation of the parabola is $\delta P=-0.07\delta\theta^2+1.4\delta\theta$. In fact, $\delta P$ and $\delta\theta$ are related quantities; \citet{serkowski1962} gives $\delta\theta = 28.65^\circ\delta P/P$. In the following analyses we use only the data obeying $\delta P<-0.07\delta\theta^2+1.4\delta\theta$ (hereafter called $\delta P$-$\delta\theta$ cut). Since the 850 $\mu$m polarized emission from LkH$\alpha$ 101 region is weak, possibly the weakest in comparison with other regions surveyed by BISTRO, after several attempts, we found that the best compromise between keeping the highest quality data and having good statistics is to use the $\delta P$-$\delta\theta$ cut which can still assure the robustness of the results and conclusions of the current studies.

Figure \ref{fig6} shows the distributions of $I/\delta I$, $PI/\delta PI$, and $P/\delta P$ for the observed region before and after the $\delta P$-$\delta\theta$ cut is applied. These signal-to-noise-ratios (S/N) measure the quality of the measurements of the corresponding quantities $I$, $PI$, and $P$. Their dimensionless values (mean, RMS) are (12.1, 19.8), (2.1, 1.9), and (1.7, 2.0) for $I/\delta I$, $PI/\delta PI$, and $P/\delta P$ respectively. When applying the $\delta P$-$\delta\theta$ cut these numbers become (38.5, 22.7), (4.0, 2.4), and (3.9, 2.4): the $\delta P$-$\delta\theta$ cut eliminates the parts of the distributions having low S/N. Some more detailed information about the raw data set can be found in Appendix \ref{appendix}.

Figure \ref{fig7} shows the map of the inferred B-fields of the LkH$\alpha$ 101 region keeping only pixels obeying the $\delta P$-$\delta\theta$ cut. The line segments are 90$^\circ$ rotated from the polarization vectors to follow the magnetic field in the plane of the sky. These are called half-vectors since there is an ambiguity in their directions. After application of the $\delta P$-$\delta\theta$ cut, the number of remaining line segments is 419. This map is a direct result from the observations and will be used for further analysis and discussions in the next sections.

\begin{table*}[!htb]
\begin{centering}
\caption{Dependence of angle dispersion and number of remaining half-vectors, $N_{\rm rm}$, on $\delta I$-cuts.}\label{tab2}
\begin{tabular}{c c c c c c c c c}
\hline
\hline
&$\delta I$ (mJy beam$^{-1}$) $<$&1.0&1.5&2.0&2.5&3.0&4.0\\
\hline
Central Region&$\sigma_{\theta}(^\circ)/N_{\rm rm}$&16.0/94&17.1/135&17.5/141&17.6/142&17.3/144&17.2/146\\
Dust Lane&$\sigma_{\theta}(^\circ)/N_{\rm rm}$&5.8/12&17.1/192&17.1/195&17.1/195&17.1/195&17.1/195\\ \hline
\end{tabular}
\end{centering}
\end{table*}

\begin{figure*}[!ht]
\centering
    \includegraphics[trim=1.7cm 2.1cm 2.6cm 2.2cm,clip,width=18cm]{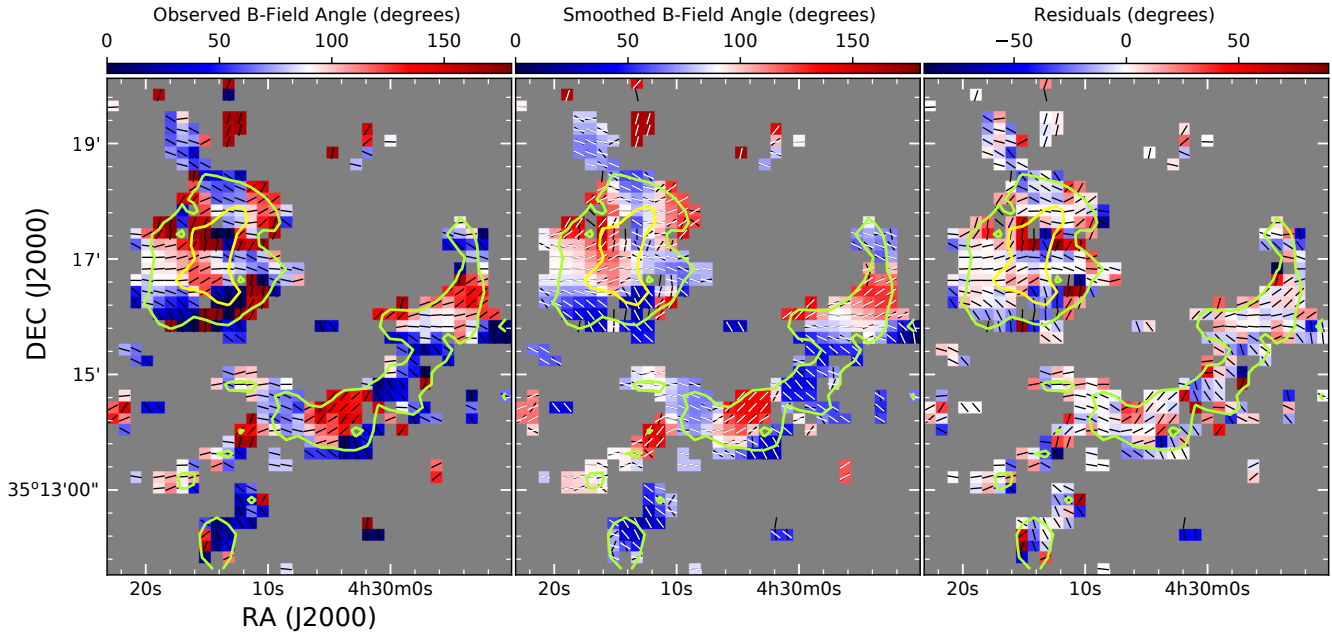}
\caption{B-field angle maps: original (left), smoothed (center), and residual (right). The green contours correspond to 50 \mbox{mJy beam$^{-1}$} and the yellow contours to 250 mJy beam$^{-1}$. In all panels, the orientation of the measured magnetic field half-vectors is plotted in black. In addition, in the central panel, the orientation of the smoothed half-vectors is shown in white.} \label{fig8}
\end{figure*}

\section{Polarization Angle Dispersion, Velocity Dispersion and Densities} \label{sec:dis}
\subsection{Davis-Chandrasekhar-Fermi Method}
It was shown by \citet{davis1951strength} and \citet{chandrasekhar1953} that turbulent motions generate irregular magnetic fields. Based on the analysis of the small-scale randomness of magnetic field lines, assuming that the dispersion in the magnetic field angles is proportional to the Alfv\'en Mach number, the field strength can be estimated. This is called the Davis-Chandrasekhar-Fermi (DCF) method. A variant of the method has been proposed by \cite{crutcher2012}; it gives an estimate of the magnitude of the magnetic field in the plane of the sky, $B_{\rm POS}$, as
\begin{equation}\label{eq7}
B_{\rm POS}=Q\sqrt{4\pi\rho}\frac{\sigma_V}{\sigma_\theta}\approx9.3\sqrt{n(\rm H_2)}\frac{\Delta V}{\sigma_\theta}\quad(\mu G)
\end{equation}
where $Q$ is the factor used to correct for the line-of-sight and beam-integration effects \citep{ostriker2001}, $\rho$ is the gas density, $\sigma_V$ the one-dimensional non-thermal velocity dispersion in km s$^{-1}$, $\Delta V=2.355\sigma_V$, $\sigma_\theta$ is the dispersion of the polarization position angles about a mean B-field in degrees, and $n(\rm H_2)$ is the number density of molecular hydrogen in units of cm$^{-3}$.

In the next sections, we evaluate the magnetic field angle dispersion, $\sigma_\theta$, a measurable of BISTRO, using two different methods.

\begin{figure*}[!htb]
\centering
\begin{tabular}{cc}
\includegraphics[trim=1.6cm 6.2cm 1.cm 7cm,clip,width=8.5cm]{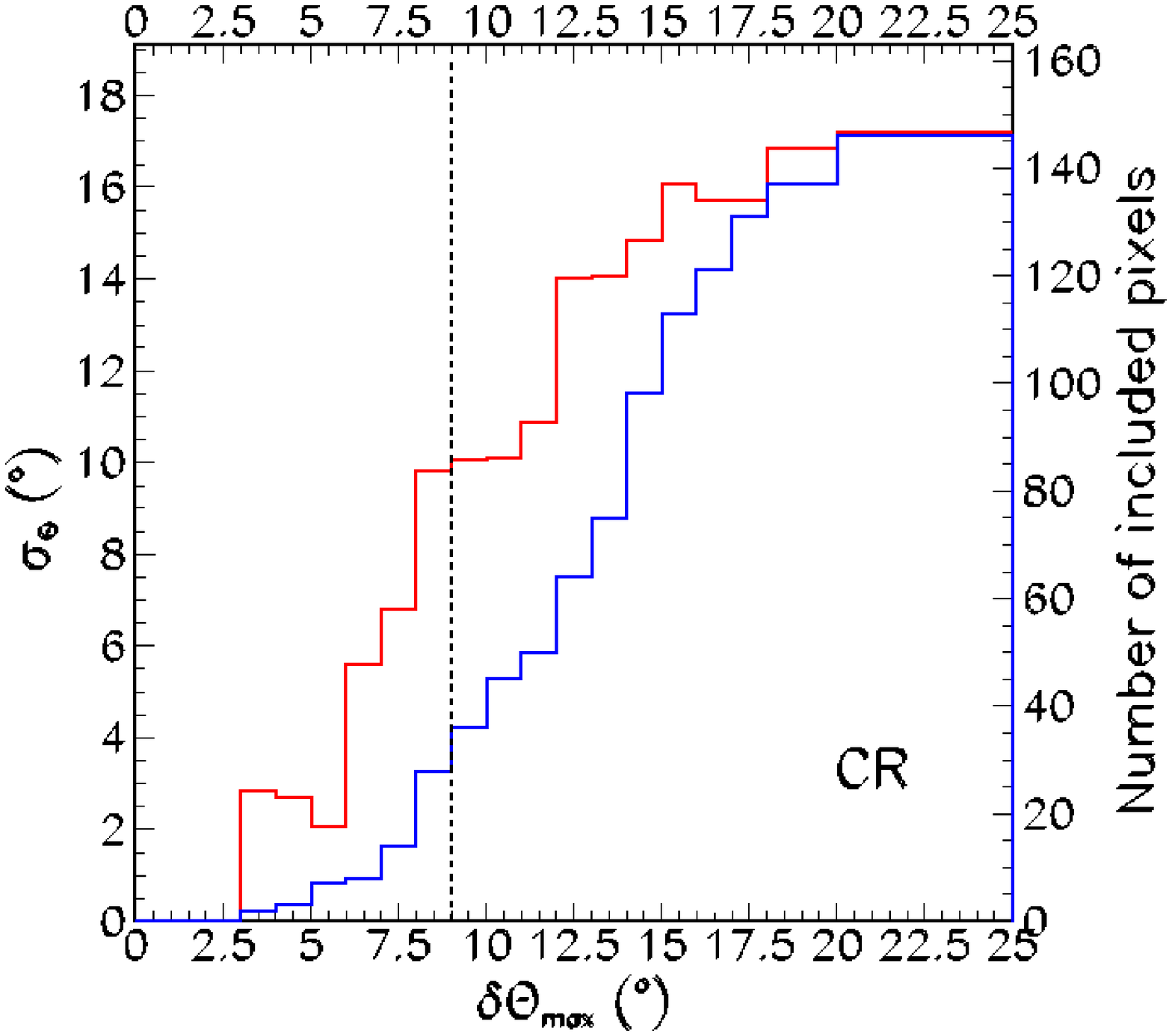}&
\includegraphics[trim=1.6cm 6.2cm 1.cm 7cm,clip,width=8.5cm]{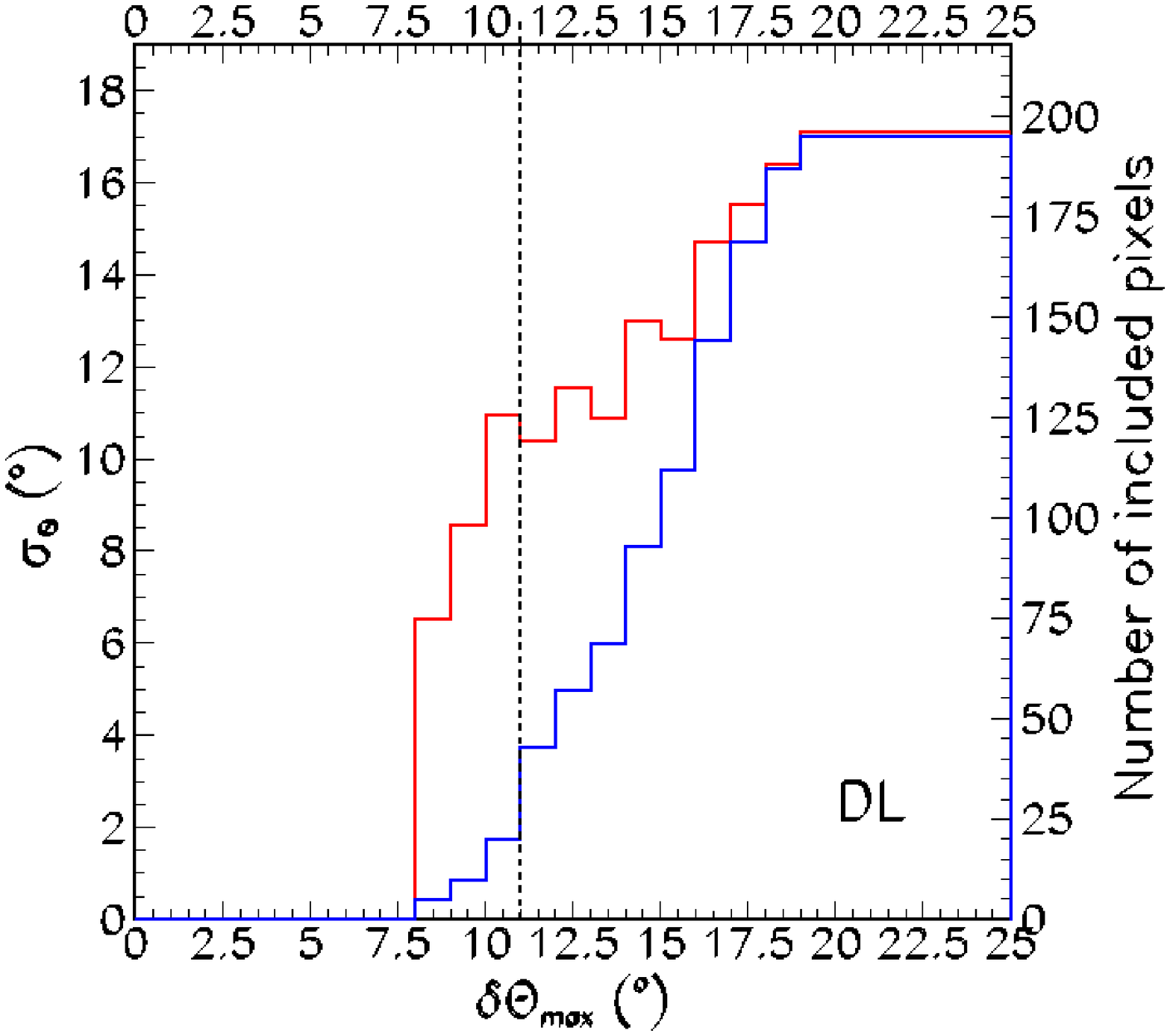}
\end{tabular}
\caption{Dependence of the polarization angle dispersion (red) and number of included pixels (blue) on maximum allowed uncertainties $\delta\theta_{max}$ for the central region (left) and the dust lane (right). Black dashed lines show the lower limit of $\delta\theta_{max}$ over which the average angle dispersion is calculated.} \label{fig9}
\end{figure*}

\subsection{Polarization Angle Dispersion: Unsharp Masking Method}\label{subsec:unsharp}
To estimate the dispersion of the polarization angle we first use the “unsharp masking” method which was introduced by \citet{pattle2017}. The principle of the method is to look for the turbulent component of the magnetic field by removing a mean field using a boxcar filter. In practice, this is done by looping over all the pixels of the map, calculating the difference between the original polarization angle map and the intensity weighted mean map consisting of a $3\times3$ pixel box centered on the considered pixel. The standard deviation of the distribution of deviation angles, $\Delta\theta=\theta_{\rm meas}-\langle\theta\rangle$, represents the turbulent component of the field and gives the angular dispersion of the region. This process is illustrated in Figure \ref{fig8}.

We apply this method to calculate the angle dispersion for the central region and the south-western dust lane (hereafter called CR for the central region and DL for the dust lane). The central region is defined as having $R<120''$ from the map center and the dust lane is the region surrounded by an ellipse having a center at $\rm{RA\sim4h30m4.7s}$ and $\rm{DEC\sim+35^\circ14'47.8''}$, major$\times$minor axes of $300''\times 108''$ and a position angle of $135^\circ$ (see Figure \ref{fig1} right and the white curves in Figure \ref{fig11}). The angle dispersion obtained by requiring $|\Delta\theta|<90^\circ$ (to avoid the effect of the $\pm$180$^\circ$ ambiguity of the magnetic field lines) are $\sigma_\theta=17.2^\circ\pm0.4^\circ$ and 17.1$^\circ\pm0.5^\circ$ for the central region and the dust lane respectively. To understand how the dispersion depends on the measurement uncertainties, we list in Table \ref{tab2} the polarization angle dispersion as a function of cuts applied on $\delta I$, in addition to the $\delta P$-$\delta\theta$ cut. Exploring the $\delta I$-dependence is sufficient since $\delta I$ and $\delta PI$ are strongly correlated (Figure \ref{fig5} left). We note that $\delta I$ ranges from $\sim$0.6 to $\sim$5.0 \mbox{mJy beam$^{-1}$} (Figure \ref{figa2}).

We see from Table \ref{tab2} that the angle dispersion is robust while applying additional cut requiring higher S/N measurements of the total and polarized intensities. This is an indication that $\delta P$-$\delta\theta$ is a robust cut. Only in the case of the most rigorous cut requiring $\delta I < 1$ \mbox{mJy beam$^{-1}$} for the dust lane does the number of remaining half-vectors drastically decrease from $\sim$190 to 12 with the angle dispersion decreasing from $\sigma_\theta\sim17^\circ$ to 5.8$^\circ$.

The uncertainties of the polarization angle dispersion are also studied following the procedure of this same method \citep{pattle2017}. Figure \ref{fig9} displays the dependence of the angle dispersion on the maximum allowed uncertainties, $\delta\theta_{max}$, for each boxcar filter. In practice, this means that the angle dispersion of the map is calculated by requiring that the maximum uncertainty of all the pixels in each $3\times3$ boxcar filter be smaller than $\delta\theta_{max}$. The polarization angle dispersion is expected to increase with the maximum allowed uncertainty of the boxcar filters. Indeed, as can be seen from Figure \ref{fig9} (left for the central region and right for the dust lane), the angle dispersion increases as $\delta\theta_{max}$ increases. As was done by \cite{soam2018} and \cite{liu2019}, from the angle dispersion for different $\delta\theta_{max}$ (Figure \ref{fig9}), we calculate the mean angle dispersion$\pm$uncertainties of $\sigma_\theta=14.9^\circ\pm2.7^\circ$ and $14.7^\circ\pm2.6^\circ$ for the central region and the dust lane respectively. These values are kept as the final polarization angle dispersion for further analysis. We note that the results are obtained excluding a first few bins of $\delta\theta_{max}$ where the numbers of included pixels are smaller than 20 and with the application of the $\delta P$-$\delta\theta$ cut.

\subsection{Polarization Angle Dispersion: Structure Function} \label{subsec:sf}
\begin{figure*}[!htb]
\centering
\begin{tabular}{cc}
    \includegraphics[trim=.6cm 6.cm 3.cm 7cm,clip,width=8.5cm]{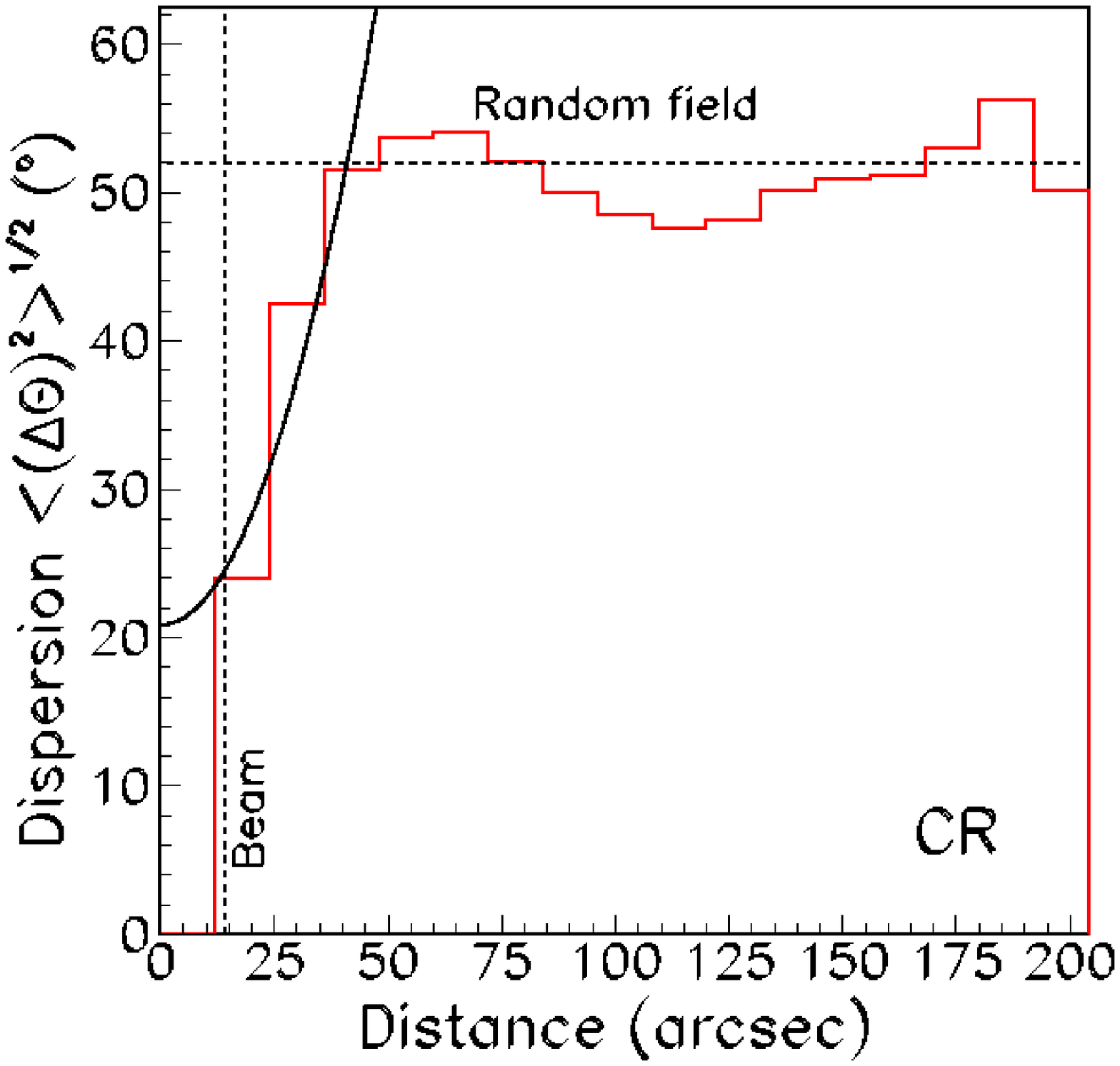}&
    \includegraphics[trim=.6cm 6.cm 3.cm 7cm,clip,width=8.5cm]{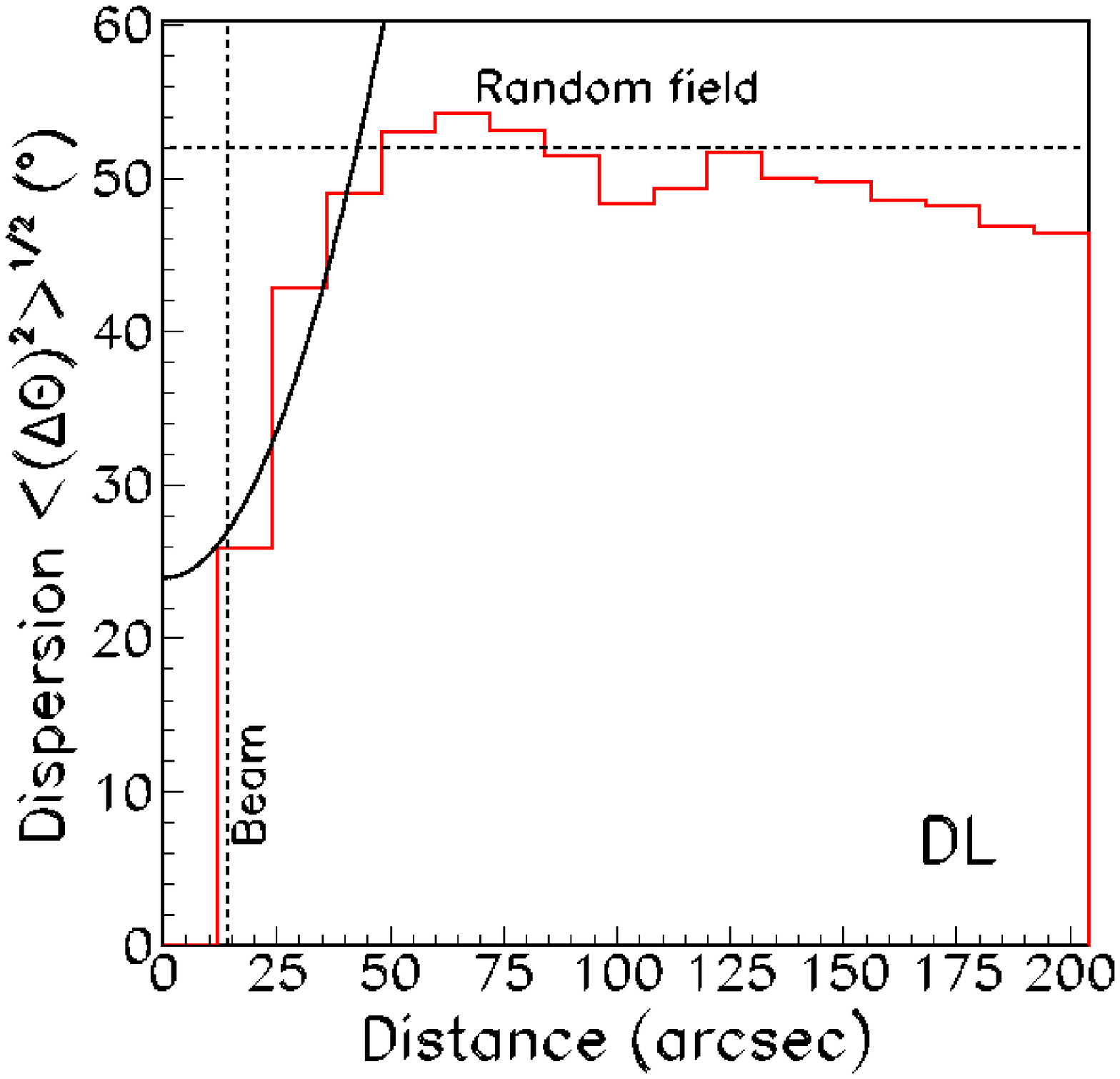}
\end{tabular}
\caption{The structure functions calculated for the central region (left) and for the dust lane (right). The horizontal dashed lines are the expected value for random fields. The solid black curves are the results of fits to Equation (\ref{eq11}) for the estimation of the polarization angle dispersion of the regions. The vertical dashed lines are at the distances equal to the beam size of the JCMT telescope.} \label{fig10}
\end{figure*}

Another method to estimate the dispersion of polarization position angles makes use of the so-called structure function \citep{hildebrand2009} which is defined as follows. We consider pairs of pixels, at locations $\vec{x}$ and $\vec{x}+\vec{l}$, each associated with a polarization position angle $\theta$. For a given pixel separation $l$ the structure function is defined as the mean square angle between the polarization half-vectors of the pair:

\begin{equation}\label{eq10}
\langle\Delta\theta^2(l)\rangle=\frac{1}{N(l)}\sum_{i=1}^{N(l)}[\theta(\vec{x})-\theta(\vec{x}+\vec{l})]^2,
\end{equation}
where $N(l)$ is the number of such pairs. We assume that the magnetic field $\vec{B}$ can be approximated by the sum of a large-scale structured field of mean amplitude $B_0$, and a turbulent component, $\delta B$; we also assume that the correlation length of the turbulent component is much smaller than the distance over which $\vec{B}$ varies significantly. For small separations $l$, we can write

\begin{equation}\label{eq11}
\langle\Delta\theta^2(l)\rangle=b^2+m^2l^2+\sigma^2_M(l),
\end{equation}
where $b$ is the root mean square contribution of the turbulent component and $m$ measures the contribution of the gradient of $\vec{B}$; in addition, the contribution of the measurement uncertainties, $\sigma_M(l)$, is included. All these contributions add in quadrature. Neglecting the contribution of the large-scale structured field in Equation (\ref{eq11}) and from the definition of the polarization angle dispersion, $\sigma_\theta$, we have $\sigma_\theta^2=b^2/2$. 

\citet{hildebrand2009} expressed the ratio of the turbulent field, $\delta B$, to the large scale underlying field (i.e., mean field), $B_0$ as
\begin{eqnarray}\label{eq9}
\frac{\delta B}{B_0}=\frac{b}{\sqrt{2-b^2}},
\end{eqnarray}
where $b=\sqrt{2}\sigma_\theta$. We note that Equations \ref{eq11} and \ref{eq9} hold only if the correlation length of the turbulent component satisfies $\delta<l$ \citep{hildebrand2009}.

We calculate the structure functions for both the central region and the dust lane; the result is displayed in Figure \ref{fig10}. At large distances, the structure functions tend to the random field value of $\sim52^\circ$ \citep{serkowski1962}. A fit to Equation \ref{eq11} for short distances, $12''<l<36''$, gives $b=20.9^\circ\pm6.8^\circ$ and $23.9^\circ\pm7.1^\circ$, namely $\sigma_\theta=14.8^\circ\pm4.8^\circ$ and $16.9^\circ\pm5.0^\circ$ for the central region and the dust lane, respectively. When applying additional cuts on $\delta I$, $\delta I<1$ ($\delta I<1.5$) mJy beam$^{-1}$ we obtain values of $\sigma_\theta=14.3^\circ\pm6.1^\circ$ ($14.5^\circ\pm4.9^\circ$) for the central region. For the dust lane, this cannot be done with the 1 mJy beam$^{-1}$ cut because too few half-vectors are retained. With the 1.5 mJy beam$^{-1}$ cut, we obtain $\sigma_\theta=14.6^\circ\pm3.1^\circ$. All the angle dispersion obtained when applying additional $\delta I$-cuts is in agreement within one standard deviation with the values obtained when applying the $\delta P$-$\delta\theta$ cut only. This once again confirms that the choice of the $\delta P$-$\delta\theta$ cut is robust.

\subsection{Column and Number Densities} \label{subsec:dens}
\begin{figure}
\centering
\includegraphics[trim=0.9cm 6.5cm 0.5cm 6.8cm,clip,width=8cm]{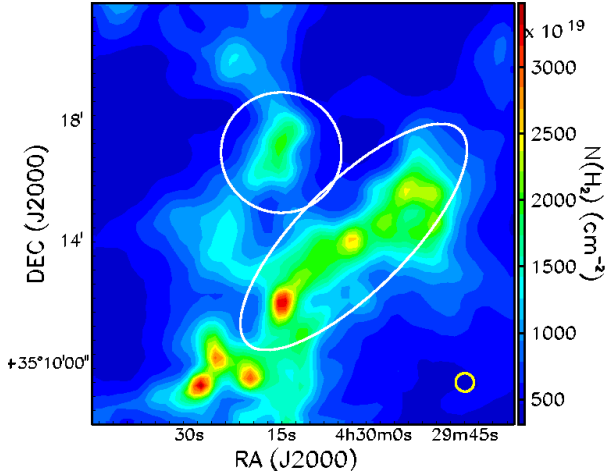} 
\caption{Positions of the central region (white circle) and the dust lane (white ellipse) superposed on the \textit{Herschel} column density map \citep{harvey2013}. The \textit{Herschel} beam size (36.6$''$) is shown in the lower right corner of the map.} \label{fig11}
\end{figure}

The dust column density of the Auriga-California molecular cloud has been constructed by \citet{harvey2013} using four \textit{Herschel} wavebands at 160, 250, 350, and 500 $\mu$m. Using the KOSMA 3-m telescope to detect the CO(J=2$-$1) and (J=3$-$2) line emissions, \citet{li2014} studied the morphology of the L1482 molecular filament of the Auriga-California molecular cloud, of which our studied region is part. The column and number densities of 23 identified clumps along L1482 were measured. Our central region, which fits in a circle of $120''$-radius, encloses their Clump 10 (an ellipse of 228$''\times$110$''$ major$\times$minor axes); the dust lane, extending over 600$''\times$216$''$, overlaps with their Clump 12 (an ellipse of 195$''\times$68$''$ major$\times$minor axes), but is significantly larger. We use the column density map (Figure \ref{fig11}) from \citet{harvey2013} to calculate the average column densities and number density; we find $N(\rm H_2)=(0.96\pm0.39)\times10^{22}$ cm$^{-2}$ and $n(\rm H_2)=1.22\times10^4$ cm$^{-3}$ over the central region, and \mbox{$N(\rm H_2)=(1.44\pm0.53)\times10^{22}$ cm$^{-2}$} and $n(\rm H_2)=1.25\times10^4$ cm$^{-3}$ over the dust lane. The number densities are calculated following the same strategy used by \citet{li2014} in their Section 3.3.2. The masses of the central region and the dust lane are calculated using $M=\beta m_{\rm H_2}N_{\rm total,H_2}(D\Delta)^2$, where $\beta=1.39$ is a factor that takes into account the contribution of He in addition to H$_2$ to the total mass, $m_{\rm H_2}$ is the mass of a hydrogen molecule, $N_{\rm total,H_2}$ is the total column density, and $\Delta=14''$ is the pixel size of \textit{Herschel} data. For consistency, as was done by \citet{harvey2013}, we use the same distance to the LkH$\alpha$ 101 region of \mbox{$D=450$ pc} to calculate the column densities. The radius of the central region is $120''$, and the radius of the dust lane is assumed to be $R=\sqrt{(ab)/2}$ where $a$ and $b$ are the major and minor axes of the dust lane ellipse. Then, the number density of hydrogen molecules is obtained using $n_{H2} = 3M/(4\pi R^3m_{\rm H_2})$. The relative uncertainties in the number densities are taken to be equal to those of the column densities (42\% and 37\% for the central region and the dust lane, respectively).

\subsection{Velocity Dispersion}\label{subsubsec:vedis}
\begin{figure*}
\centering
    \includegraphics[trim=0.9cm 6.3cm 1.5cm 5.9cm,clip,width=8.2cm]{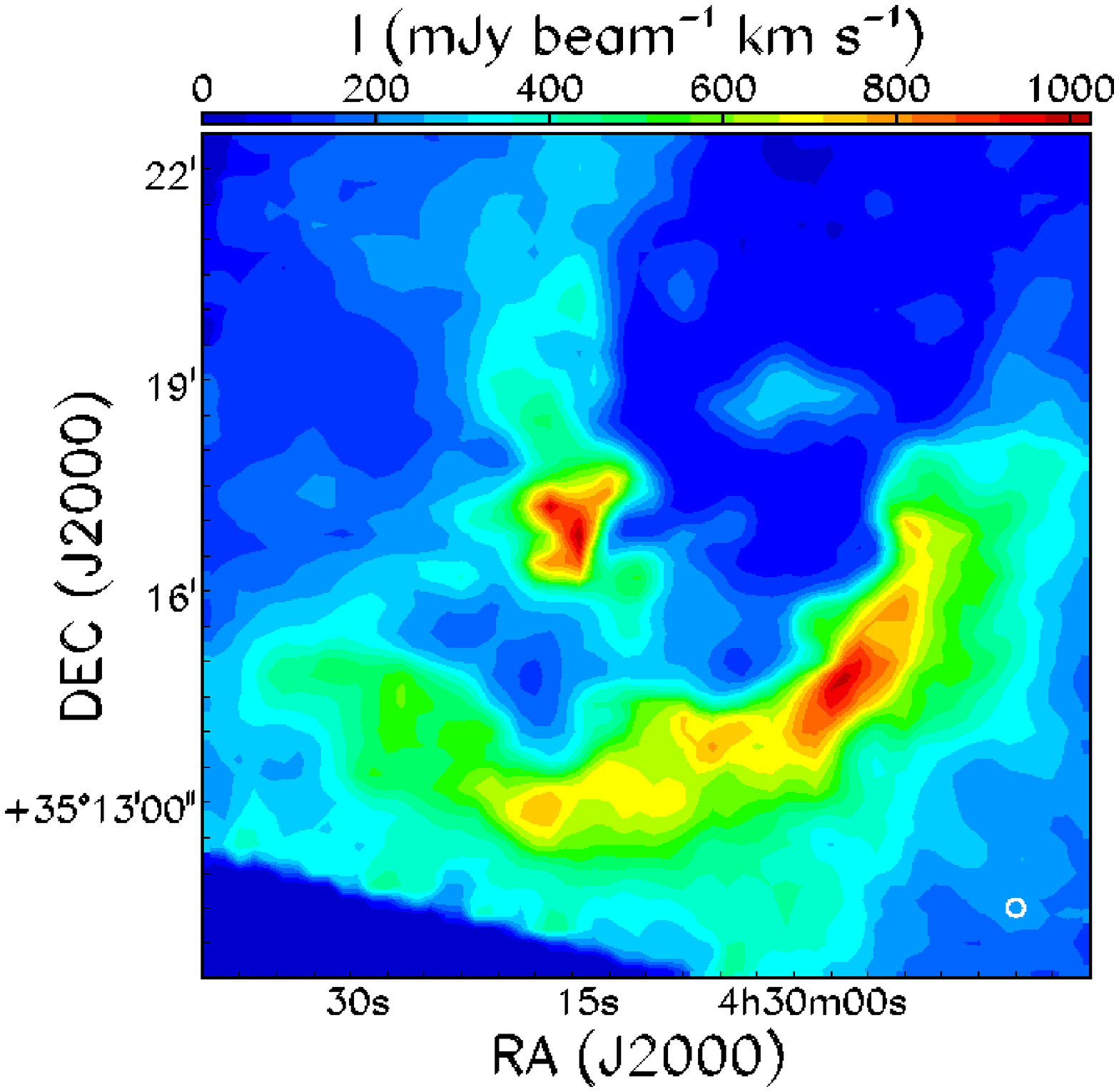}
    \hspace{1cm}
    \includegraphics[trim=0.9cm 6.3cm 1.5cm 5.9cm,clip,width=8.2cm]{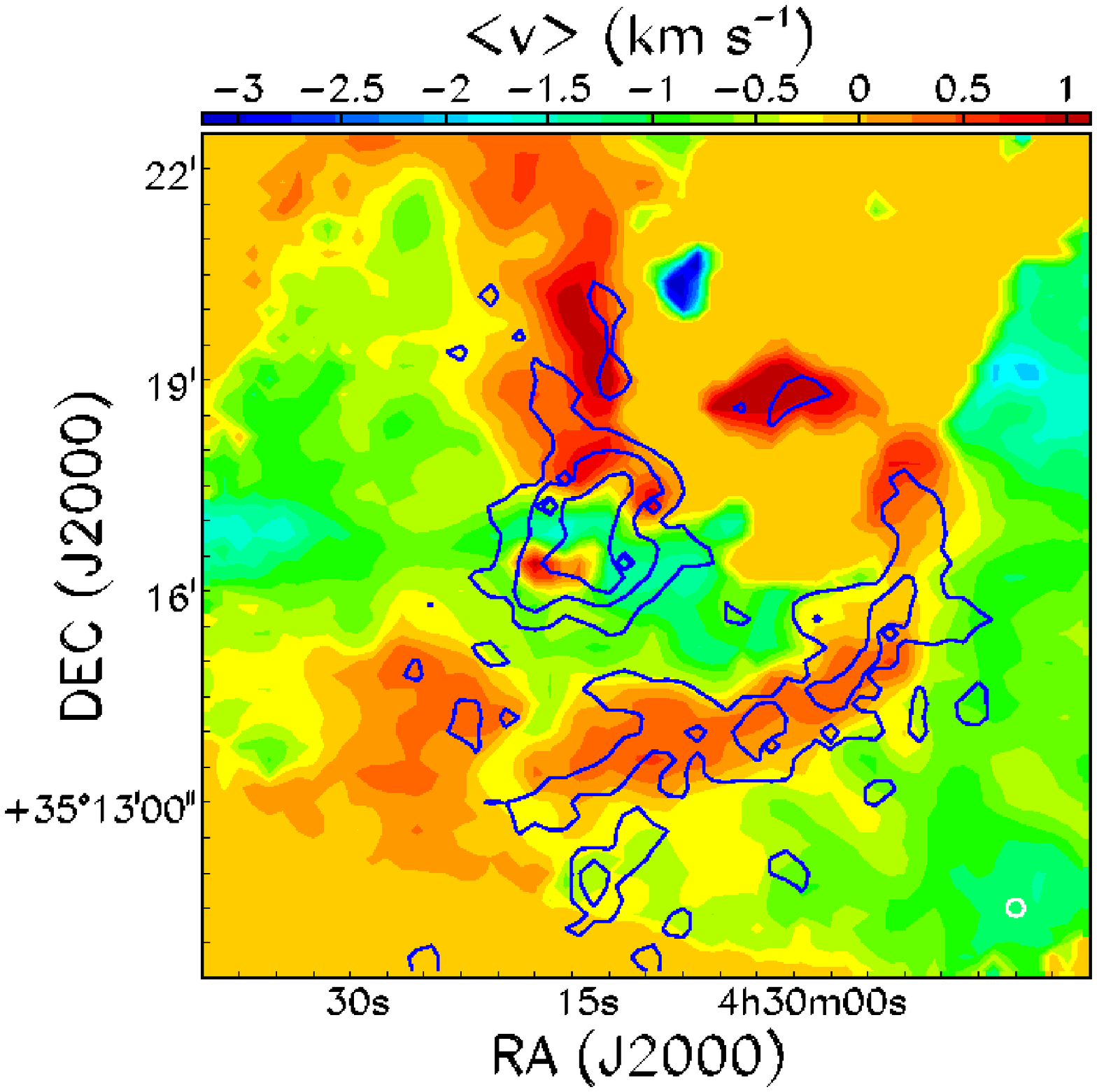}
\caption{HARP data. Left: CO(3$-$2) intensity map integrated over $\pm$5.5 km s$^{-1}$; Right: map of the intensity-weighted mean Doppler velocity relative to the mean VLSR of the cloud. The blue contours are the intensity map at 15, 100 and 250 mJy beam$^{-1}$ levels of the 850 $\mu$m POL-2 polarized emission (the same as that of Figure \ref{fig1} right). The HARP beam size (14$''$) is shown in the lower right corner of the map.} \label{fig12}
\end{figure*}

\begin{figure*}[!htb]
\centering
 \begin{tabular}{ccc}
    \includegraphics[trim=1.3cm 6.3cm 3.4cm 5.75cm,clip,width=5.9cm]{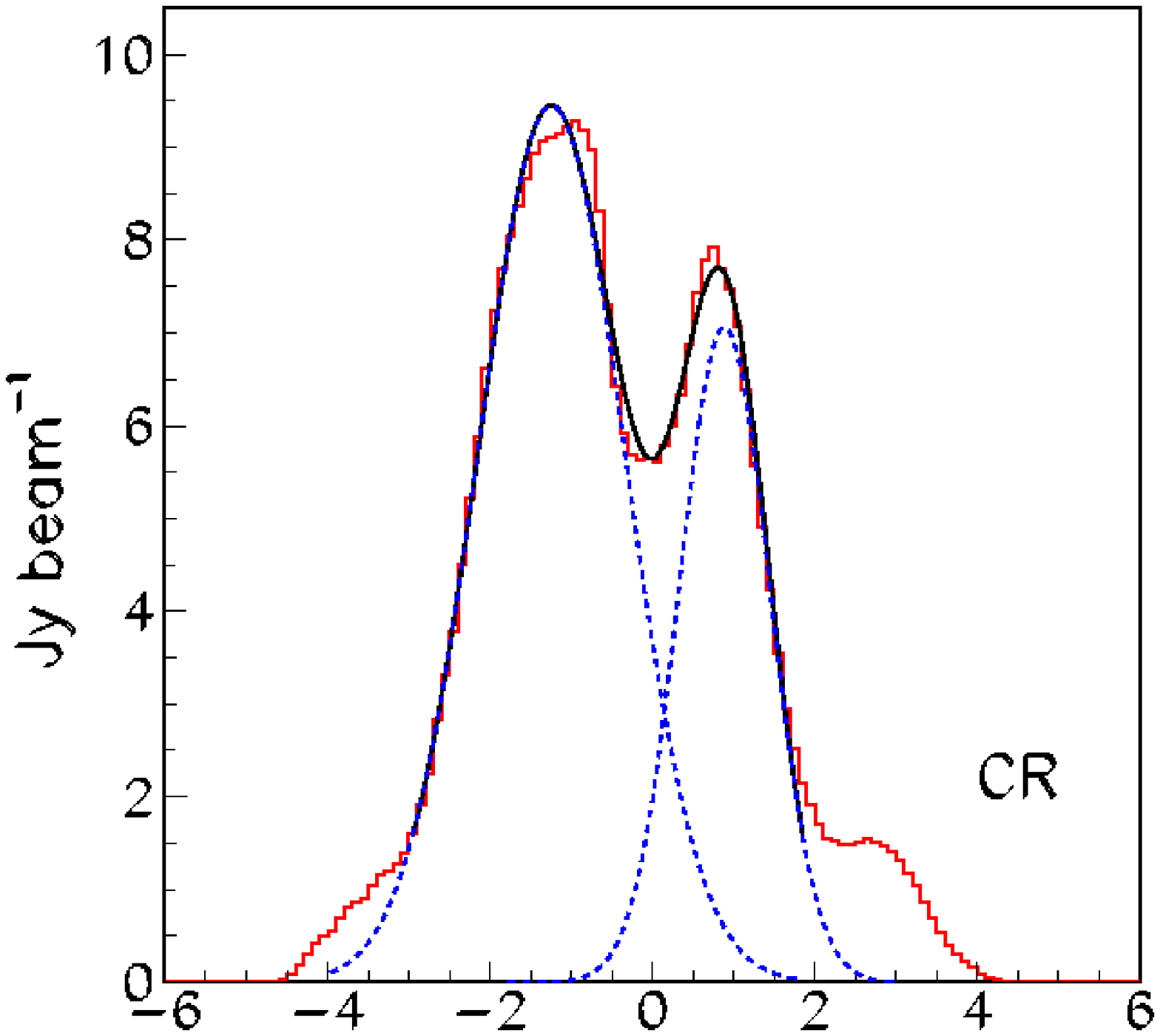}&
    \includegraphics[trim=1.3cm 6.3cm 3.4cm 5.75cm,clip,width=5.9cm]{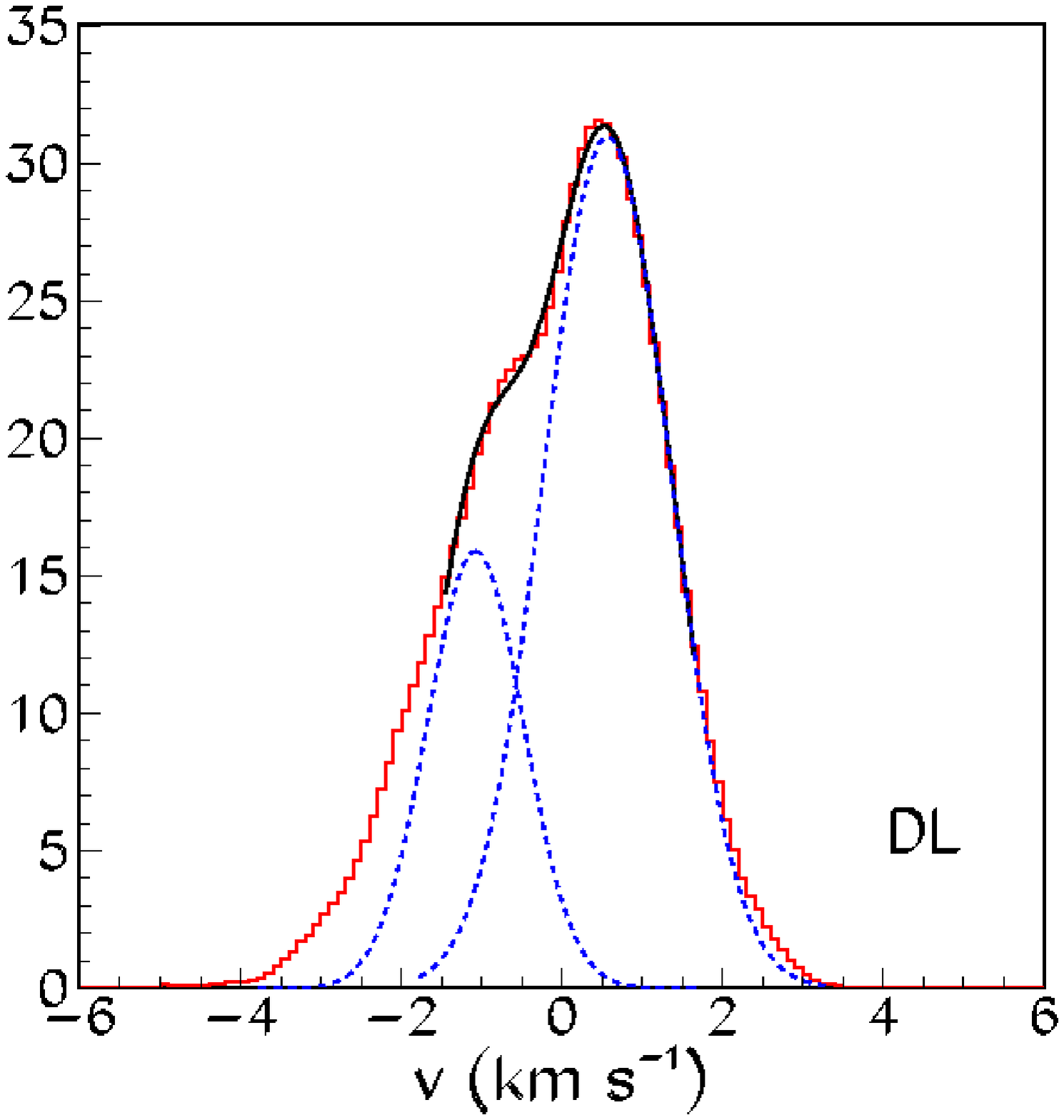}&
    \includegraphics[trim=1.3cm 6.3cm 3.4cm 5.75cm,clip,width=5.9cm]{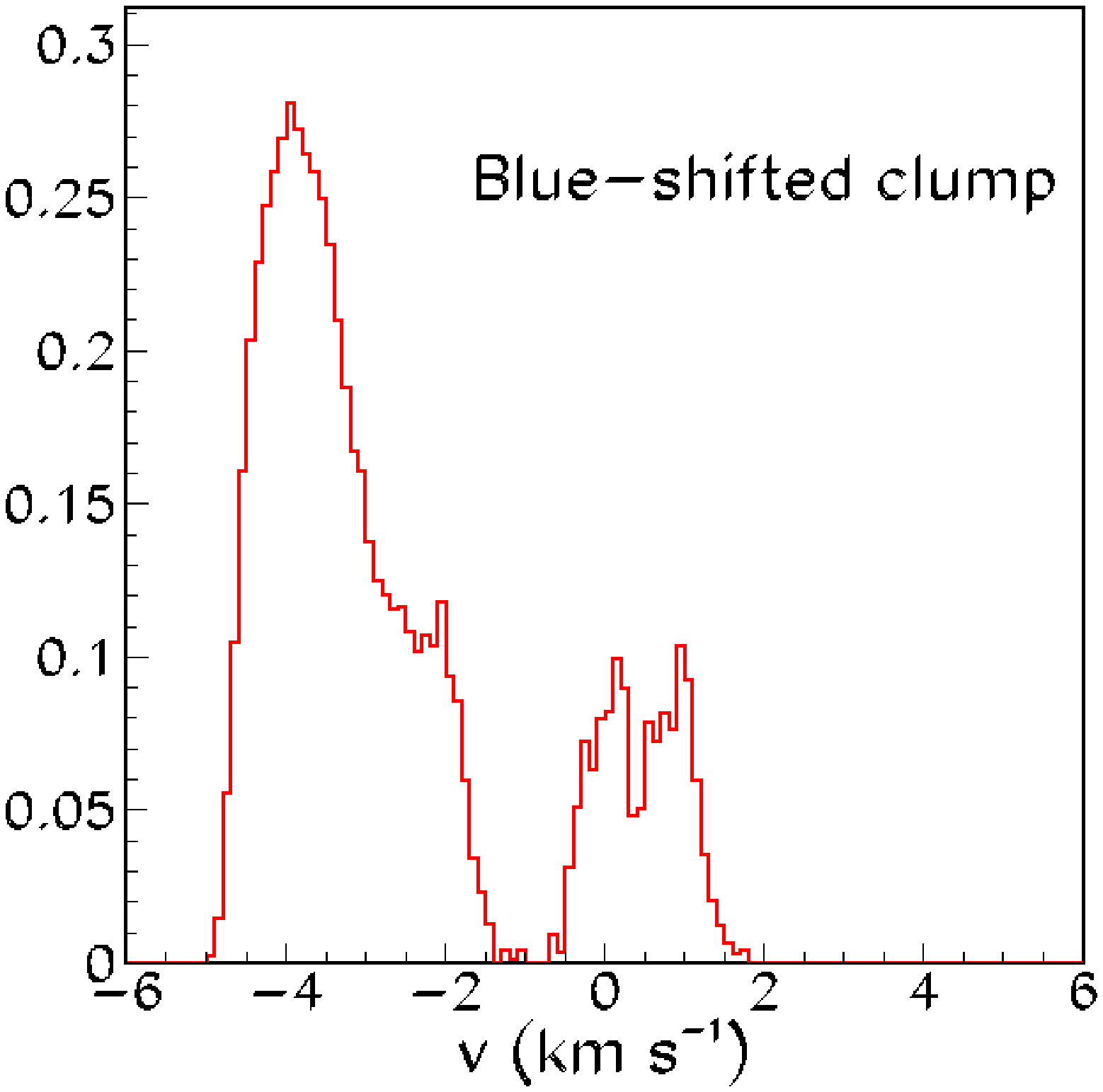}
  \end{tabular}
\caption{HARP CO(3$-$2) integrated spectra. Left: central region; Center: dust lane. Right: blue-shifted clump at $\rm{RA\sim4h30m07s}$ and $\rm{DEC\sim+35^\circ13'20.5''}$. Blue dashed curves are the Gaussian components shown separately. Black curves are the results of the two-Gaussian fits (see text in Section \ref{subsubsec:vedis}).}\label{fig13} 
\end{figure*}

The non-thermal velocity dispersion along the line-of-sight has been measured by the Purple Mountain Observatory (PMO) 13.7-m radio telescope using the $^{13}$CO(1$-$0) transition \citep{li2014}. The results are $\sigma_V=0.52$ km s$^{-1}$ (meaning \mbox{$\Delta V=2.355\sigma_V=1.22$ km s$^{-1}$}) for Clump 10 and $\sigma_V=0.64$ km s$^{-1}$ (meaning $\Delta V=2.355\sigma_V=1.52$ km s$^{-1}$) for Clump 12.

We also analyzed archival CO(3$-$2) JCMT/HARP\footnote{Heterodyne Array Receiver Program, a single sideband array receiver with 16 mixers of the JCMT. HARP can be tuned between 325 and 375 GHz and has a instantaneous bandwidth of $\sim$2 GHz.} data \citep{buckle2009} in order to evaluate the velocity dispersion. Figure \ref{fig12} (left) displays the velocity-integrated intensity map and Figure \ref{fig12} (right) displays the intensity-weighted mean Doppler velocity of the region observed by HARP. We note that when we superimpose the intensity map of the 850 $\mu$m emission collected by POL-2 (the blue contours in Figure \ref{fig12} right) on the mean velocity map we find that the location of the emission matches very well that of the red-shifted arc of the cloud. It suggests a cloud emitting polarized light and moving away from us over a blue-shifted background (Figure \ref{fig12} right). A small clump in the northern part of the map is seen to account for most of the blue-shifted emission beyond \mbox{$-2$ km s$^{-1}$}; its spectrum is shown in the right panel of Figure \ref{fig13}.

The left panels of Figure \ref{fig13} show the integrated spectra for the central region (left) and the dust lane (center); they display a two-component structure. The fits of these spectra to two Gaussians give standard deviations of the blue- and red-shifted components ($\sigma_{B},\sigma_R$) $=$ ($0.90\pm0.02$, $0.56\pm0.01$) km s$^{-1}$ and \mbox{($0.60\pm0.03$, $0.79\pm0.02$) km s$^{-1}$} with mean values of ($-$1.25, 0.89) km s$^{-1}$ and \mbox{($-$1.08, 0.57) km s$^{-1}$} for the central region and the dust lane respectively. The average temperatures are 29.7 K for the central region and 20.8 K for the dust lane \citep{harvey2013}. The corresponding velocity dispersion caused by thermal turbulence are only at per mil level and therefore negligible: the non-thermal FWHM line-widths are equal to ($\Delta V_{B},\Delta V_R$) $=$ ($2.12\pm0.05$, $1.32\pm0.02$) km s$^{-1}$ and ($1.41\pm0.07$, $1.86\pm0.05$) km s$^{-1}$ for the central region and the dust lane, respectively. Since the red-shifted parts of the cloud trace well the 850 $\mu$m polarized emission from the region we use their velocity dispersion to calculate the B-field strength in the next section. However, we conservatively use uncertainties on velocity dispersion estimated from the combination of the three independent measurements added in quadrature: $^{13}$CO(1$-$0) emission \citep{li2014}, HARP blue-shifted, and HARP red-shifted spectra instead of using only the uncertainties from the fits to the HARP red-shifted spectra. The final velocity dispersion retained for further analysis are $1.32\pm0.40$ km s$^{-1}$ and $1.86\pm0.19$ km s$^{-1}$ for the central region and dust lane, respectively. Though the velocity dispersion obtained from CO(3$-$2) (red-shifted part) and from $^{13}$CO(1$-$0) are agreed within $\sim$20\% which supports the use the CO(3$-$2) line emission for calculating the gas velocity dispersion, we note that the CO(3$-$2) line available to us may not be optically thin. Therefore, the results regarding the magnetic field strength in the current paper are obtained under the assumption that the CO(3$-$2) line traces the observed dust volume.

\section{Results and Discussions} \label{sec:rslt}
\subsection{Magnetic Field Morphology}
The B-fields are measured over a region of $\sim$1.6 pc at a spatial resolution of $\sim$0.03 pc. With the irregular mass distribution and the presence of several protostar candidates and an early B star in the region, the magnetic field morphology of the observed region is expected to be complex. In fact, it is the case of the B-field patterns at the central region of the map, the highest emission region, which shows a drastic change of the plane-of-the-sky field directions (see Figure \ref{fig7} and Figure \ref{fig13b}). The field lines are perpendicular to each other running north-south and east-west. This is an indication of the existence of important field turbulence or of the magnetic field tangling in the dense region. The measured polarization angle dispersion of the order of $\sigma_\theta\sim15^{\circ}$ also supports the existence of the field turbulence. However, it is identified from the measurement of the ratio of the turbulent field to the large scale underlying field that the B-fields of the whole LkH$\alpha$ 101 region is still generally dominated over the turbulence (see Section \ref{subsec4.3}). Going farther away from the center of the map where the density is lower the B-fields tend to follow the periphery of the matter distribution. However, for the outermost parts, in particular, where the contour level (green curve in Figure \ref{fig13b}) is highly curved the B-fields are perpendicular to the curve. 

In the dust lane, the main orientation of the B-fields has the tendency to follow the filamentary structure (northwest-southeast) of the dust lane.

Similar tendency of the field running along the filamentary structure is also found in the low-density clumps enclosed by the green contours in Figure \ref{fig13b} scattered around the central region and the dust lane. This trend is better seen with the elongated clumps.

Prestellar cores, protostars, and then low mass stars are believed to form in filaments \citep{andre2014}. This paradigm is supported by simulations (e.g. \cite{2012ApJ...759...35I,soler2013}). B-fields help funneling matter onto the filaments. Since the dust lane is a subcritical filament (see Section \ref{subsec4.3}), the overall magnetic field direction parallel to its structure is in agreement with the popular picture of the B-field evolution in star-forming regions. Indeed, at large scale, B-fields are typically perpendicular to the main structure of filaments (e.g. \cite{mathews2014, 2015A&A...576A.104P}). At the scale of the core- and filament-size, it is found roughly that magnetic field runs perpendicular to a filament when the filament is gravitationally supercritical, but parallel when it is subcritical \citep{2013A&A...550A..38P,ward2017}. However, we note that criticality is not the only parameter to decide on the configurations of field vs. structure. {\it Planck} and BLASTPol\footnote{a balloon-borne polarimeter} data show a parallel-to-perpendicular transition at visual extinction $A_V\sim3$ mag \citep{2016A&A...586A.138P,2017A&A...603A..64S}. In addition, using SOFIA data in Serpens South \cite{2020NatAs.tmp..159P} found another transition from perpendicular back to parallel at $A_V\gtrsim21$ mag. Our dust lane has an average column density of \mbox{$1.44\times10^{22}$ cm$^{-2}$} (Table \ref{tab3}) which corresponds to $A_V\sim15.3$ mag. The visual extinction is obtained using the standard conversion factor between column densities and visual extinction \mbox{$N({\rm H_2})=9.4\times10^{20}$ cm$^{-2}$ $A_V$ mag} \citep{1978ApJ...224..132B}. Comparing with what is found in Serpens South, the dust lane lies in the region where the field is perpendicular to the filament (see Figure 3 of \cite{2020NatAs.tmp..159P}) and close to the region where the median relative orientation of the field and filament crosses $45^\circ$. This suggests that the perpendicular-to-parallel transition may vary depending not only on visual extinction but also on other parameters of a cloud. The central region is an example of this. It is also subcritical but the fields are complex with the presence of protostar candidates and an early B star. More statistics will help settling down this issue.

\subsection{Magnetic Field Strength}
\begin{table*}[!htb]
\begin{center}
\caption{Summary of physical parameters estimated for the central region and the dust lane.} 
\label{tab3}
\begin{tabular}{c c c c}
\hline
\hline
&&Central Region& Dust Lane\\
\hline
\multirow{2}{*}{\textit{Herschel}} & Number density, $n(\rm H_2)$ (cm$^{-3}$) & $(1.22\pm0.50)\times10^4$ & $(1.25\pm0.46)\times10^4$\\
&Column density, $N(\rm H_2)$ (cm$^{-2}$) & $(0.96\pm0.39)\times10^{22}$ & $(1.44\pm0.53)\times10^{22}$\\
\hline
HARP & Dispersion velocity, $\Delta V$ (km s$^{-1}$) & $1.32\pm0.40$ & $1.86\pm0.19$\\
\hline
\multirow{4}{*}{Unsharp Masking} & Polarization angle dispersion, $\sigma_\theta$ ($^\circ$) & $14.9\pm2.7$ & $14.7\pm2.6$\\
&$B_{\rm POS}$ ($\mu$G) & $91\pm32$ & $132\pm27$\\
&$\delta B/B_0$ & 0.269 & 0.265\\
&Mass-to-flux ratio, $\lambda$ & $0.27\pm0.15$ & $0.28\pm0.12$\\
\hline
\multirow{4}{*}{Structure Function} & Polarization angle dispersion, $\sigma_\theta$ ($^\circ$) & $14.8\pm4.8$ & $16.9\pm5.0$\\
& $B_{\rm POS}$ ($\mu$G) & $92\pm42$ & $144\pm36$\\
& $\delta B/B_0$ & 0.267 & 0.309\\
&Mass-to-flux ratio, $\lambda$ & $0.27\pm0.16$ & $0.32\pm0.15$\\
\hline\hline
\end{tabular}
\end{center}
\end{table*}

\begin{figure*}[!htb]
\centering
    \includegraphics[trim=1.cm .2cm 2.5cm 2.5cm,clip,width=8.7cm]{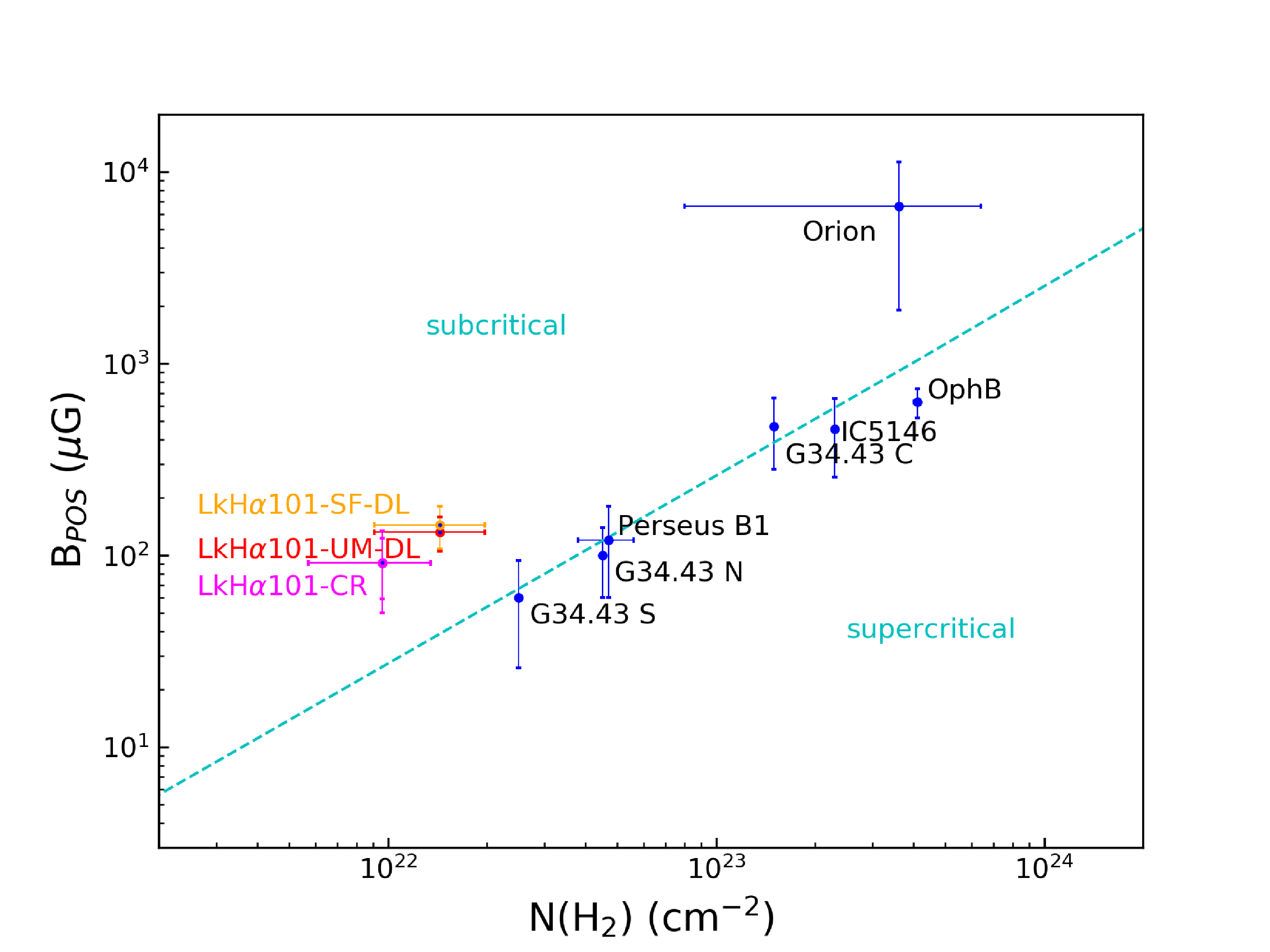}
    \includegraphics[trim=1.cm .2cm 2.5cm 2.5cm,clip,width=8.7cm]{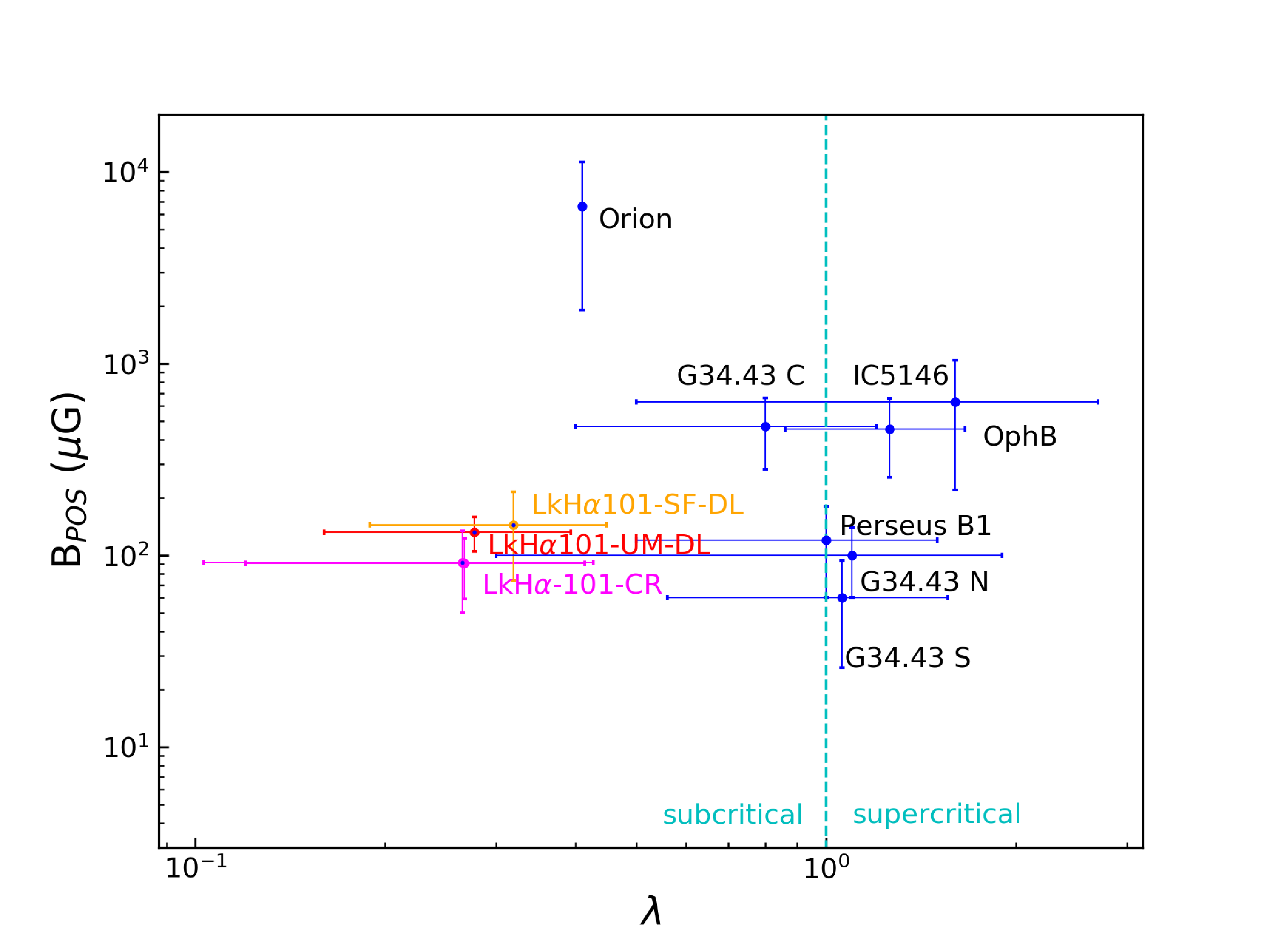}
\caption{Dependence of $B_{\rm POS}$ on column density, $N(\rm H_2)$, (left) and mass-to-magnetic-flux ratio, $\lambda$, (right). Red and yellow are for the dust lane and for unsharp masking (UM) and structure function (SF) methods, respectively. Purple are for the central region. The dashed lines separate the super- and sub-critical conditions (the one in the left panel is obtained from Equation \ref{eq8}).} \label{fig14}
\end{figure*}

\begin{figure*}
\centering
\includegraphics[trim=1.5cm 0.5cm 2.cm 0.cm,clip,width=11cm]{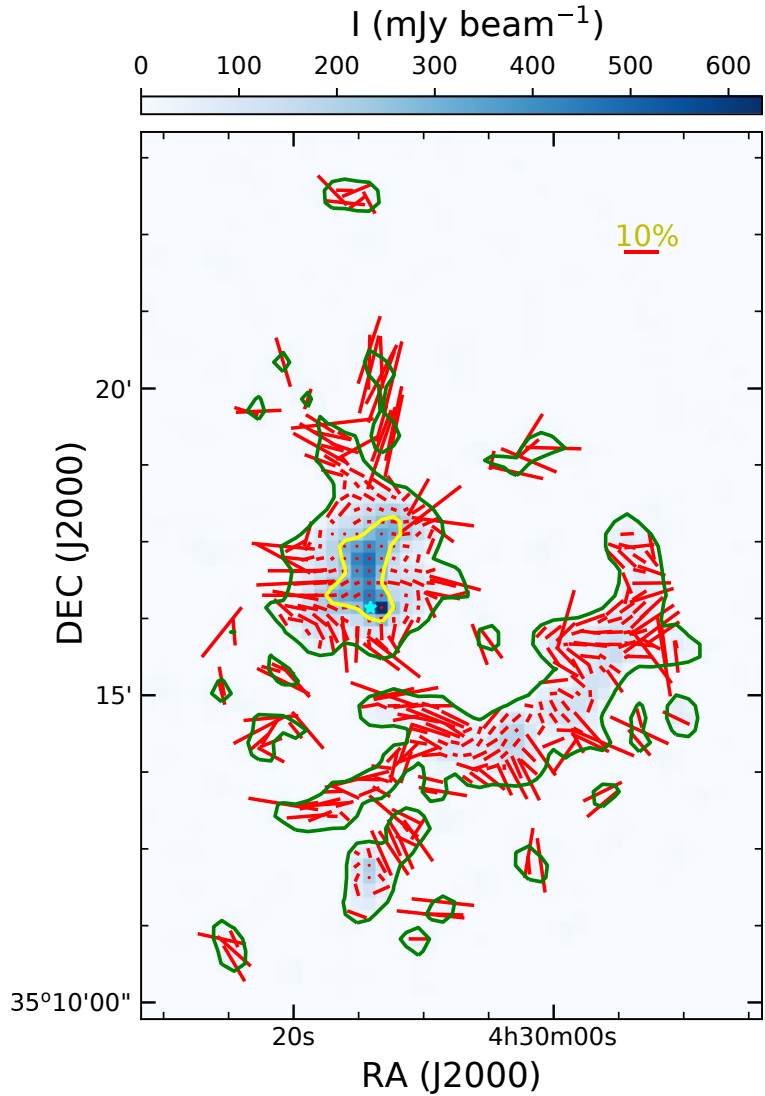}
\caption{Same as Figure \ref{fig7} but the length of the line segments is now proportional to the polarization fraction, $P$(\%). A 10\% line segment is shown for reference.} \label{fig13b}
\end{figure*}

\begin{figure}[!htb]
\centering
    \includegraphics[trim=.5cm 0.1cm 1.5cm 1.5cm,clip,width=8.3cm]{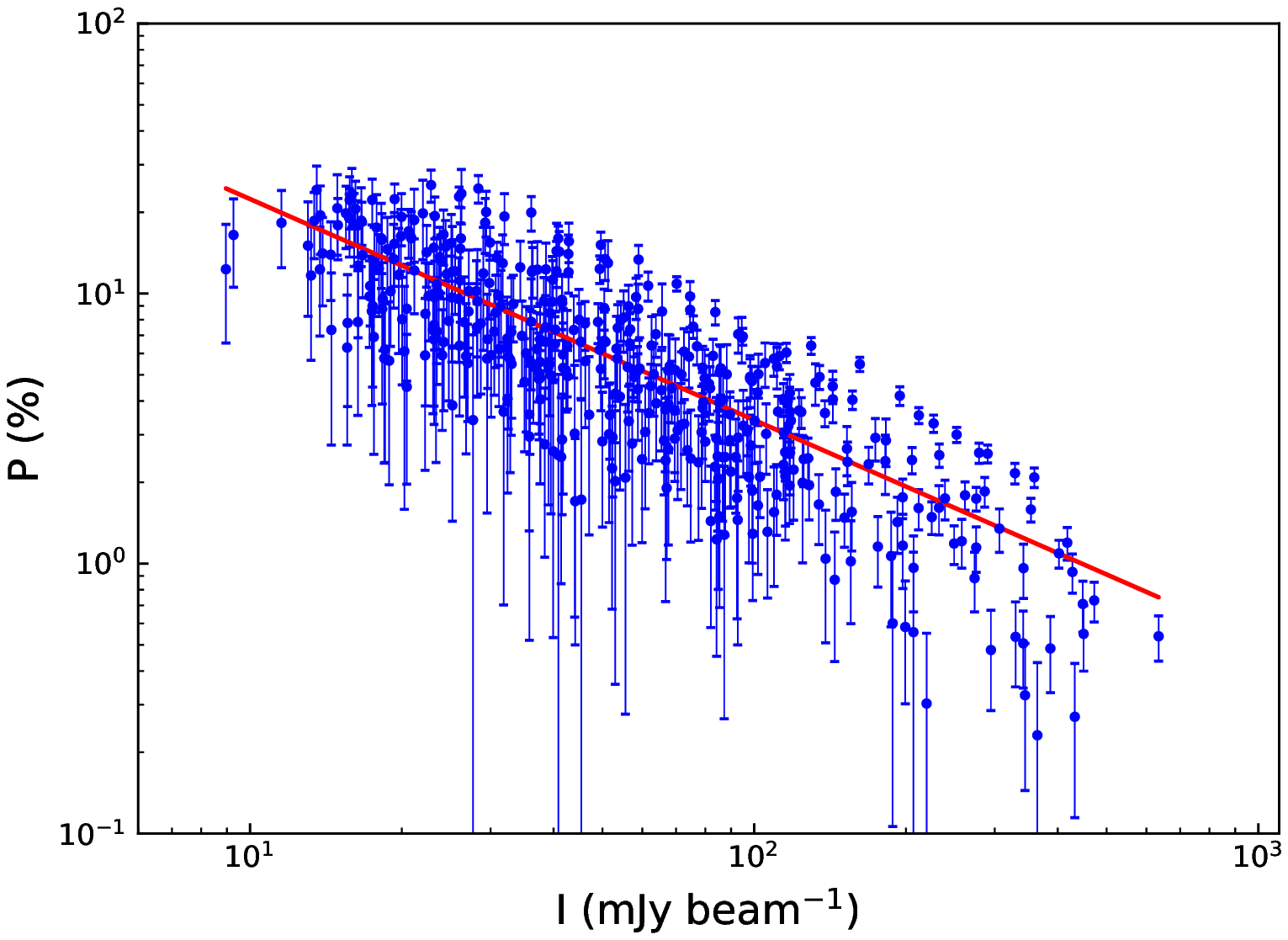}
\caption{Dependence of polarization fraction on the total intensity. $P$ tends to decrease with increasing $I$. The solid line is the best fit to a power law function (see Section \ref{subsec:grainalign}).} \label{fig15}
\end{figure}

For the DCF method to be applicable, \citet{ostriker2001} suggested that the polarization angle dispersion should be smaller than $\sim25^\circ$, which is the case of the present data set. Using the H$_2$ number densities and the non-thermal velocity dispersion along the line-of-sight given in Sections \ref{subsec:dens} and \ref{subsubsec:vedis}, Equation \ref{eq7} gives the magnitude of the magnetic field in the plane of the sky, $B_{\rm POS}$. The uncertainties of $B_{\rm POS}$ are propagated using the following relation:
\begin{equation}
\frac{\delta B_{\rm POS}}{B_{\rm POS}}=\sqrt{\left(\frac{1}{2}\frac{\delta n(\rm H_2)}{n(\rm H_2)}\right)^2+\left(\frac{\delta\Delta V}{\Delta V}\right)^2+\left(\frac{\delta\sigma_\theta}{\sigma_\theta}\right)^2} ,
\end{equation}
where $\delta n(\rm H_2)$, $\delta\Delta V$, and $\delta\sigma_\theta$ are the uncertainties of $n(\rm H_2)$, $\Delta V$, and $\sigma_\theta$, respectively. 

The magnetic field strengths obtained from two methods are listed in Table \ref{tab3} for the central region and the dust lane. As shown, the two methods yield the similar magnetic field strength, of $B_{\rm POS}\sim 91$ $\mu$G for the central region and $B_{\rm POS}\sim 138$ $\mu$G for the dust lane.

It is very interesting to note that the mean measured field strength of $\sim$115 $\mu$G is very close to the value of 100 $\mu$G adopted by \citet{zhang2020} explain the observed fragmentation length-scale (core spacing) of star-forming filaments in the X-shape Nebula of the California molecular cloud.

\subsection{Mass-to-flux ratio}\label{subsec4.3}
The relative importance of gravity to magnetic fields is usually described by the mass-to-flux ratio, $M/\Phi$. In the units of the critical value, the mass-to-flux ratio is given by the formula from \cite{crutcher2004},
\begin{eqnarray}\label{eq8}
\lambda=\frac{(M/\Phi)_{\rm observed}}{(M/\Phi)_{\rm critical}}=7.6\times10^{-21}\frac{N(\rm H_2)}{B_{\rm POS}}
\end{eqnarray} where $(M/\Phi)_{\rm critical}=1/(2\pi\sqrt{G})$, $G$ is the gravitational constant, $N(\rm H_2)$ is the gas column density measured in cm$^{-2}$, and $B_{\rm POS}$ is the strength in $\mu$G. As $B_{\rm POS}$ is the magnetic field component in the plane of the sky, a factor of 3 is introduced to correct for geometrical biases \citep{crutcher2004}.

Plugging $B_{\rm POS}$ and $N({\rm H_2})$ obtained for the central region and the dust lane into Equation \ref{eq8}, we obtain $\lambda$ for these regions, which are listed in Table \ref{tab3}.

The left panel of Figure \ref{fig14} shows the measured values of $B_{\rm POS}$ as a function of the column density $N(\rm H_2)$ for the different regions using the results from Table \ref{tab3} together with those of previous studies \citep{pattle2017,soam2018,coude2019,soam2019,wang2019} surveyed by POL-2. The dashed line corresponds to the separation between sub-critical and super-critical regions. The line is obtained from Equation \ref{eq8} setting $\lambda=1$ where we have equal contribution of mass and magnetic flux. Compared to other regions, LkH$\alpha$ 101 has particularly low values of $N(\rm H_2)$. The Auriga-California region lies well above the dashed line, which is sub-critical.

The right panel of Figure \ref{fig14} shows the variation of $B_{\rm POS}$ with the mass-to-flux ratio, $\lambda$, for the different regions as in the left panel. Compared to the other regions, the Auriga-California has a rather low value of $\lambda$ and lies in the sub-critical regime. 

Being sub-critical at the same time as being the densest region of Auriga-California may help to explain the very low star formation efficiency in comparison with that of the OMC as being discussed in Section \ref{sec:intro}. 

However, the star formation efficiency of a cloud depends on several parameters other than $\lambda$, such as matter distribution, evolutionary stage, and turbulence. More detailed studies are required to understand why the star formation efficiency in Auriga-California is lower than that of the OMC. Moreover, the measured ratio of the turbulent component to the large-scale component of the magnetic field is $\delta B/B_0\sim$0.3, suggesting that the effect of B-fields is dominant over the turbulence in the region.

\subsection{Dust Polarization and Grain Alignment}\label{subsec:grainalign}
We now analyze the spatial variation of the polarization fraction within the Auriga-California and explore grain alignment physics.

\begin{figure}[!htb]
\centering
\includegraphics[trim=0.5cm 0.2cm 1.cm 1.5cm,clip,width=8.5cm]{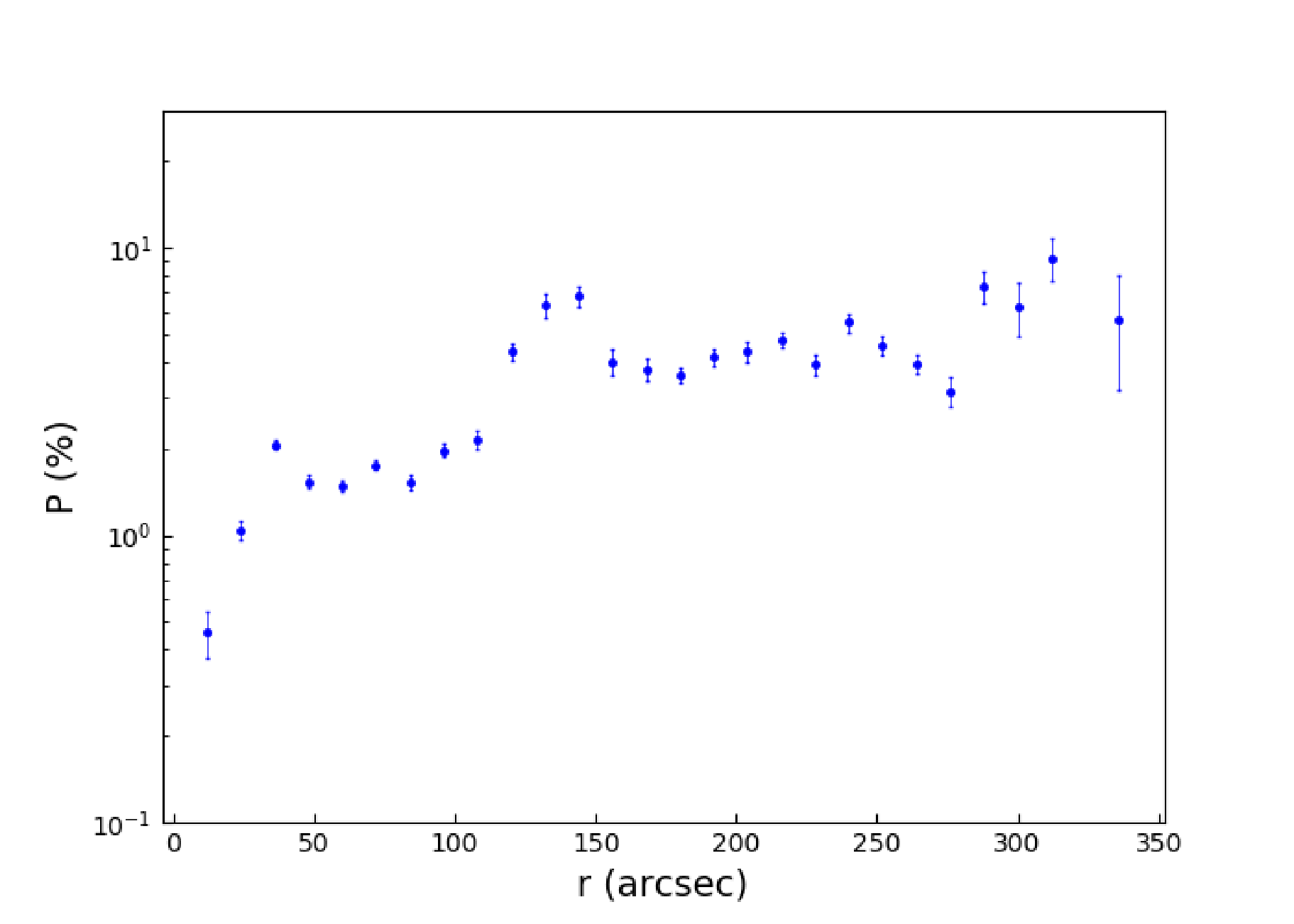} 
\caption{Dependence of $P$ on the distance from LkH$\alpha$ 101. The polarization fraction decreases with $r$ for $r<150''$.} \label{fig16}
\end{figure}

\begin{figure*}[!htb]
\centering
\begin{tabular}{cc}
\includegraphics[trim=.5cm 6.cm 4.cm 4.cm,clip,width=7.cm]{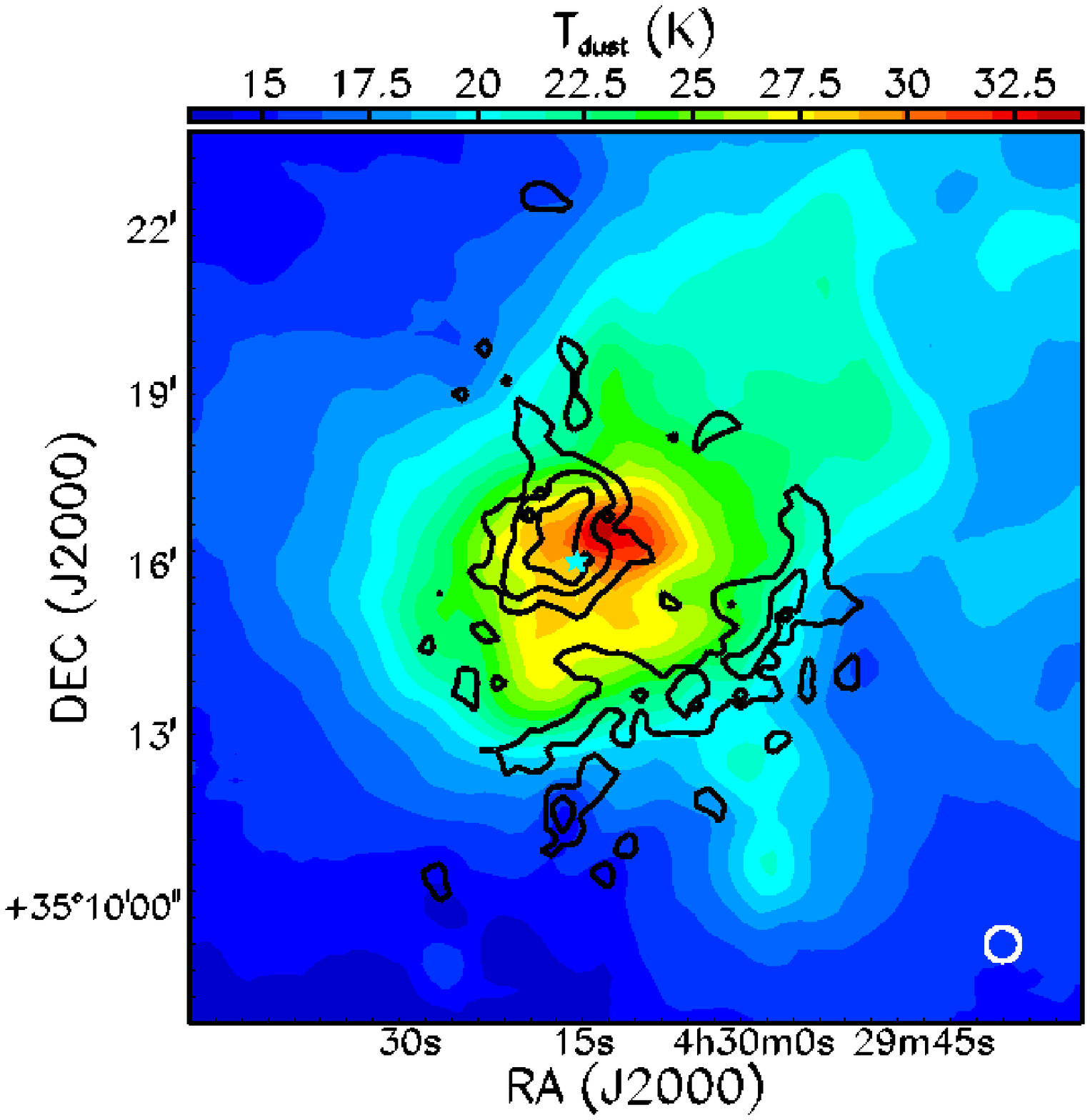}&
\includegraphics[trim=.5cm -0.3cm 1.2cm 1.4cm,clip,width=8.5cm]{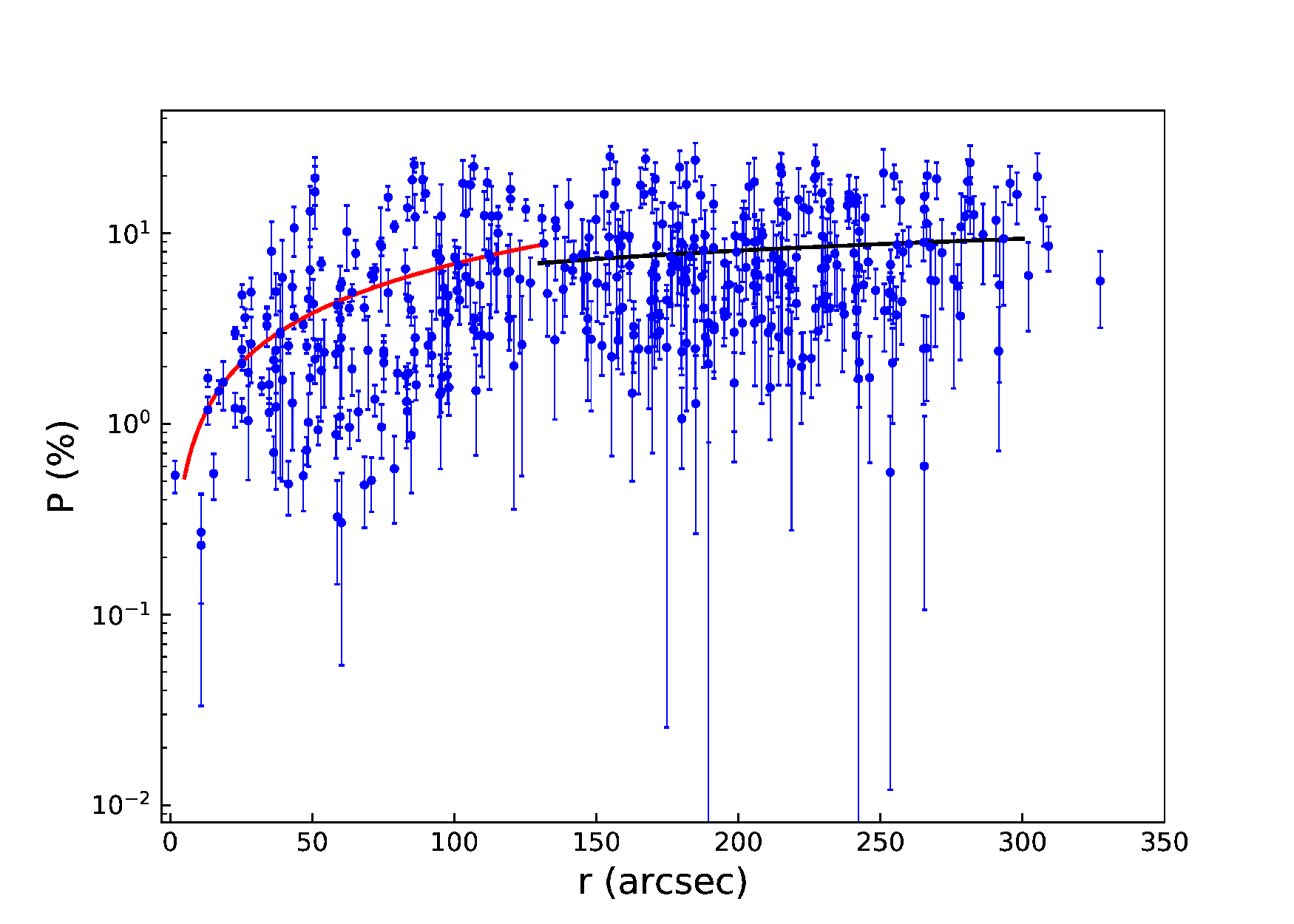}\\
\includegraphics[trim=.5cm 0.35cm 1.2cm 1.cm,clip,width=8.cm]{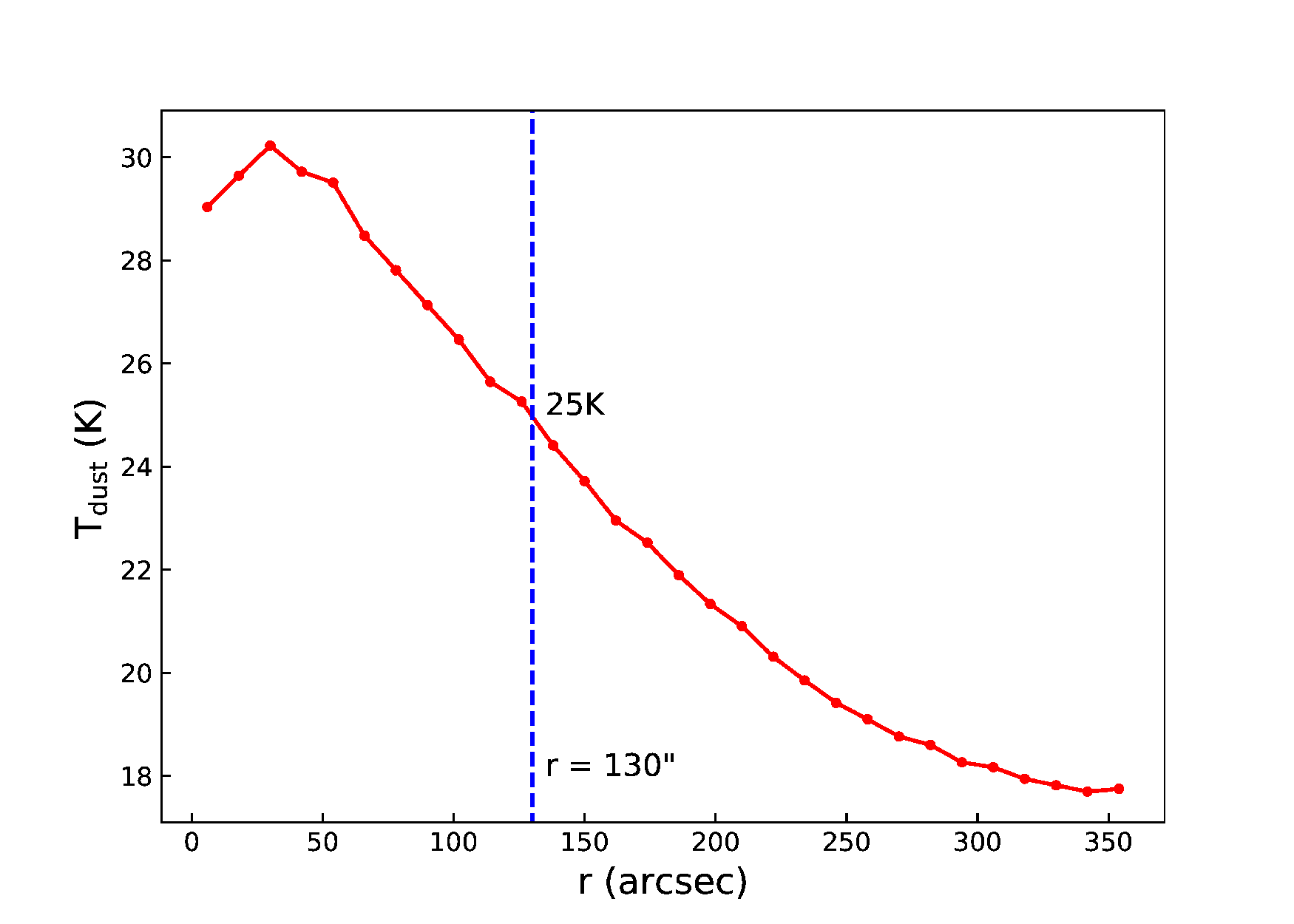}&
\includegraphics[trim=.5cm 0.35cm 1.2cm 1.cm,clip,width=8.cm]{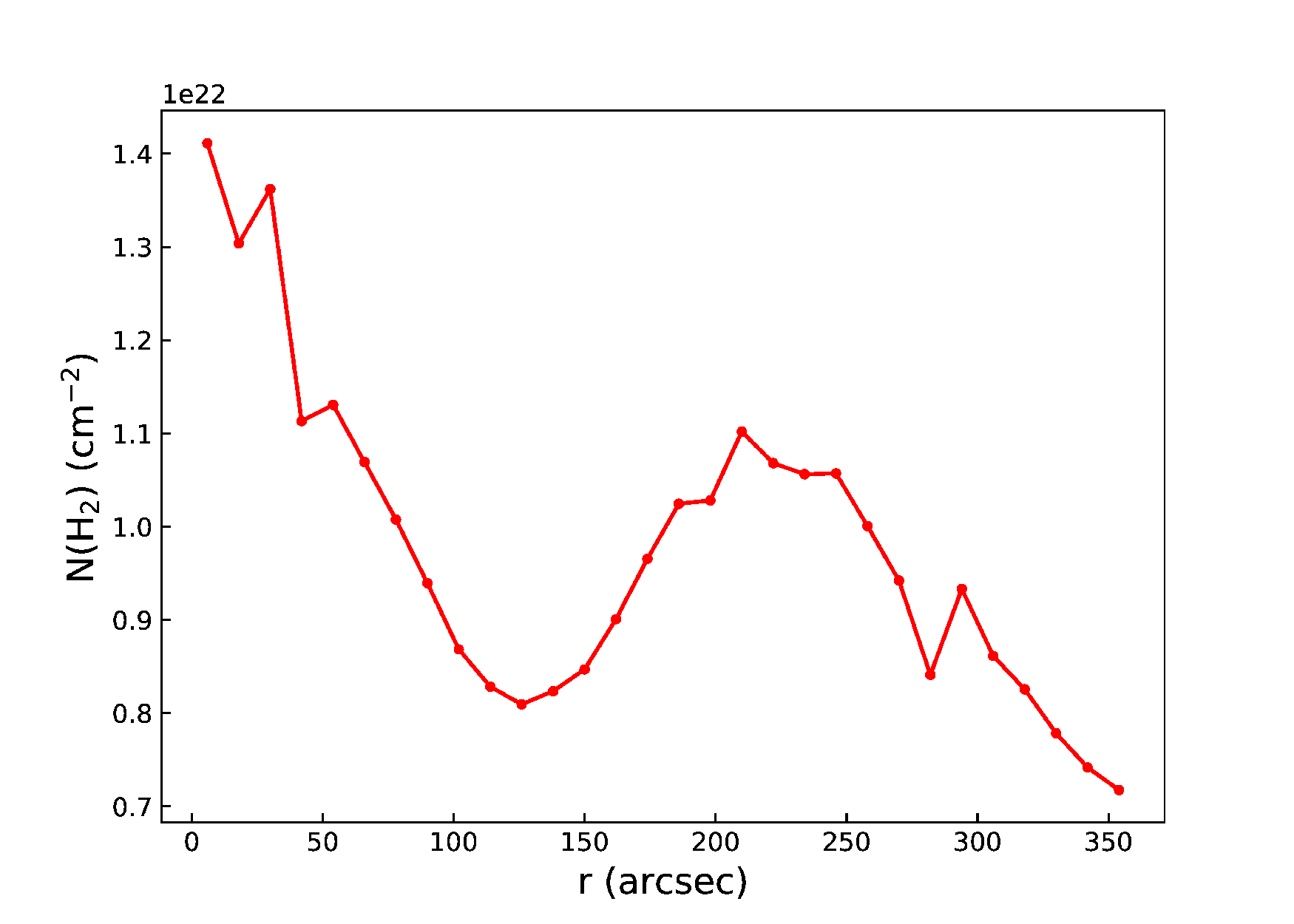}
\end{tabular}     
\caption{Upper left panel: Dust temperature map (color scale) overlaid by 850 $\mu$m emission (contours). The \textit{Herschel} beam size (36.6$''$) is shown in the lower right corner of the map. Location of the early B star, LkH$\alpha$ 101, is at the cyan star marker. Upper right panel: Dependence of $P$ and on $r$. The $r$-dependence of $P$ is fitted to a power law function for $r<130''$ (red curve) and $r>130''$ (black curve) (see text). Lower left panel: Dependence of $T_{\rm dust}$ on $r$. The blue vertical dashed line shows the distance, $r=130''$, from the star at which the average dust temperature is $\sim25$ K. Lower right panel: Dependence of $N(\rm H_2)$ on $r$.} \label{fig17}
\end{figure*}

Figure \ref{fig13b} displays the map of the inferred magnetic line segments similar to Figure \ref{fig7} but the length of the line segments is now proportional to the polarization fraction. It is clearly seen from the figure that the degree of polarization is higher in the more diffuse regions, but it drops significantly in the dense central region with maximum emission intensity (yellow contour). 
To see explicitly how the polarization fraction, $P$, changes with the total intensity $I$, in Figure \ref{fig15} we show the variation of $P$(\%) with $I$ (the $\delta P$-$\delta\theta$ cut has been applied on the data). The observed polarization fraction tends to decrease with increasing the total intensity, which is usually referred to as {\it polarization hole}. A power law fit of the form $P\propto I^{-\alpha}$ gives the power index $\alpha=0.82\pm0.03$. The uncertainty in $\alpha$ is obtained from the fit and does not include systematic errors. 
The value of $\alpha$ found for the LkH$\alpha$ 101 region is in the expected range for molecular clouds, between 0.5 and 1. In other regions surveyed by BISTRO, the estimated values are $\alpha=0.8$ for $\rho$ Ophiuchus A \citep{kwon2018}, 0.9 for $\rho$ Ophiuchus B \citep{soam2018}, 1.0 for $\rho$ Ophiuchus C \citep{liu2019}, and 0.9 for Perseus B1 \citep{coude2019}. Using a different approach working with non-debiased data, \citet{pattle2019} found $\alpha=0.34$ for Oph A and $\alpha=0.6-0.7$ for Oph B and C, which are significantly smaller than those obtained by the previous authors. 

We note that $\alpha$ is widely used as an indicator of dust grain alignment efficiency. One expects to have $\alpha=0$ for constant efficiency of grain alignment and $\alpha=1$ for grain alignment that only occurs in the outer layer of the cloud, with complete loss of grain alignment inside the cloud (\citealt{2008ApJ...674..304W}). Various observations report the loss of grain alignment (i.e., $\alpha\sim1$) at large visual extinction of $A_V \sim 20$ toward starless cores \citep{alves2014, jones2014grain, santos2019}, which is consistent with the prediction by RAT theory \citep{Hoang:2020vg}. Therefore, the best-fit value $\alpha=0.82$ here reveals that grain alignment still occurs inside the cloud, but with a decreasing efficiency. This is consistent with the prediction from Radiative Torque (RAT) alignment theory (\citealt{lazarianhoang2007,hoang2016}) that grains inside the cloud can still be aligned due to stellar radiation from the LkH$\alpha$ 101 star.

Since LkH$\alpha$ 101 is the only early B star in the region, we display in Figure \ref{fig16} the radial dependence of the polarization fraction averaged over 12$''$-wide rings centered on the star. There is evidence for a rapid decrease of the polarization fraction at distances from the star smaller than $\sim150''$. LkH$\alpha$ 101 is 15 times more massive than the Sun with a luminosity of 8$\times$10$^3$ \(\textup{L}_\odot\) \citep{herbig2004}. The angular distance from LkH$\alpha$ 101 at which the mean energy density in empty space is equal to that of the interstellar radiation field, 2.19$\times$10$^{-12}$ erg cm$^{-3}$ \citep{draine2010}, is $\sim3300''$. This distance is significantly larger than the distance where we find $P$ to decrease.

To better understand the decreasing feature of $P$, in particular, the drastic decrease at $r<150''$, we study the behavior of the polarization fraction $P$, column density $N(\rm H_2)$, and dust temperature $T_{\rm dust}$ as a function of distance $r$ from LkH$\alpha$ 101. The two latter quantities are taken from \textit{Herschel} \citep{harvey2013}. The upper left panel of Figure \ref{fig17} shows the dust temperature map overlaid by 850 $\mu$m emission. Figure \ref{fig17} (upper right) shows the observational data (symbols) and our power-law fit, $P = ar^b$, with two slopes of $b=0.86\pm0.04$ for $r<130''$ and $b=0.35\pm0.05$ for $r>130''$. Here, the radius $r\sim 130''$ is chosen to be similar to the location of the separation between the central region and the dust lane (the valley in $N(\rm H_2)$ vs $r$ at $r\sim130''$ in Figure \ref{fig17} lower right). For the outer region ($r>130''$), the polarization degree decreases slowly with decreasing $r$, which implies a slow decline in the grain alignment efficiency. However, the polarization degree decreases rapidly when approaching the location of LkH$\alpha$ 101 for $r<130''$, whereas $T_{\rm dust}$ increases as expected from stronger heating by the star (Figure \ref{fig17} lower left), except for only two data points close to the star at $r<25''$ with decreased $T_{\rm dust}$. Note that LkH$\alpha$ 101 is located at the location of the highest column density (Figure \ref{fig17} lower right, for the column density map see Figure \ref{fig11}).

In theory, the rapid decrease of $P$ for $r<130''$ can arise from (1) significant loss of grain alignment, and (2) strong variation of the magnetic field. According to the popular RAT alignment theory (\citealt{lazarianhoang2007,hoang2016}), the loss of grain alignment is induced by the decrease of the incident radiation field that can align grains, and/or the increase of the gas density that enhances grain randomization. In our situation, grains are subject to increasing radiation flux from the LkH$\alpha$ 101 star when $r$ decreases, and the gas density in the central region is not very high of \mbox{$n(\rm H_2)\sim 10^4$ cm$^{-3}$}. As a result, the degree of grain alignment is expected to increase with increasing the local radiation energy density described by $T_{\rm dust}$ \citep{Hoang:2020vg}, which would result in the increase of dust polarization \citep{lee2020physical}. Therefore, the rapid decline of $P$ for $r<130''$ (Figure \ref{fig17} upper right), or $T_{\rm dust}>25$ K (Figure \ref{fig17} lower left) may challenge the popular theory of grain alignment based on RATs.

Recently, \cite{hoang2019rotational} suggested that RATs from an intense radiation field can spin grains up to extremely fast rotation. As a result, the centrifugal stress can exceed the maximum tensile strength of grain material, resulting in the disruption of large grains into smaller fragments. A detailed modeling of grain disruption toward a dense cloud with an embedded source is presented in \cite{Hoang:2020vg}. Since such large grains dominate dust polarization at far-IR/submm, the degree of dust polarization is found to decrease with increasing the local radiation energy density \citep{lee2020physical}. For the hydrogen density of $n(\rm H_2)$ $\sim 10^{4}$ cm$^{-3}$ listed in Table \ref{tab3}, numerical modeling in \cite{lee2020physical} implies that the polarization degree first increases with grain temperature due to increase of the radiation flux, then it decreases when $T_{\rm dust}$ exceeds $\sim 25$ K. This can explain the decrease of $P$ at small $r$ (or high $T_{\rm dust}$) observed in Figure \ref{fig17}. We note that previous observations by {\it Planck} (\citealt{guillet2018dust}) and SOFIA/HAWC+ (\citealt{tram2020understanding}) also report the decrease of $P$ when $T_{\rm dust}$ exceeds some value, which were explained by means of rotational disruption by RATs. A detailed modeling to understand the dependence of $P$ vs $r$ is beyond the scope of this paper.

The tangling of magnetic fields is usually invoked to explain the decrease of the polarization fraction, $P$, with the emission intensity, $I$, or column density (usually referred to as polarization hole, see \citealt{2019FrASS...6...15P} for a review). However, there is no quantitative study which addresses the role of field tangling in causing the polarization holes at the scales of JCMT observations (or at smaller scales, i.e., CARMA\footnote{Combined Array for Research in Millimeter-wave Astronomy}/SMA\footnote{Submillimeter Array} and ALMA\footnote{Atacama Large Millimeter/submillimeter Array}). The fact that we observe relatively ordered B-field in the region supports the RATD effect, but cannot rule out the role of the field tangling. Single dish vs. interferometric observations of Class 0 protostellar cores frequently show ordered fields on multiple spatial scales, even if the fields have very different morphologies on large (JCMT $\sim$10,000 au resolution) and small (ALMA $\sim$100 au resolution) scales. JCMT, CARMA, and ALMA maps of Ser-emb 8 \citep{hull2017a} and Serpens SMM1 \citep{hull2017b} show ordered fields on all scales. The case of SMM1 is particularly striking, where a well ordered East-West field at JCMT scales (at least on the outskirts of the location of SMM1) is seen, and then a very well ordered field North-South at CARMA scales, and then an highly complex field with multiple plane-of-sky components at ALMA scales. There are several other examples (see \citet{sadavoy2018a, sadavoy2018b} on IRAS 16293 and VLA 1623).

Analyses of the Planck polarization data (in particular, \citet{2015A&A...576A.104P} and \citet{planck2018}) found that the polarization hole effect can be attributed entirely to turbulent tangling of the magnetic field along the line-of-sight, and that the dust grain-alignment efficiency is constant across a wide range of column densities. However, those spatial scales tend to be significantly larger than what we are dealing with in BISTRO observations. Similar conclusions are also reached in a work performing similar statistical analyses of ALMA data, at spatial scales closer to the JCMT scales \citep{le2020statistical}.
 
In summary, polarization holes have been observed in many star-forming regions (see \citealt{2019FrASS...6...15P}). In the absence of field tangling, polarization holes observed toward protostars are inconsistent with the RAT alignment theory, but can be explained by the joint effect of grain alignment and rotational disruption by RATs. There are several possible solutions that are being explored to explain the polarization holes, including RATD, magnetic field tangling. Detailed modeling of dust polarization taking into account grain alignment, disruption, and realistic magnetic fields is required to understand the origins of the polarization hole.

\section{Conclusions} \label{sec:conclusions}
Using POL-2, we have measured the morphology and strength of the magnetic field of the LkH$\alpha$ 101 region for the first time. While the magnetic field is generally parallel to the filamentary structure of the dust lane it is quite complex in the central region. In low density clumps, in particular the elongated ones, the field is more aligned with the matter structure. The field strength is $\sim91$ $\mu$G for the central region and \mbox{$\sim138$ $\mu$G} for the dust lane. The polarization angle dispersion obtained from both unsharp masking and structure function methods are in good agreement.

HARP data are used to evaluate the velocity dispersion of the regions which show that the red-shifted component of the cloud matches the region where we observed dust polarization; it matches the dust lane particularly well.

The power-law index of the dependence of the polarization fraction on total intensity was found to be $0.82\pm0.03$ which is in the expected range for molecular clouds. The mass-to-magnetic-flux-ratios in units of the critical value are $\lambda=0.27$ for the central region and $\lambda=0.30$ for the dust lane, smaller than unity and the smallest among regions surveyed by POL-2. The regions are sub-critical, i.e. the B-fields are strong enough to be able to resist gravitational collapse. The ratio of the turbulent field to the underlying field $\delta B/B_0\sim0.3$ means that the underlying field is dominated over turbulent field. LkH$\alpha$ 101 is the densest region of Auriga-California. This gives supporting arguments for the low star forming efficiency of Auriga-California in comparison with, in particular, the OMC. However, further study is required to explain the contrasting star formation efficiency of the AMC and the OMC.

Finally, we found that the polarization fraction decreases with increasing proximity to the B star, LkH$\alpha$ 101, which is also the highest density region of the observed field. This effect is similar to many previous observations. The rapid decrease of $P$ with distance from the only B star in the region for $r<130''$ is inconsistent with the popular RAT alignment theory, but could be explained by the joint effect of grain alignment and rotational disruption by RATs. Other effects such as the geometry of the magnetic fields due to turbulence could potentially explain the polarization hole. More studies are required to understand the nature of polarization holes.

\acknowledgments{
The James Clerk Maxwell Telescope is operated by the East Asian Observatory on behalf of The National Astronomical Observatory of Japan, Academia Sinica Institute of Astronomy and Astrophysics in Taiwan, the Korea Astronomy and Space Science Institute, the National Astronomical Observatories of China and the Chinese Academy of Sciences (grant No. XDB09000000), with additional funding support from the Science and Technology Facilities Council of the United Kingdom and participating universities in the United Kingdom and Canada. Additional funds for the construction of SCUBA-2 and POL-2 were provided by the Canada Foundation for Innovation. The data taken in this paper were observed under project code M16AL004. We thank L.H.M. Ngan for contributions to this work in its earlier phase. We are grateful to Prof. Pierre Darriulat and other VNSC/DAP members for their useful comments and discussions. This research is funded by Vietnam National Foundation for Science and Technology Development (NAFOSTED) under grant number 103.99-2019.368. T.H. acknowledges the support by the National Research Foundation of Korea (NRF) grant funded by the Korea government (MSIT) through the Mid-career Research Program (2019R1A2C1087045). C.L.H.H. acknowledges the support of the NAOJ Fellowship and JSPS KAKENHI grants 18K13586 and 20K14527. J. D. F and D.J. are supported by the National Research Council of Canada and by individual NSERC Discovery Grants. C.W.L. is supported by Basic Science Research Program through the National Research Foundation of Korea (NRF) funded by the Ministry of Education, Science and Technology (NRF-2019R1A2C1010851). M.T. is supported by JSPS KAKENHI grant Nos.18H05442,15H02063, and 22000005. JK is supported JSPS KAKENHI grant No.19K14775. A.S. acknowledge the support from the NSF through grant AST-1715876. The authors wish to recognize and acknowledge the very significant cultural role and reverence that the summit of Maunakea has always had within the indigenous Hawaiian community. We are most fortunate to have the opportunity to conduct observations from this mountain.
}
\software{Starlink \citep{currie2014}, Astropy \citep{robitaille2013astropy,price2018astropy}}
\facility{James Clerk Maxwell Telescope (JCMT)}

\bibliographystyle{aasjournal}
\bibliography{Auriga_1}{}

\appendix
\counterwithin{figure}{section}
\counterwithin{table}{section}
\section{Characteristics of the raw data} \label{appendix}
\begin{figure*}[!htb]
\centering
  \begin{tabular}{cc}
    \includegraphics[trim=1.3cm 6.3cm 3.3cm 5.75cm,clip,width=5.8cm]{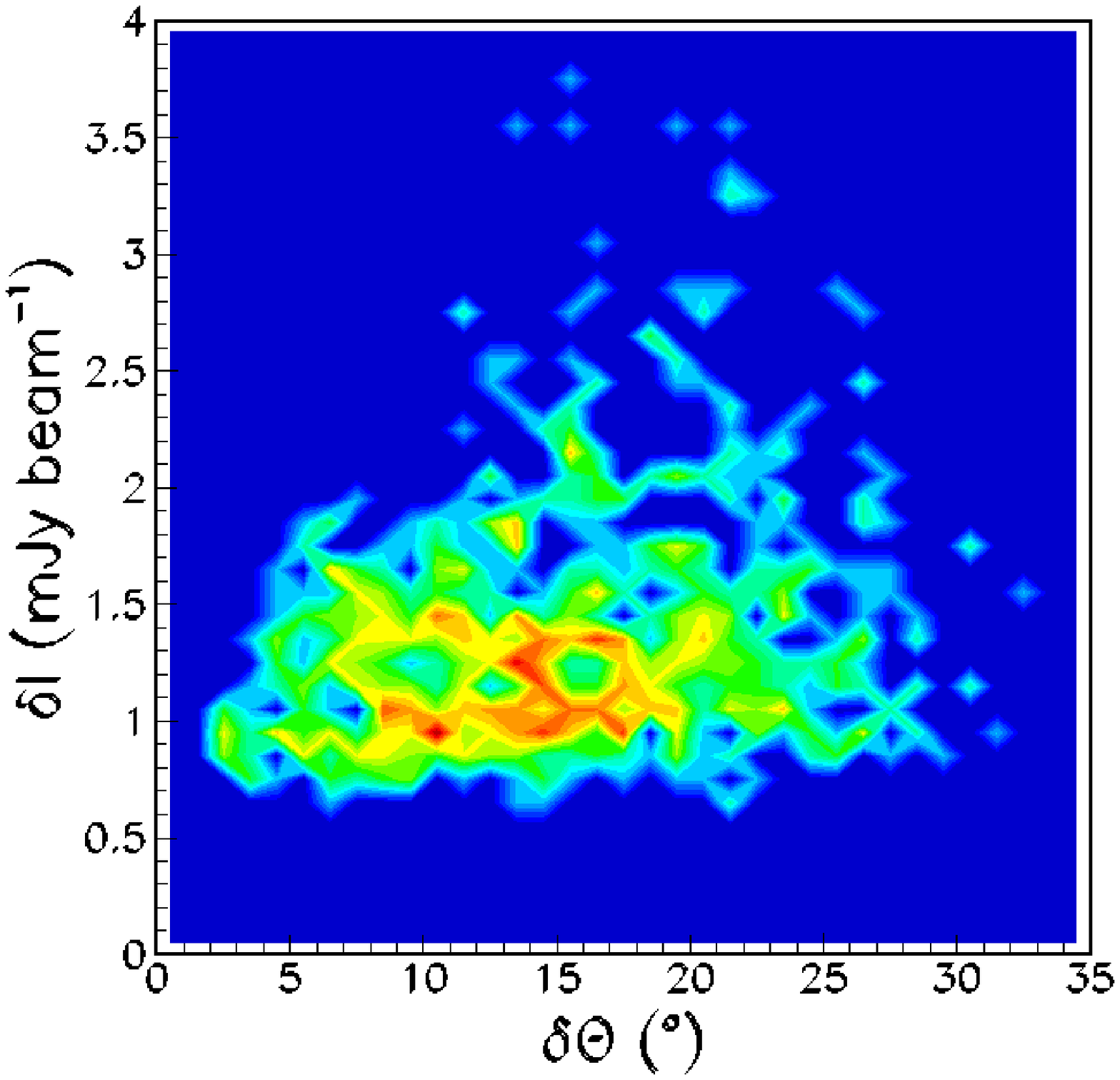}&
    \includegraphics[trim=1.3cm 6.3cm 3.3cm 5.75cm,clip,width=5.8cm]{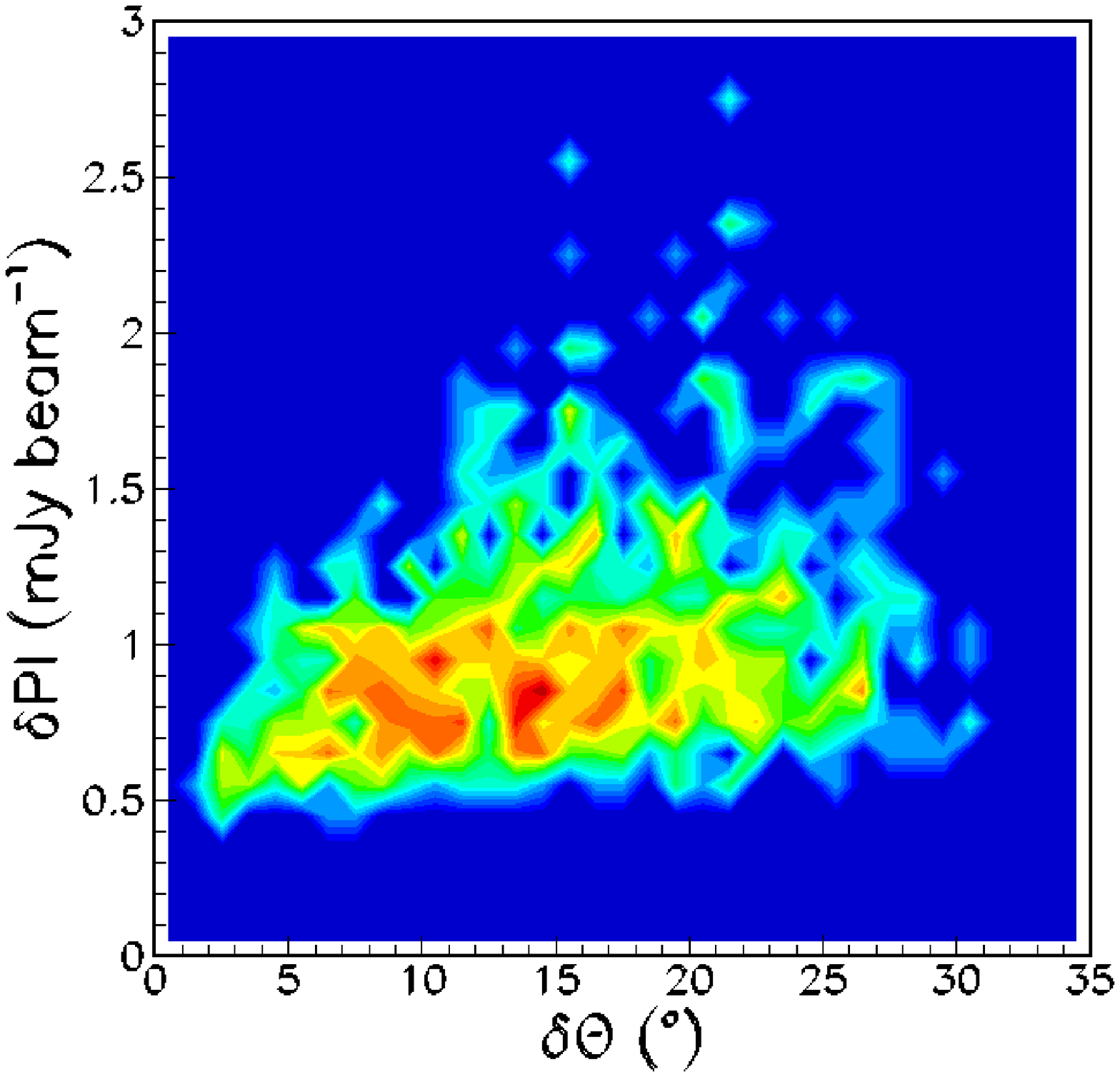}\\
    \includegraphics[trim=1.3cm 6.3cm 3.3cm 5.75cm,clip,width=5.8cm]{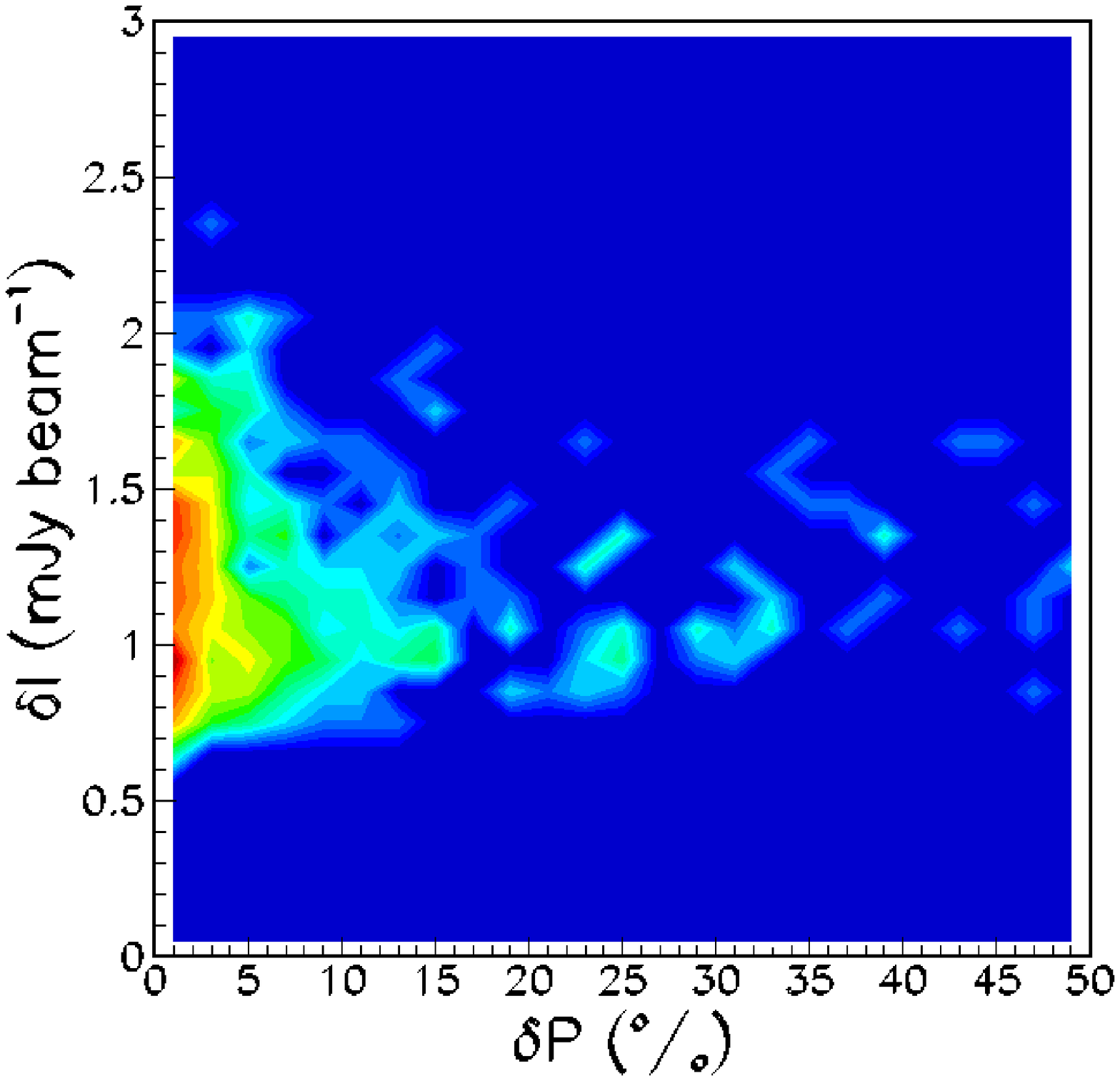}&
    \includegraphics[trim=1.3cm 6.3cm 3.3cm 5.75cm,clip,width=5.8cm]{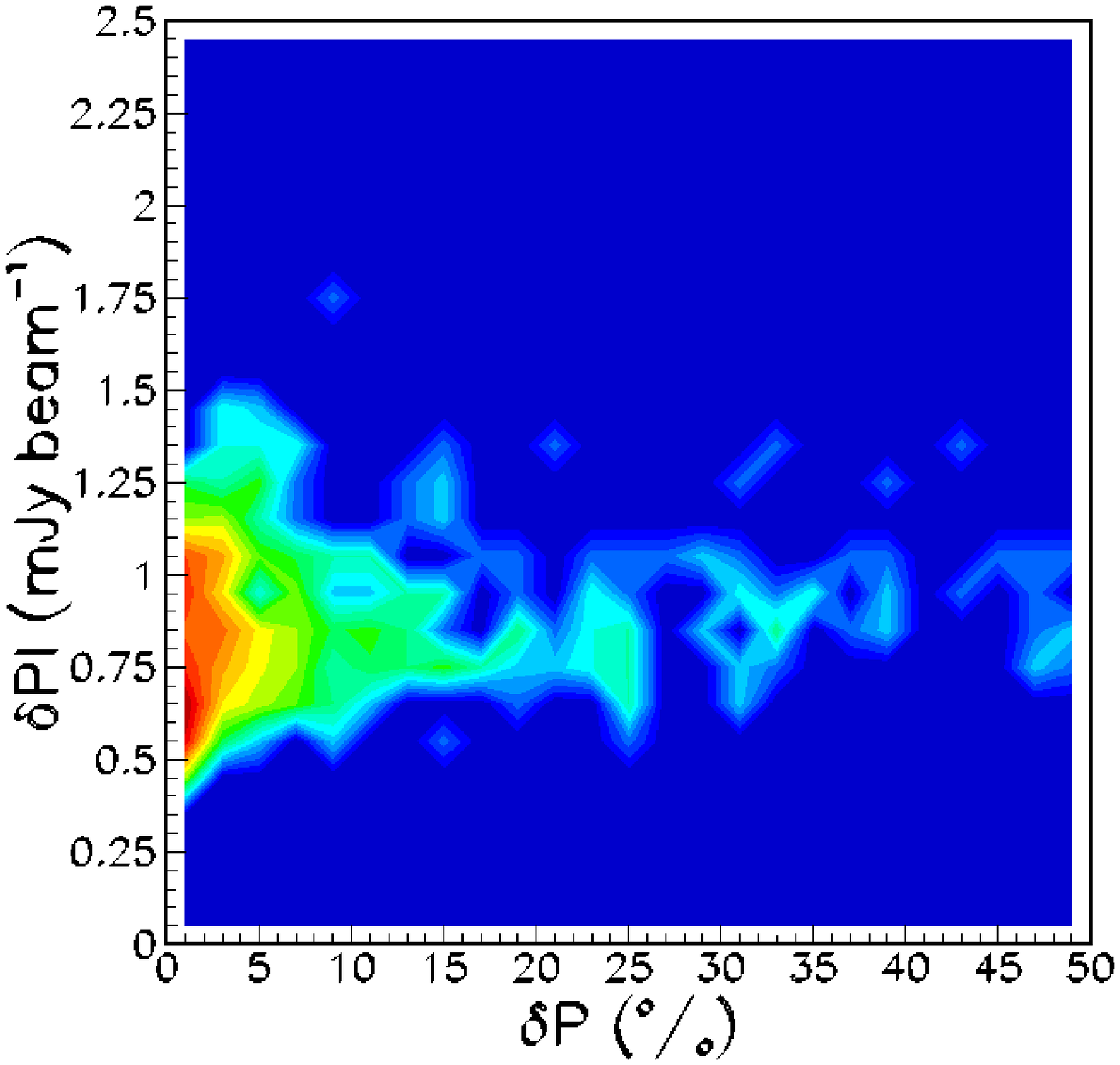}
  \end{tabular}
\caption{Correlations of measurement uncertainties from left to right, top to bottom (the color represents the number of data points in logarithmic scales): $\delta I$ vs $\delta\theta$, $\delta PI$ vs $\delta\theta$, $\delta I$ vs $\delta P$, and $\delta PI$ vs $\delta P$.} \label{figA3}
\end{figure*}

\begin{table*}[!htb]
\begin{centering}
\caption{Means and RMS values of $I$, $\delta I$, $PI$, $\delta PI$, $P$, $\delta P$, $\theta$, and $\delta\theta$.}\label{tab1}
\begin{tabular}{c c c c c c c c c}
\hline
\hline
{}&$I$&$\delta I$&$PI$&$\delta PI$&$P$&$\delta P$& $\theta$&$\delta\theta$\\
{}&(mJy beam$^{-1}$)&(mJy beam$^{-1}$)&(mJy beam$^{-1}$) &(mJy beam$^{-1}$)& (\%)&(\%)&($^\circ$)&($^\circ$)\\
\hline
Mean&24.2&1.6&2.0&1.1&129.2&220.6&89.6&15.7\\ \hline
RMS&44.2&0.8&1.5&0.5&311.4&524.0&50.2&6.8\\ \hline
\end{tabular}
\end{centering}
\end{table*}

\begin{figure*}[!htb]
\centering
  \begin{tabular}{cc}
    \includegraphics[trim=1.3cm 6.2cm 3.cm 7.75cm,clip,width=5.9cm]{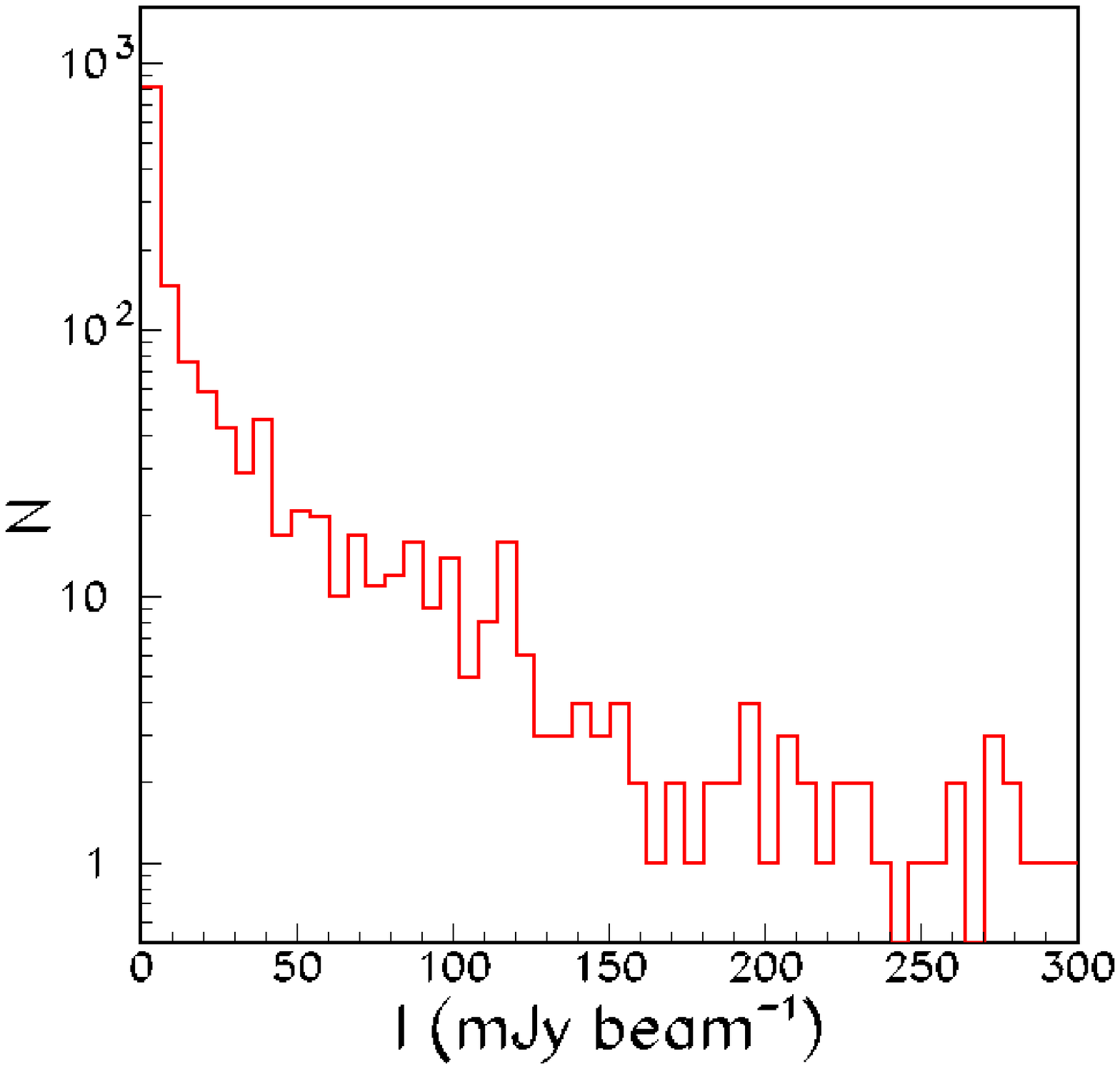}&
    \includegraphics[trim=1.3cm 6.3cm 3.cm 7.75cm,clip,width=5.9cm]{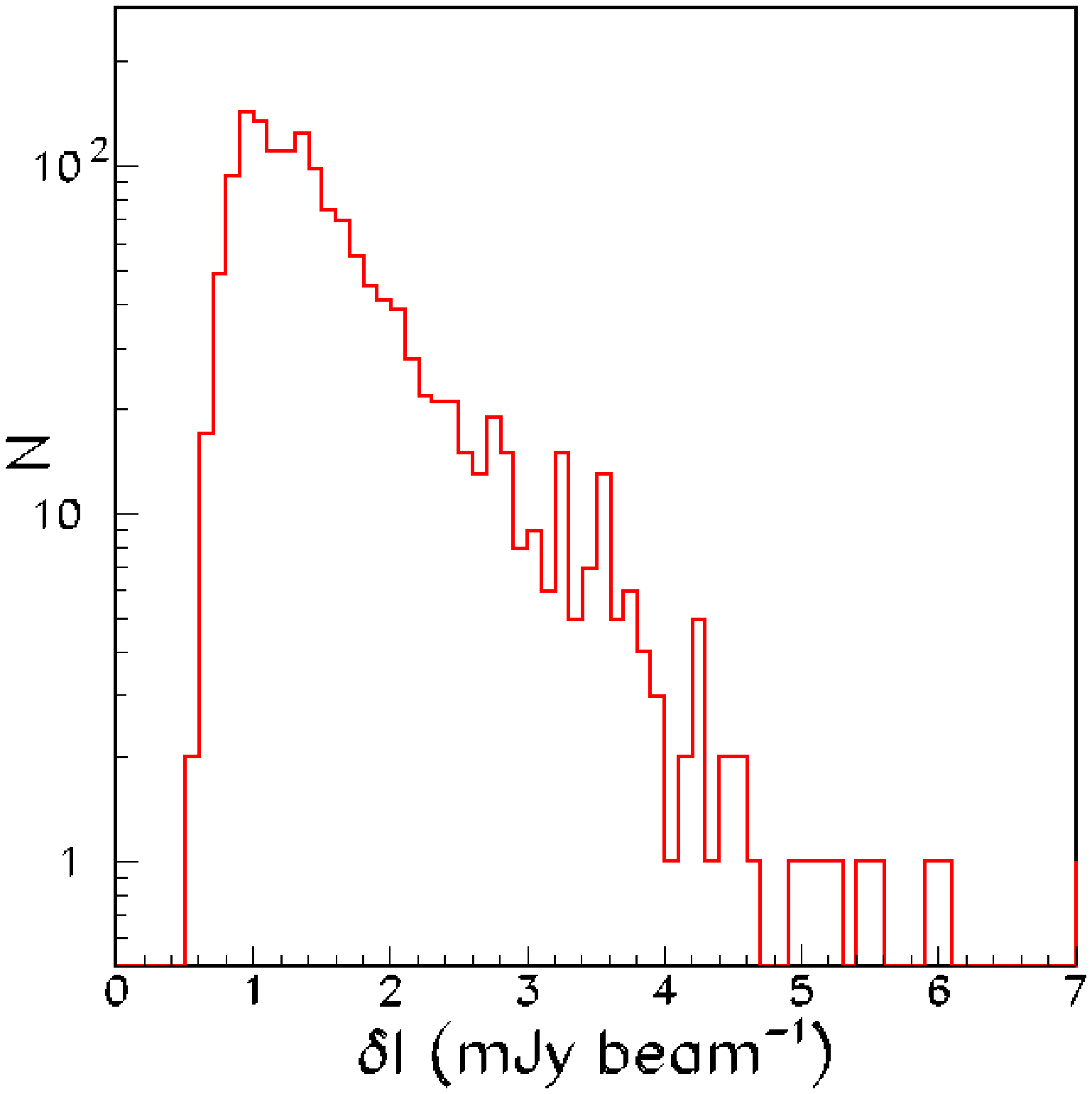}\\
    \includegraphics[trim=1.3cm 6.2cm 3.cm 7.75cm,clip,width=5.9cm]{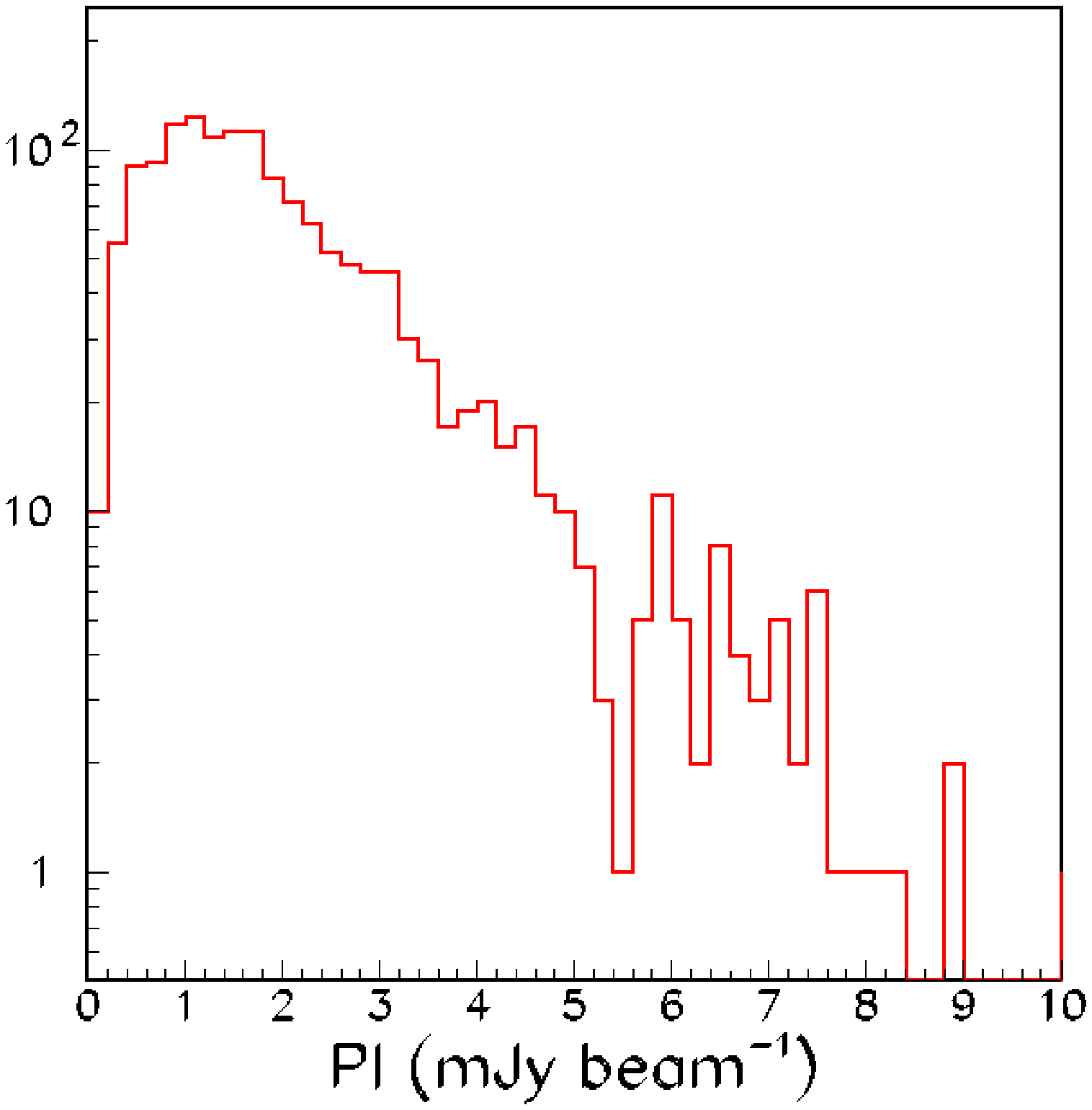}&
    \includegraphics[trim=1.3cm 6.3cm 3.cm 7.75cm,clip,width=5.9cm]{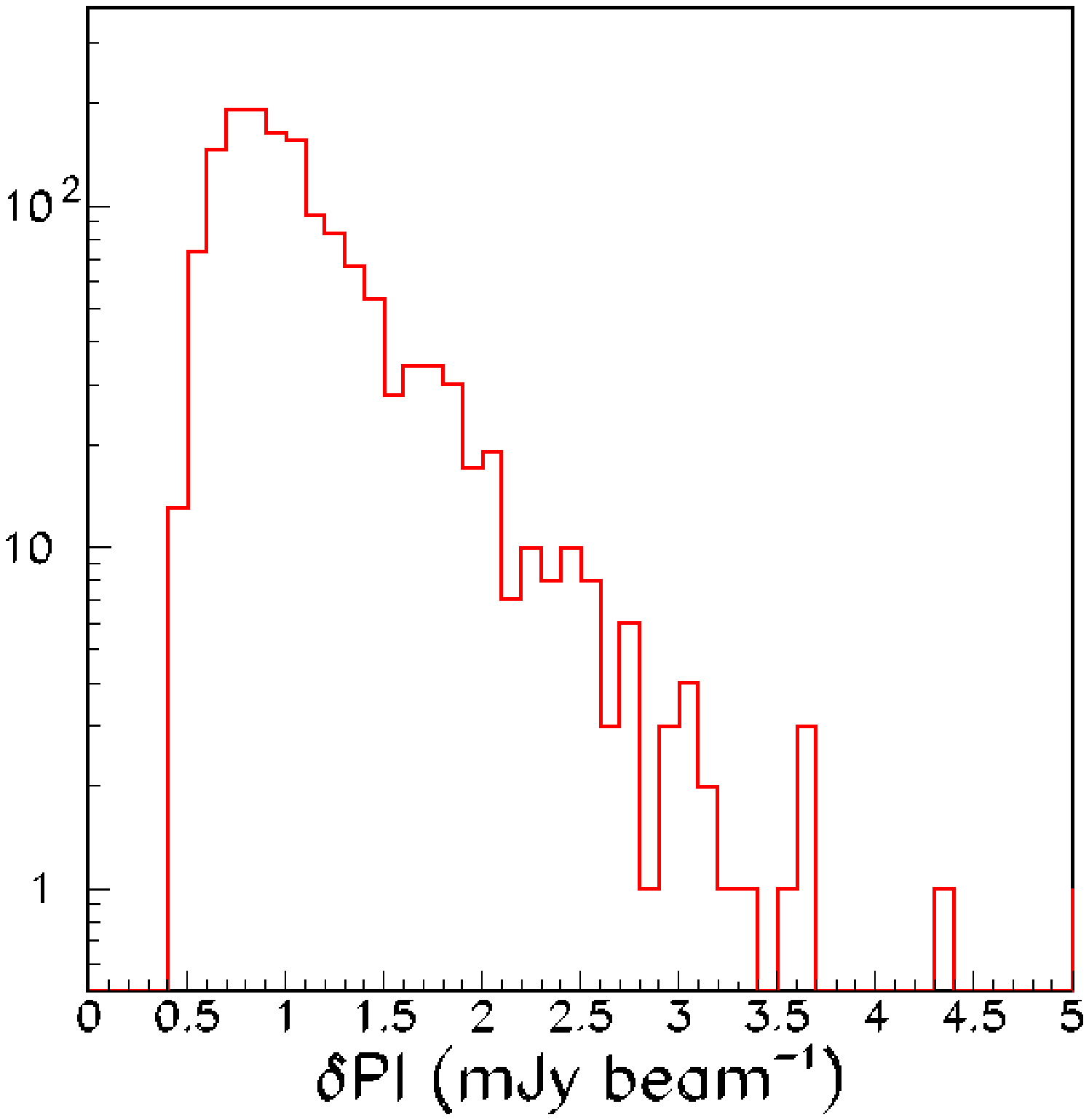}\\
    \includegraphics[trim=1.4cm 6.2cm 2.9cm 7.75cm,clip,width=5.9cm]{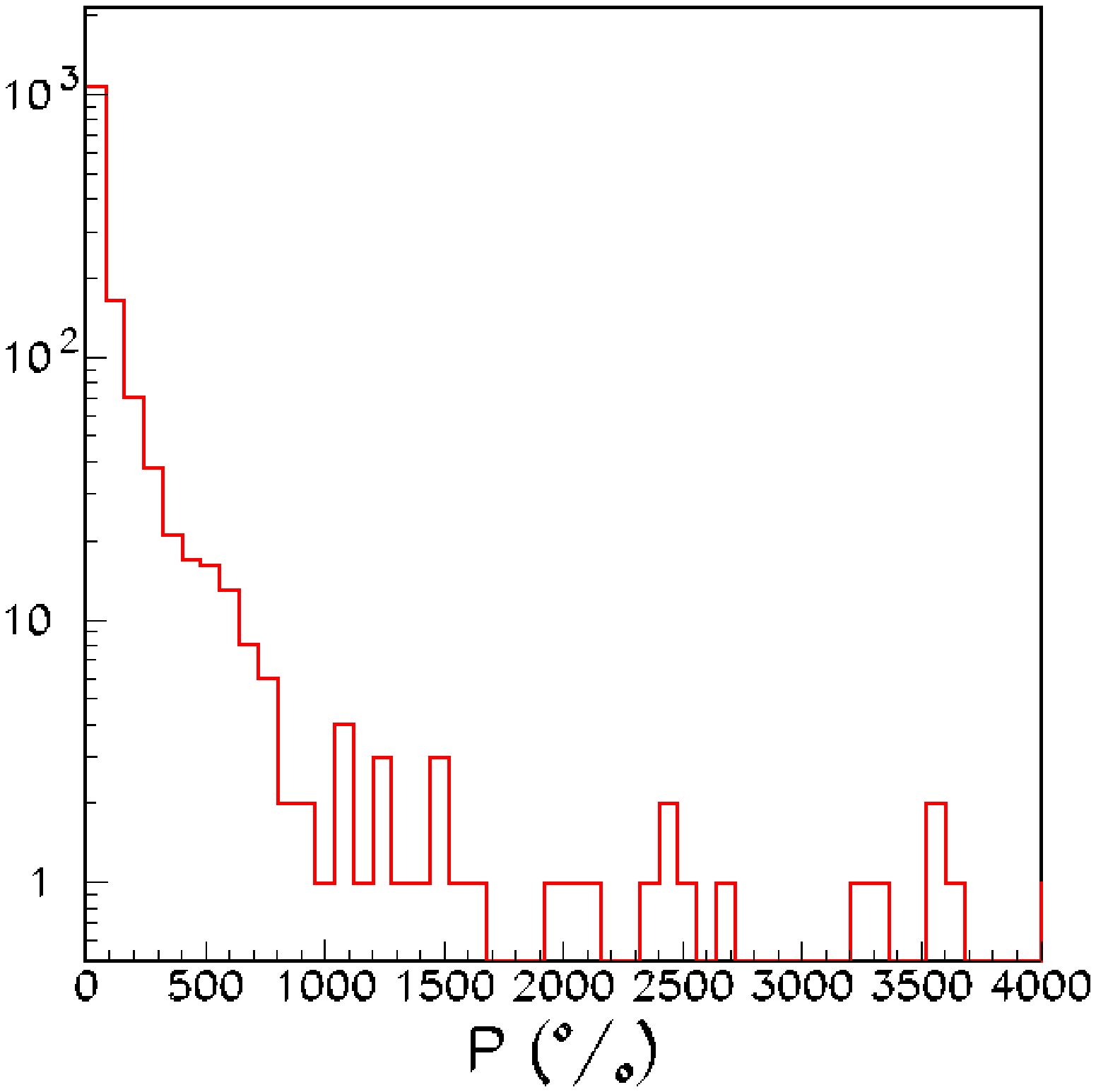}&
    \includegraphics[trim=1.4cm 6.3cm 2.9cm 7.75cm,clip,width=5.9cm]{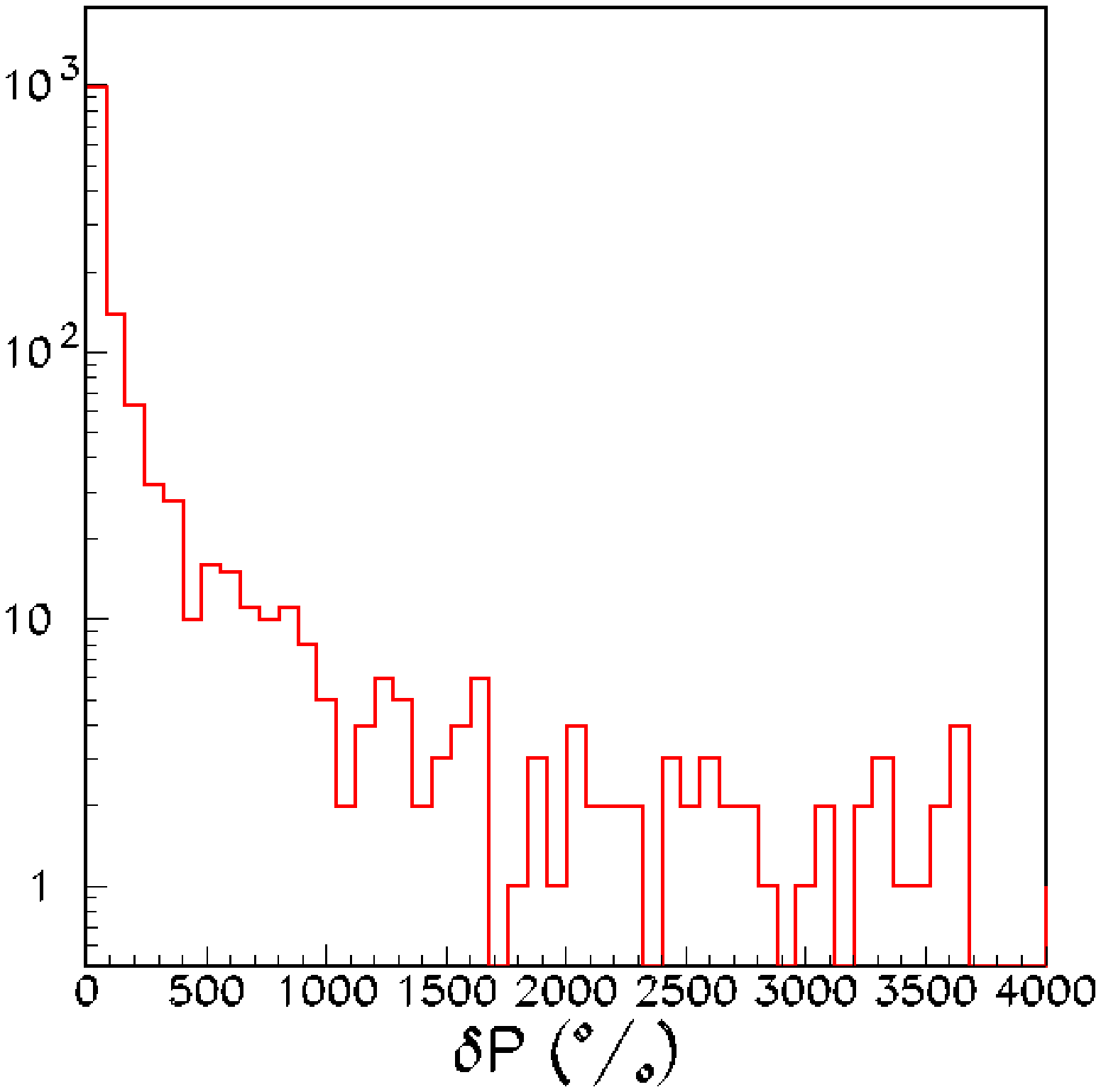}\\
    \includegraphics[trim=1.3cm 6.3cm 3.cm 7.75cm,clip,width=5.9cm]{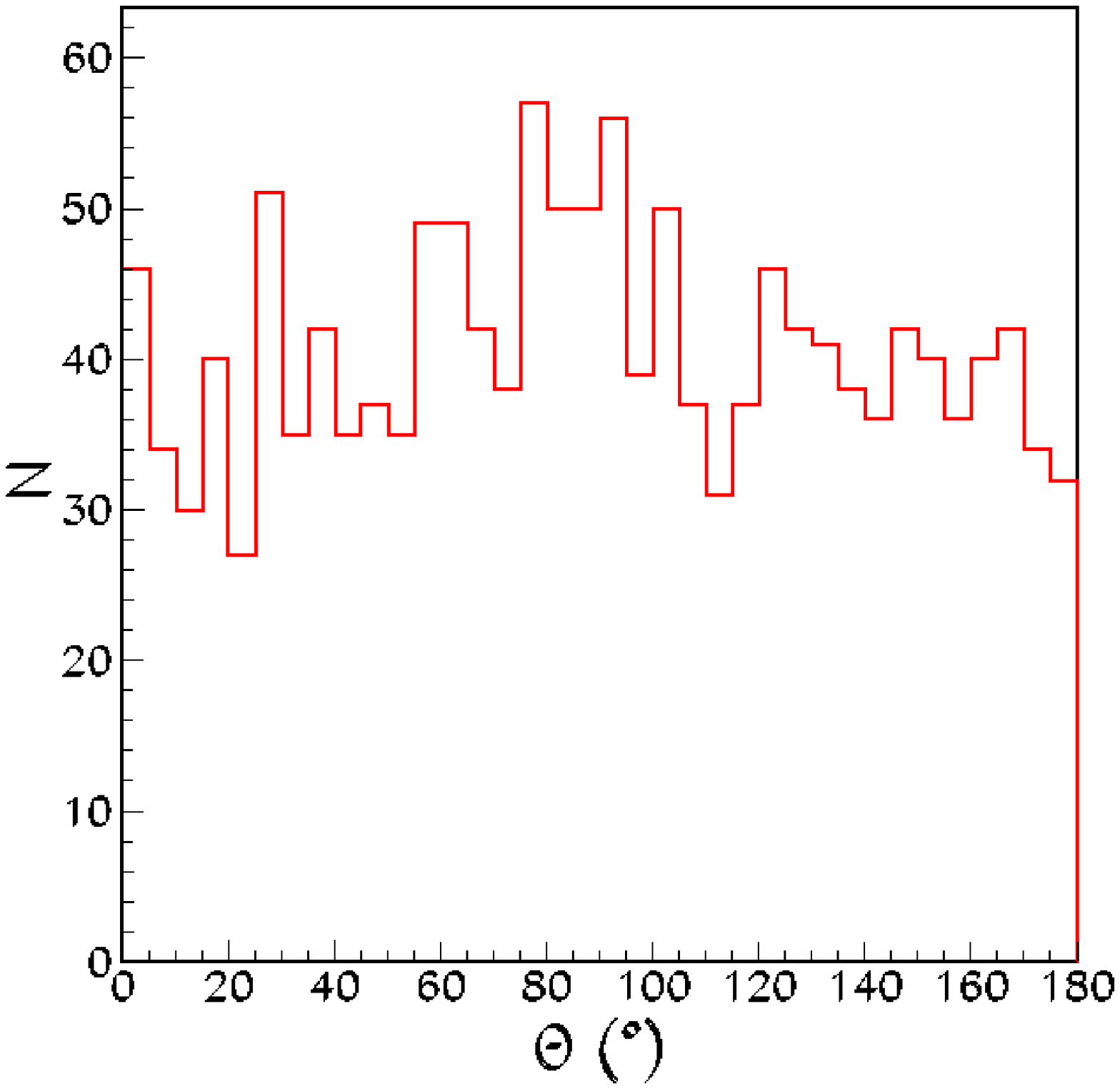}&
    \includegraphics[trim=1.3cm 6.3cm 3.cm 7.75cm,clip,width=5.9cm]{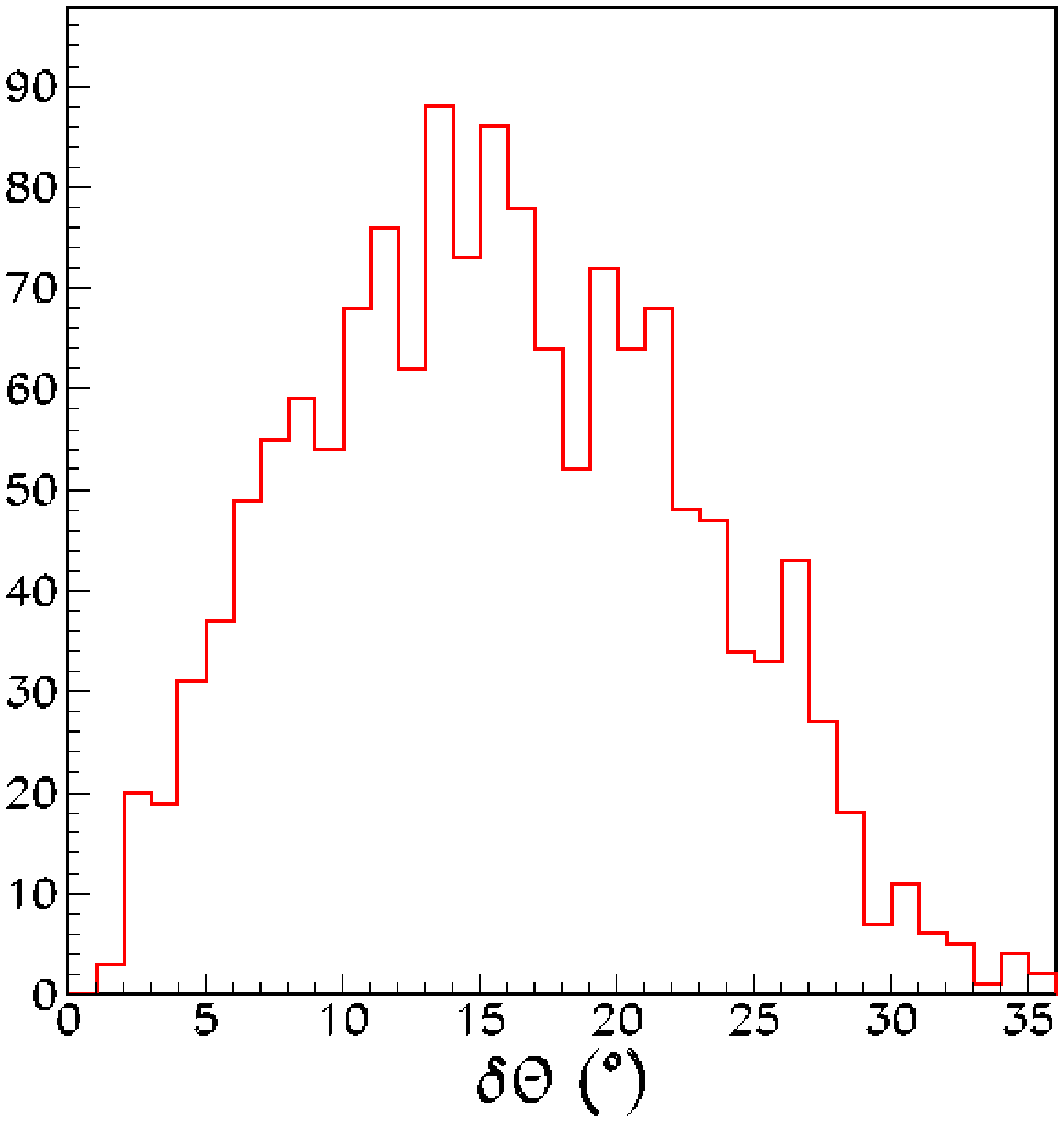}\\
  \end{tabular}
\caption{From left to right, top to bottom: distributions of $I$, $\delta I$, $PI$, $\delta PI$, $P$, $\delta P$, $\theta$, and $\delta\theta$.}\label{figa2} 
\end{figure*}

Figure \ref{figa2} displays the distributions of $I$, $PI$, $P$, $\delta I$, $\delta PI$, $\delta P$, $\theta$, and $\delta\theta$ for the raw data set, namely no cut has been applied. Their respective means and RMS values are listed in Table \ref{tab1}. Note that the presence of $P$ value larger than 100\% is caused by noise in $PI$ and $I$ (i.e. random noise spikes can result in $PI$ being larger than $I$). The total number of pixels containing data, namely pixels having $I>0$ \& $PI>0$ \& $P>0$, is 1466 out of $90\times90=8100$ pixels of the whole map.

\begin{figure*}
\centering
  \begin{tabular}{ccc}
    \includegraphics[trim=1.3cm 6.3cm 3.6cm 5.75cm,clip,width=5.5cm]{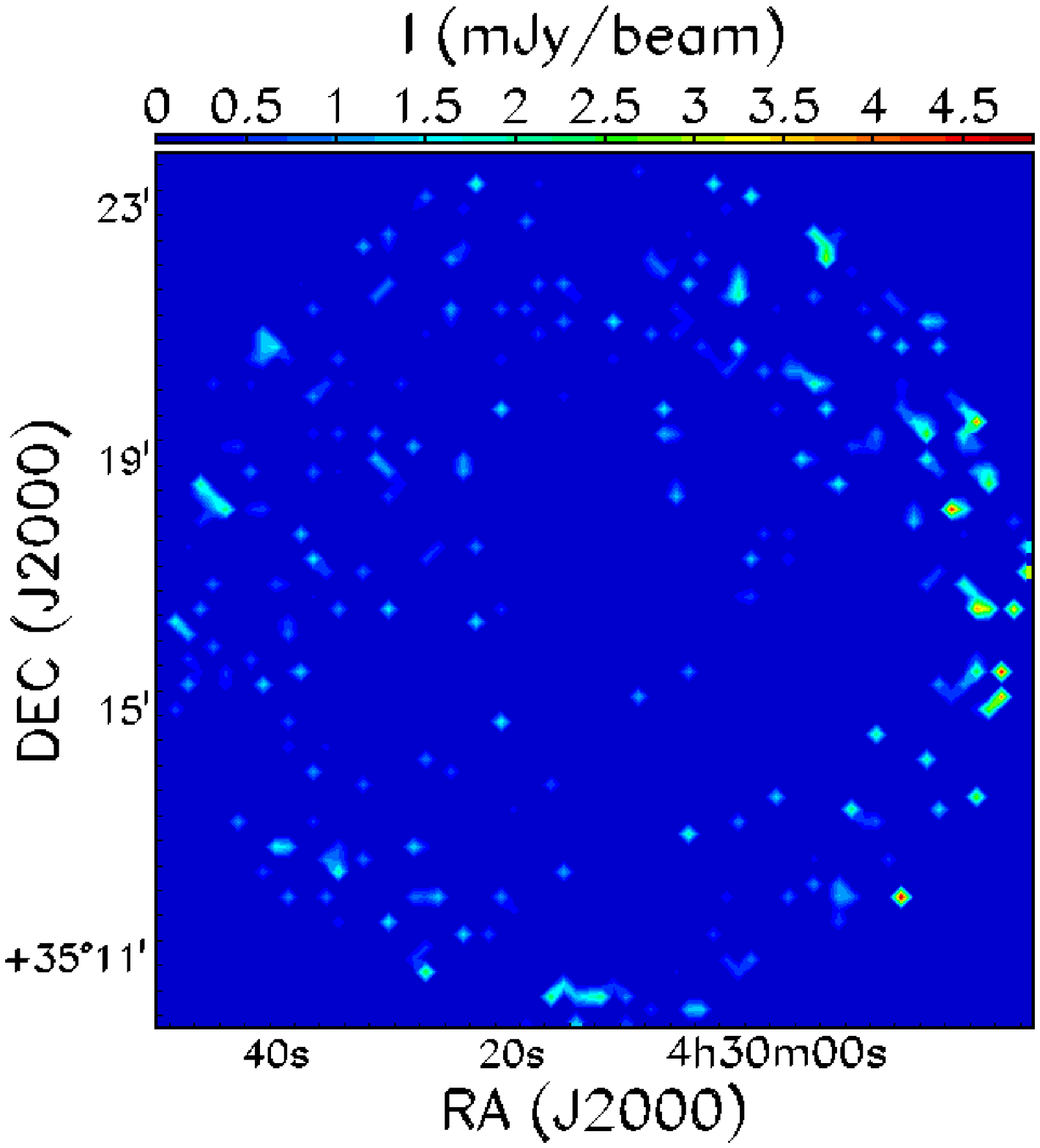}&
    \includegraphics[trim=2.cm 6.3cm 3.25cm 4.1cm,clip,width=5.5cm]{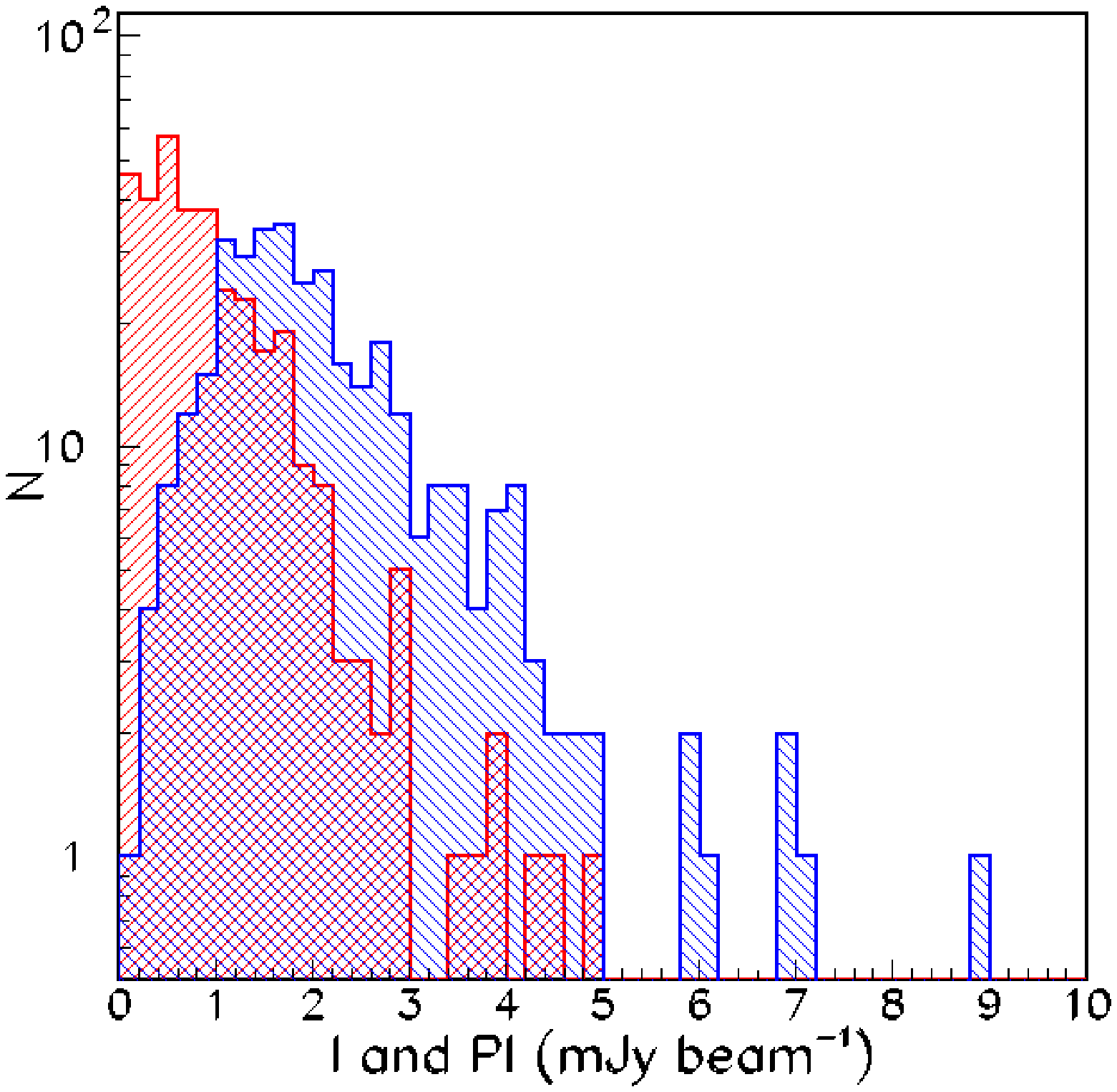}&
    \includegraphics[trim=2.5cm 6.3cm 3.45cm 4.1cm,clip,width=5.3cm]{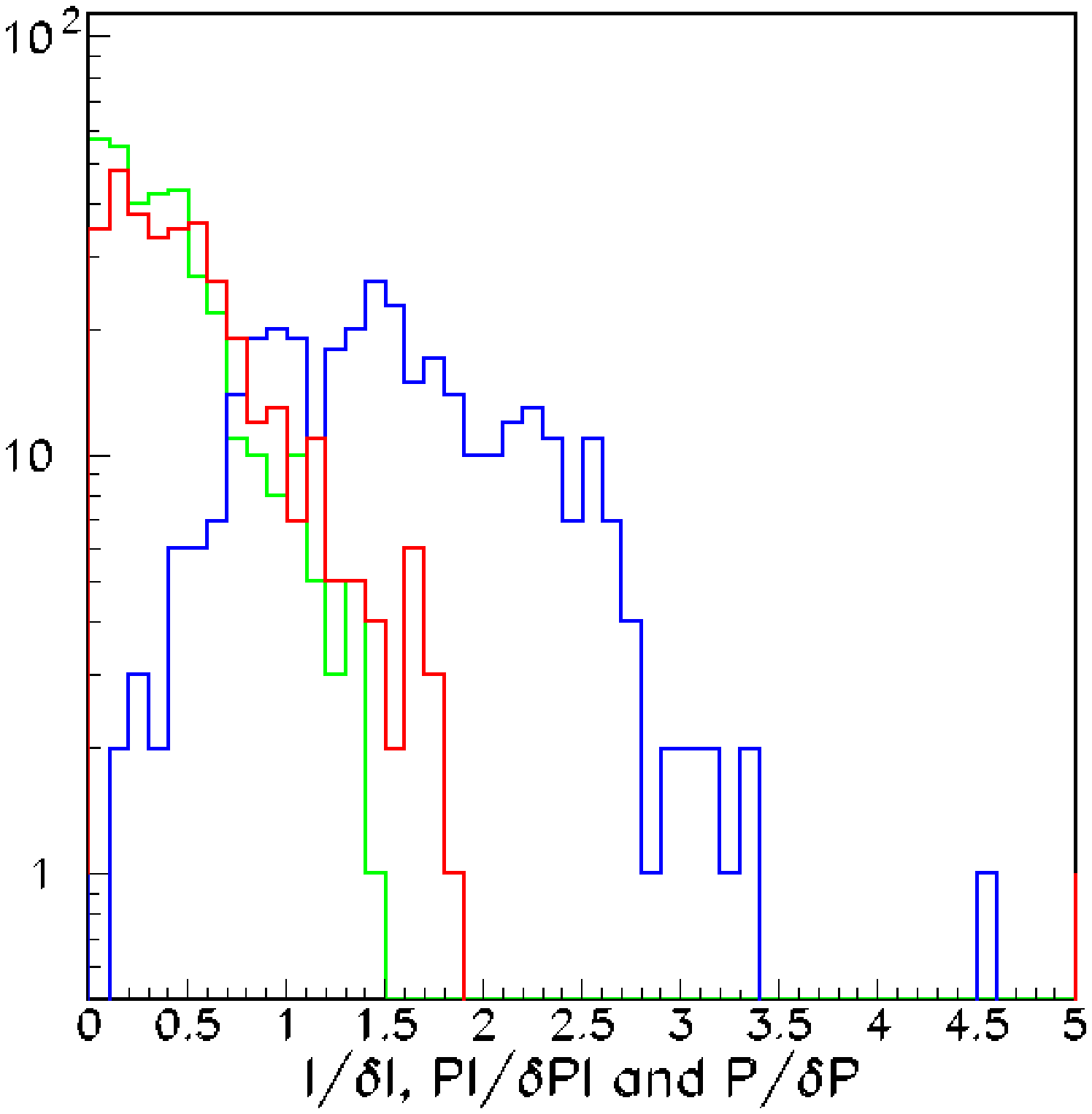}
  \end{tabular}
\caption{Bad pixels (pixels having $P>100\%$ due to low S/N or having short exposure time). Left: $I$-map; Center: distributions of $I$ (red) and $PI$ (blue); Right: distributions of $I/\delta I$ (red), $PI/\delta PI$ (blue), $P/\delta P$ (green). Bad pixels are located at the map edges (left) and have low $I$, $PI$ (center), and low S/N (right).} \label{fig4}
\end{figure*}

The mean uncertainties of the polarization angle of $\langle\delta\theta\rangle\sim16^\circ$ is average in comparison with other regions surveyed by BISTRO. It is better for Orion with $\langle\delta\theta\rangle\sim4^\circ$, Perseus B1 with $\langle\delta\theta\rangle\sim6^\circ$, \mbox{IC 5146} with $\langle\delta\theta\rangle\sim9^\circ$, Ophiuchus A with $\delta\theta<12^\circ$ but worse for Ophiuchus B and C with $20^\circ<\delta\theta<80^\circ$ and $12^\circ<\delta\theta<47^\circ$ respectively. The main factors dictating the precision of the polarization angle measurements are the weather conditions and whether the region has bright polarized emission. Our observed region is rather polarization-faint. We note that $\langle\delta\theta\rangle$ for the sources mentioned here are calculated with some data selection criteria which naturally bring the value of $\langle\delta\theta\rangle$ down. For example, in the case of Orion only pixels having $P/\delta P>5$ were kept. From the right panel of the upper row of the same figure, we can see from the distribution of $P$ that there are a number of pixels having $P>100\%$, which is unphysical. Investigating these pixels in more details, we find that most of the pixels come from low emission regions where the edge effect -high noise at the edges of the map- is important (Figure \ref{fig4} left). Moreover, they have small values of $I$ (0.93, 0.79) mJy beam$^{-1}$ and $PI$ (2.08, 1.02) mJy beam$^{-1}$ (Figure \ref{fig4} center) as well as S/Ns whose means and RMSs are (0.53, 0.40) for $I/\delta I$, (1.55, 0.67) for $PI/\delta PI$, and (0.40, 0.32) for $P/\delta P$ itself (Figure \ref{fig4} right). This result is expected and due to errors on polarization fraction (Figure \ref{figa2} middle right) being Ricean-, rather than Gaussian-distributed \citep{pattle2019}. In POL-2 observations, the central $3'$-radius region is designed to have flat S/N and coverage drops off sharply towards the map edges at radius of $6'$ \citep{friberg2016}.

\end{document}